\documentclass[preprint, superscriptaddress,amsmath, nofootinbib]{revtex4-1}
\usepackage{graphicx}
\usepackage{latexsym}
\usepackage{slashed}
\usepackage{bm}
\usepackage{color}
\usepackage[colorlinks,citecolor=blue,urlcolor=blue,linkcolor=blue]{hyperref}
\usepackage{diagbox}
\usepackage{makecell}
\pdfoutput=1
\usepackage{dcolumn}% Align table columns on decimal point
\usepackage{bm}% bold math
\usepackage{multirow}

\begin{document}

\title{Probing Dark QCD Sector through the Higgs Portal with Machine Learning at the LHC}
\def\slash#1{#1\!\!/}

\author{Chih-Ting Lu}
\email{ctlu@njnu.edu.cn}
\affiliation{Department of Physics and Institute of Theoretical Physics, Nanjing Normal University, Nanjing, 210023, P. R. China}
\affiliation{CAS Key Laboratory of Theoretical Physics, Institute of Theoretical Physics, Chinese Academy of Sciences, Beijing 100190, P. R. China}

\author{Huifang Lv}
\email{lvhf@njnu.edu.cn}
\affiliation{Department of Physics and Institute of Theoretical Physics, Nanjing Normal University, Nanjing, 210023, P. R. China}

\author{Wei Shen}
\email{shenwei@itp.ac.cn}
\affiliation{CAS Key Laboratory of Theoretical Physics, Institute of Theoretical Physics, Chinese Academy of Sciences, Beijing 100190, P. R. China}
\affiliation{School of Physical Sciences, University of Chinese Academy of Sciences, Beijing 100049, P. R. China}	

\author{Lei Wu}
\email{leiwu@njnu.edu.cn}
\affiliation{Department of Physics and Institute of Theoretical Physics, Nanjing Normal University, Nanjing, 210023, P. R. China}

\author{Jia Zhang}
\email{jiazhang@njnu.edu.cn}
\affiliation{Department of Physics and Institute of Theoretical Physics, Nanjing Normal University, Nanjing, 210023, P. R. China}

\begin{abstract}

The QCD-like dark sector with GeV-scale dark hadrons has the potential to generate new signatures at the Large Hadron Collider (LHC). In this paper, we consider a singlet scalar mediator in the tens of GeV-scale that connects the dark sector and the Standard Model (SM) sector via the Higgs portal. We focus on the Higgs-strahlung process, $q\overline{q}'\rightarrow W^{\ast}\rightarrow WH $, to produce a highly boosted Higgs boson. Our scenario predicts two different processes that can generate dark mesons: (1) the cascade decay from the Higgs boson to two light scalar mediators and then to four dark mesons; (2) the Higgs boson decaying to two dark quarks, which then undergo a QCD-like shower and hadronization to produce dark mesons. We apply machine learning techniques, such as Convolutional Neural Network (CNN) and Energy Flow Network (EFN), to the fat-jet structure to distinguish these signal processes from large SM backgrounds. We find that the branching ratio of the Higgs boson to two light scalar mediators can be constrained to be less than about $10\%$ at $14$ TeV LHC with $\mathcal{L} = 3000 fb^{-1}$.
\end{abstract}

\maketitle

\tableofcontents
\newpage

\section{Introduction} 

The existence of dark matter (DM) is now a well-established fact, with its gravitational effects providing compelling evidence of its presence in the universe~\cite{Frenk:2012ph}. Despite decades of research, the identity of DM remains one of the greatest mysteries of modern physics. A wide variety of theoretical models have been proposed to explain the nature of DM, ranging from weakly interacting massive particles (WIMPs)~\cite{Arcadi:2017kky} to axions~\cite{Chadha-Day:2021szb} and other exotic candidates~\cite{Battaglieri:2017aum}. The experimental search for DM has also been a major focus of many research efforts, utilizing a wide range of techniques from direct detection and collider experiments to indirect detection methods~\cite{Schumann:2019eaa,Boveia:2018yeb,Gaskins:2016cha}.

Despite many theoretical models that have been proposed, the search for DM has primarily focused on the WIMP paradigm. This approach has been guided by the observation that WIMPs naturally arise in many well-motivated extensions of the standard model (SM) of particle physics~\cite{Feng:2010gw,Roszkowski:2017nbc}. However, the lack of any conclusive evidence of WIMP DM has led to increasing interest in alternative models. 
Especially, more complex dark sectors can exhibit rich phenomenology such that they attract more and more people's attention~\cite{Asadi:2022njl,Albouy:2022cin}. The dark sector is an intriguing field of study in the context of DM models, and the dark QCD model with gauge symmetry $SU(N)_d$ (where $N\geq 2$) is one such example~\cite{Strassler:2006im,Bai:2013xga,Hochberg:2014kqa}. In dark QCD models, DM is composed of particles that interact strongly with a new dark gauge sector, which is modeled after the familiar QCD sector. This kind of models have garnered significant interest as they have the potential to connect to matter-antimatter asymmetry~\cite{Bai:2013xga,Lonsdale:2018xwd,Zhang:2021orr,Bottaro:2021aal,Ibe:2021gil} and trigger a first-order phase transition in the early universe~\cite{Schwaller:2015tja,Tsumura:2017knk,Aoki:2017aws,Reichert:2021cvs,Hall:2019rld,Hall:2021zsk}. Additionally, the strong interaction between particles in the dark sector can provide insights into astrophysical issues at small scales~\cite{Hochberg:2014kqa,Tulin:2017ara}. In this study, the dark QCD sector, which is not charged under the SM gauge symmetry, can be connected to the SM sector via the Higgs portal~\cite{Strassler:2006ri}, which is a simple real singlet scalar. Other portals that connect the SM sector with the dark QCD sector include the dark photon~\cite{Hochberg:2014kqa}, $Z'$~\cite{Strassler:2006im}, $t$-channel mediators~\cite{Bai:2013xga} and so on.

The dark QCD model can be roughly classified into two well-known types based on the mass spectrum of the mediator and the dark QCD sector. The first is the strongly interacting massive particle (SIMP) model~\cite{Hochberg:2014kqa}, where the mediator and dark mesons belong to the sub-GeV scale. In this scenario, the DM annihilation process primarily occurs through the $3\rightarrow 2$ number-changing process or the forbidden channel~\cite{Hochberg:2014kqa,Berlin:2018tvf,Bernreuther:2019pfb}. The second model is the Hidden Valley model~\cite{Strassler:2006im}, where the mediator is much heavier than dark hadrons. If the heavy mediator is produced at the Large Hadron Collider (LHC), it can directly decay into dark quarks due to its higher mass compared to the dark confinement scale $\Lambda_d$. The energetic dark quark then undergoes a QCD-like shower and hadronization to produce dark hadrons. If the dark hadron in the final state is long-lived particle at the collider experiment scale and then decays back to SM particles, the dark shower will generate the novel emerging jet signature~\cite{Schwaller:2015gea,Renner:2018fhh,CMS:2018bvr,Mies:2020mzw,Knapen:2021eip,Linthorne:2021oiz,Archer-Smith:2021ntx,Albouy:2022cin}. If there are both stable and unstable dark hadrons in the final state, the dark shower will generate another special signature called the semivisible jet~\cite{Cohen:2015toa,Cohen:2017pzm,Beauchesne:2017yhh,Beauchesne:2018myj,Bernreuther:2019pfb,Cohen:2020afv,Bernreuther:2020vhm,Kar:2020bws,Bernreuther:2020xus,Knapen:2021eip,Beauchesne:2021qrw,Canelli:2021aps,CMS:2021dzg,Bernreuther:2022jlj,Finke:2022lsu,Albouy:2022cin,Cazzaniga:2022hxl,Kar:2022hxn,Faucett:2022zie,Buckley:2022zry,Beauchesne:2022phk,Pedro:2023sdp}. 
If all dark hadrons promptly decay into SM particles, the resulting dark shower produces a distinctive signature known as a "dark jet" \cite{Park:2017rfb, Mies:2020mzw, Knapen:2021eip, Buss:2022lxw}. This occurs when energetic dark partons undergo showering and subsequent hadronization, forming a cluster of dark mesons. These dark mesons then decay into observable particles, generating discernible signals resembling conventional jets. 
Finally, if all dark hadrons in the final state are stable, the dark shower is hard to be distinguished from the WIMP signature at the LHC. However, there is a lack of research in the literature on the tens of GeV scale mediators in the above classification. Therefore, in this work, we aim to explore this scenario and identify relevant signal signatures at the LHC.

This study focuses on the Higgs-strahlung process, $q\overline{q}'\rightarrow W^{\ast}\rightarrow WH $, to produce a highly boosted Higgs boson. This process has previously been utilized to explore the bottom quark Yukawa coupling~\cite{Butterworth:2008iy,ATLAS:2022buf,CMS:2022exo} and other Higgs boson exotic decays~\cite{Falkowski:2010hi,Bellazzini:2010uk,Englert:2011iz,Lewis:2012pf,Jung:2021tym}. Here, we investigate two distinct processes for generating dark mesons in the exotic decays of the Higgs boson. The first process involves the cascade decay of the Higgs boson to two light scalar mediators, which then decay to produce four dark mesons. The second process involves the Higgs boson decaying into two dark quarks that subsequently undergo a QCD-like shower and hadronization, resulting in the production of dark mesons. The main difference between these two signal signatures is that the second process can generate stable or unstable dark vector mesons in the final state. On the other hand, we focus on the most challenging case for the prompt decay of the unstable dark meson in the final state. For unstable dark mesons with a mass around $5$ GeV, they mainly decay into a pair of charm quarks or tau leptons. Therefore, we anticipate the presence of multiple charm quarks and tau leptons in the final state from these two processes, which we cluster together as a fat-jet originating from the highly boosted Higgs boson.

While the charged lepton from the Higgs-strahlung process serves as a good initial filter for identifying event candidates, distinguishing the signal from significant SM backgrounds remains challenging, as these backgrounds can also produce similar signatures. To address this issue, we can utilize jet substructure observables~~\cite{Butterworth:2008iy,Almeida:2008yp,Ellis:2009su,Kribs:2009yh,Chen:2010wk,Falkowski:2010hi} as one method to identify signal events and suppress background events.  However, machine learning techniques are also powerful tools for this type of analysis~~\cite{Komiske:2016rsd,Lin:2018cin,Lee:2019cad,Guo:2020vvt,Khosa:2021cyk,Ren:2021prq,Jung:2021tym,Chigusa:2022svv}. The Convolutional Neural Network (CNN) technique has previously been used for jet image analysis~\cite{Cogan:2014oua,deOliveira:2015xxd} and has proven more effective than jet substructure observables in certain situations~\cite{Komiske:2016rsd,Cogan:2014oua,Lee:2019cad,Gallicchio:2012ez,Larkoski:2017jix}. Additionally, the Energy Flow Network (EFN) is another useful technique for discriminating jet structures. The EFN can be used to distinguish the jet structure because it takes into account the detailed energy flow patterns within the jets, including both the local and global features, which provides a more comprehensive representation of the jet structure compared to traditional methods that only use kinematic variables~\cite{Komiske:2018cqr}.  To analyze the fat-jet structure, identify signal signatures, and distinguish signal events from relevant background events, we employ both the CNN and EFN techniques in this work.

This paper is organized as follows. We first introduce the Higgs portal dark QCD model and consider relevant constraints in Sec.~\ref{sec:model}. The signal and background processes, event selections as well as the machine learning techniques are discussed in Sec.~\ref{sec:ml}. We summarize and discuss our numerical results in Sec.~\ref{sec:result}. Finally, we conclude our findings in Sec.~\ref{sec:conclu}.

\section{Model and Constraints} 
\label{sec:model}
 
In the dark sector, there are multiple ways for assigning dark quark species and gauge symmetry structures. For this study, however, we opt to focus on a single flavor scalar dark quark field $\varphi_d$ with gauge symmetry $SU(3)_d$ to simplify the dark QCD sector, without compromising generality. In addition, to connect the dark QCD sector and the SM sector, we utilize both an extra real singlet scalar field $\phi_s$ and a SM-like Higgs doublet field $\phi_H$ as mediators. The Lagrangian for this model is given by

\begin{equation}
\mathcal{L} = \mathcal{L}_{\text{SM}} + \frac{1}{2} \partial_{\mu}\phi_s \partial^{\mu}\phi_s + D_{\mu}\varphi_d^{\dagger} D^{\mu}\varphi_d - V(\phi_H,\phi_s,\varphi_d),
\end{equation}
where ${\cal L}_{\text{SM}}$ represents the part for all SM interactions, and $D_{\mu} = \partial_{\mu} - ig_d G^d_{\mu}$ with $g_d$ and $G^d_{\mu}$ denoting the $SU(3)_d$ gauge coupling and gauge field, respectively. The general scalar potential $V(\phi_H,\phi_s,\varphi_d)$ can be written as, 
\begin{equation}
    V(\phi_H,\phi_s,\varphi_d) = V_{H}(\phi_H) + V_{s}(\phi_s) + V_{\varphi}(\varphi_d) + V_{Hs\varphi}(\phi_H,\phi_s,\varphi_d),  
\label{eq:potential}    
\end{equation} 
where 
\begin{eqnarray}
V_{H}(\phi_H)&=&\mu^2_H\phi_H^{\dagger}\phi_H +\frac{\lambda_H}{2}(\phi_H^{\dagger}\phi_H)^2, \\
V_{s}(\phi_s)&=&\mu_1^3 \phi_s +\frac{\mu_{s}^2}{2}\phi_s^2 +\frac{\mu_3}{3!}\phi_s^3 +\frac{\lambda_s}{4!}\phi_s^4, \\ 
V_{\varphi}(\varphi_d)&=&\mu^2_{\varphi}\varphi_d^{\dagger}\varphi_d +\frac{\lambda_{\varphi}}{2}(\varphi_d^{\dagger}\varphi_d)^2, \\ 
V_{Hs\varphi}(\phi_H,\phi_s,\varphi_d) &=&  \mu_{Hs}\phi_s\phi_H^{\dagger}\phi_H +\frac{\lambda_{Hs}}{2}\phi_s^2\phi_H^{\dagger}\phi_H + \mu_{s\varphi}\phi_s\varphi_d^{\dagger}\varphi_d \\ \nonumber &&+\frac{\lambda_{s\varphi}}{2}\phi_s^2\varphi_d^{\dagger}\varphi_d +\lambda_{H\varphi}\varphi_d^{\dagger}\varphi_d\phi_H^{\dagger}\phi_H, 
\end{eqnarray}
where $\mu_i$'s are parameters with the same dimension as mass and $\lambda_i$'s are dimensionless parameters. The $\phi_H$ and $\phi_s$ are expanded with their vacuum expectation values (VEVs) as 
\begin{equation}
\phi_H = \frac{1}{\sqrt 2} 
\left(
\begin{tabular}{c}
0
\\
$v + h$
\end{tabular}
\right)
\;\;\; , \;\;\; 
\phi_s = v_s + h_s.
\label{eq:expand}
\end{equation}
Because the scalar potential in Eq.~(\ref{eq:potential}) is invariant under a shift of the singlet scalar filed VEV by $v_s\rightarrow v'_s$, we take $v_s = 0$ without any loss of generality~\cite{Chen:2014ask}.

After the electroweak symmetry breaking (EWSB), $h$ and $h_s$ mix with each other via the mixing angle, $\theta$, to form the intermediate mass eigenstates, $h_1$ and $h_2$. Moreover, in the infrared (IR) region, we define the scalar meson $\tilde{\eta}$ as the bound state $\varphi_d\varphi^{\dagger}_d$ in the dark QCD sector. Following the mapping $\varphi_d\varphi^{\dagger}_d\rightarrow\Lambda_d\tilde{\eta}$, as described in Ref.~\cite{Knapen:2021eip}, the relevant terms in the Lagrangian involving $\tilde{\eta}$ are given by: 
\begin{align}
{\cal L}_{\text{IR}} \supset {}& 
\frac{\lambda_{\varphi}}{2}\Lambda^2_d\tilde{\eta}^2 + \left(\mu_{s\varphi}\sin\theta +\lambda v\cos\theta\right)\Lambda_d h_1\tilde{\eta} + \left(\mu_{s\varphi}\cos\theta +\lambda v\sin\theta\right)\Lambda_d h_2\tilde{\eta}
\nonumber  \\
& + \frac{1}{2}\left(\lambda_{s\varphi}\sin^2\theta +\lambda_{H\varphi}\cos^2\theta\right)\Lambda_d h^2_1\tilde{\eta} + \frac{1}{2}\left(\lambda_{s\varphi}\cos^2\theta +\lambda_{H\varphi}\sin^2\theta\right)\Lambda_d h^2_2\tilde{\eta}
\nonumber  \\ 
& + \left( \lambda_{s\varphi}-\lambda_{H\varphi}\right)\sin\theta\cos\theta\Lambda_d h_1 h_2\tilde{\eta}. 
\end{align}
It should be noted that $\tilde{\eta}$ further mixes with both $h_1$ and $h_2$ in the low-energy regime, resulting in the final mass eigenstates denoted as ${\eta_d}$, $h'_1$ and $h'_2$ in the IR region. Therefore, we will use the final mass eigenvalues, $M_{h'_1}$, $M_{h'_2}$ and $M_{\eta_d}$, as variables in the following analysis.

Due to the mixing between $\tilde{\eta}$ and both $h_1$ and $h_2$, the ${\eta_d}$ is inherently unstable. In our analysis, we incorporate the mixing effects by denoting the factor $\sin\Psi$. 
It is challenging to individually constrain each parameter within our model. Therefore, let's instead focus on directly constraining the mixing parameter $\sin\Psi$ as a whole.
While $\sin\Psi$ can be represented as a complicated function of the model parameters in Eq.~(\ref{eq:potential}), individually constraining each parameter is challenging. 
%Consequently, we directly consider the limit of $\sin\Psi$ in our analysis. 
The partial decay widths of ${\eta_d}$ are given by:
\begin{align}
 & \Gamma (\eta_d\rightarrow f\overline{f}) = \sin^2\Psi \frac{N_c M^2_f M_{\eta_d}}{8\pi v^2}\left( 1-\frac{4m^2_f}{M^2_{\eta_d}}\right)^{3/2}, 
\nonumber  \\
& \Gamma (\eta_d\rightarrow gg) = \sin^2\Psi \frac{\alpha^2_s M^3_{\eta_d}}{32\pi^3 v^2}\bigg | \sum_i\tau_i\left[ 1+\left( 1-\tau_i\right)f(\tau_i)\right]\bigg |^2,
\end{align} 
where $f$ and $g$ are the SM fermion and gluon fields, $N_c$ equals $1$ for leptons and $3$ for quarks, $\tau_i = 4M^2_i/M^2_{\eta_d}$ and the loop function $f(\tau)$ can be defined as~\cite{Gunion:1989we}: 
\begin{align}
f(\tau) = & \left[\arcsin\left(\sqrt{\frac{1}{\tau}}\right)\right]^2,\quad\quad\quad\text{for } \tau \geq 1, \\ 
& -\frac{1}{4}\left[\log\left(\frac{\eta_+}{\eta_-}\right)-i\pi\right]^2,\quad\text{for } \tau < 1, 
\end{align} 
with \(\eta_{\pm} = 1 \pm \sqrt{1-\tau}\). 

In this study, we investigate the scalar dark meson $\eta_d$ as the lightest bound state. Moreover, we assume the stability of the vector dark meson $\tilde{\omega}$ and set its mass to $M_{\tilde{\omega}} = 1.8 M_{\eta_d}$\footnote{As pointed out in Ref.~\cite{Knapen:2021eip}, we don't discuss the scalar dark meson $\tilde{\sigma}$ here because it's not included in the Hidden Valley module of Pythia8 and cannot be simulated from the dark showers and hadronization. On the other hand, we don't consider $\tilde{\omega}\rightarrow\eta_d\eta_d$ since this decay mode in our simplified model will violate Bose symmetry and angular momentum conservation.}.
Note the dark baryon can be much heavier than $\eta_d$, $\tilde{\omega}$ and it can also be the DM candidate thanks to the accidental symmetry from the global dark baryon number conservation~\cite{Bai:2013xga}.

\begin{figure}[h]
\centering 
\includegraphics[height=8cm,width=16cm]{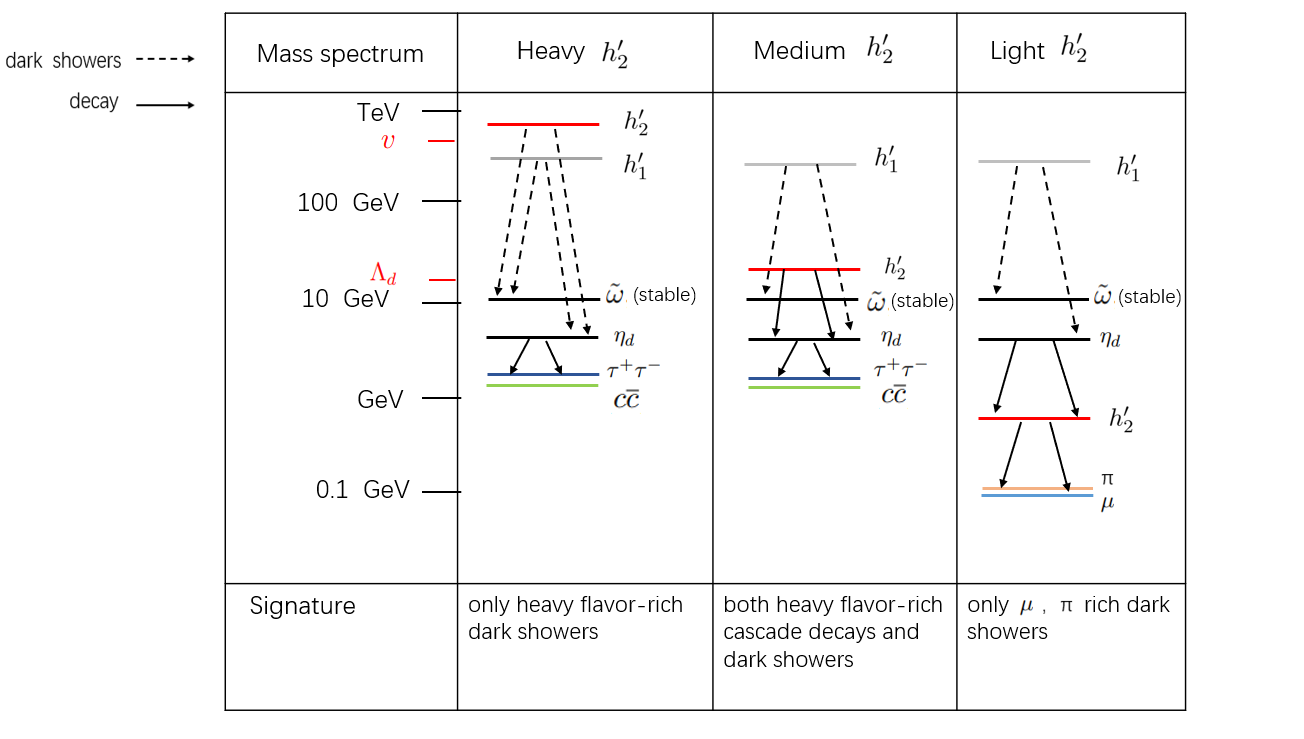}
\caption{Three different scenarios in the mass spectrum of the Higgs portal dark QCD model : Heavy $h'_2$, Medium $h'_2$ and Light $h'_2$ as well as possible signal signatures at the LHC.}
\label{diffh2}
\end{figure}

We first investigate three different scenarios in the mass spectrum as a general study, namely, Heavy $h'_2$, Medium $h'_2$, and Light $h'_2$, and their respective signal signatures at the LHC as depicted in Fig.~\ref{diffh2}. In the Heavy $h'_2$ scenario, where $M_{h'_2} > M_{h'_1}$, only heavy flavor-rich dark showers are produced from $h'_1$ and $h'_2$, as previously discussed in Ref.~\cite{Knapen:2021eip}. For the Medium $h'_2$ scenario, where $M_{h'_1} > 2M_{h'_2}$ and $M_{h'_2} > 2M_{\eta_d}$, both heavy flavor-rich cascade decays and dark showers can be generated from $h'_1$. Finally, for the Light $h'_2$ scenario, where $M_{\eta_d} > 2M_{h'_2}$, only dark showers with multiple muons and pions in the final state are produced from $h'_1$. This special signature has also been studied in Ref.~\cite{Born:2023vll}. In this work, we focus on the Medium $h'_2$ scenario as it provides a rich variety of signals at the LHC. We will provide a comprehensive analysis of the associated signatures of this scenario in terms of various kinematic distributions, background estimations, and event selection efficiencies.

\begin{table}[th!]
\begin{tabular}{|c|c|c|}
\hline
Scale &Particles & Processes \\
\hline
\multirow{2}*{$\Lambda_d < E\lesssim v$} & \multirow{2}*{$h'_1$, $h'_2$, $\varphi_d$} & $h'_1$, $h'_2$ productions \\
 & & $h'_1$ decays \\
\hline
\multirow{2}*{$M_{\eta_d}\leq E\lesssim\Lambda_d$} & \multirow{2}*{ $\tilde{\omega}$, $\eta_d$} & $\tilde{\omega}$, $\eta_d$ productions \\
 & & $h'_2$, $\eta_d$ decays \\
\hline
\end{tabular}
\caption{The SM-like higgs boson and new particles in different energy scales for the scalar-mediated dark QCD model.}\label{Tab:scale}
\end{table}

\begin{table}[h!]
\vspace{1.0mm}
\begin{ruledtabular}
 \begin{tabular}{ l c c c c }
 ~$M_{\eta_d}$ (GeV)~ & $Br(c\overline{c})$ & $Br(\tau^+\tau^-)$ & $Br(gg)$ & $\tau_{\eta_d}$ (s) \\ \hline
 ~$4$~ & $ 36.44\% $ & $ 46.93\% $ & $ 16.74\% $ & ~$3.85\times 10^{-13}$~ \\ \hline
 ~$6$~ & $ 56.88\% $ & $ 40.08\% $ & $ 3.03\% $ & ~$4.05\times 10^{-14}$~ \\
\end{tabular}
\end{ruledtabular} 
\caption{\small  \label{tab:BR-eta}
The branching ratio for the most three dominant decay modes of $\eta_d$ for $M_{\eta_d} = 4$, $6$ GeV. Here the lifetime of $\eta_d$ is calculated by using $\sin\Psi = 10^{-3}$ which is still allowed from the existing bounds. }
\end{table}

To consider a mediator at the tens of GeV scale, we focus on the light singlet-like scalar with a mass range of $10$ GeV $< M_{h'_2}\lesssim 60$ GeV and its potential detection at the LHC. By setting $M_{h'_2} > 10$ GeV, we ensure that direct detection of $h'_2$ can only occur at high energy colliders, rather than at B factories~\cite{BaBar:2001yhh,Belle-II:2018jsg} or BESIII~\cite{Asner:2008nq}. We assume that $M_{h'_2} = 5M_{\eta_d}$, where $M_{\eta_d}\geq 2$ GeV. 

This assumption ensures that $h'_2$ can decay on-shell to a pair of $\eta_d$. Moreover, we also assume $M_{h'_2}\sim\Lambda_d$ such that $h'_2$ will decay to dark mesons, rather than to dark gluons or dark quarks in the final state. This is a unique feature of this work in comparison with previous studies on heavy mediators in the Hidden Valley models~\cite{Strassler:2006im}. Additionally, we set $M_{\eta_d}\geq 2$ GeV to ensure that $\eta_d$ decays to quarks and gluons instead of SM hadrons in the final state. We summarize the SM-like Higgs boson and new particles in different energy scales in Tab.~\ref{Tab:scale}.
In this study, we concentrate on investigating two specific benchmark points, namely $M_{\eta_d} = 4$ and $6$ GeV. The decay channels of $\eta_d\rightarrow c\overline{c}, \tau^+\tau^-$ are considered as dominant decay modes, as shown in Tab.~\ref{tab:BR-eta}. The obtained values for the branching ratio and lifetime of $\eta_d$ are in accordance with the case of a light scalar presented in Refs.~\cite{Winkler:2018qyg,Li:2022zgr}. We will discuss the constraints on $\sin\Psi$ later, and note that some parameter space remains permissible for sufficiently short lifetimes of $\eta_d$ with $M_{\eta_d} = 4$, $6$ GeV. Therefore, we will focus on investigating the prompt decay of $\eta_d$ in this study, which can result in either dark jet or semi-visible jet signatures at the LHC. 
Note that we leave the $\eta_d\rightarrow gg$ decay mode with $M_{\eta_d}\sim 2.5$ GeV for future study, as such a light $\eta_d$ would become a long-lived particle, causing an emerging jet or a multi-displaced vertex plus missing energy at the LHC. When $M_{\eta_d}\gtrsim 10$ GeV, the primary decay mode of $\eta_d$ is $b\bar{b}$. As the analysis of highly boosted $h'_1$ to multiple $b$ quarks is already presented in Ref.~\cite{Jung:2021tym}, we do not re-examine this possibility.

\begin{figure}[h]
\centering 
\includegraphics[height=8cm,width=8cm]{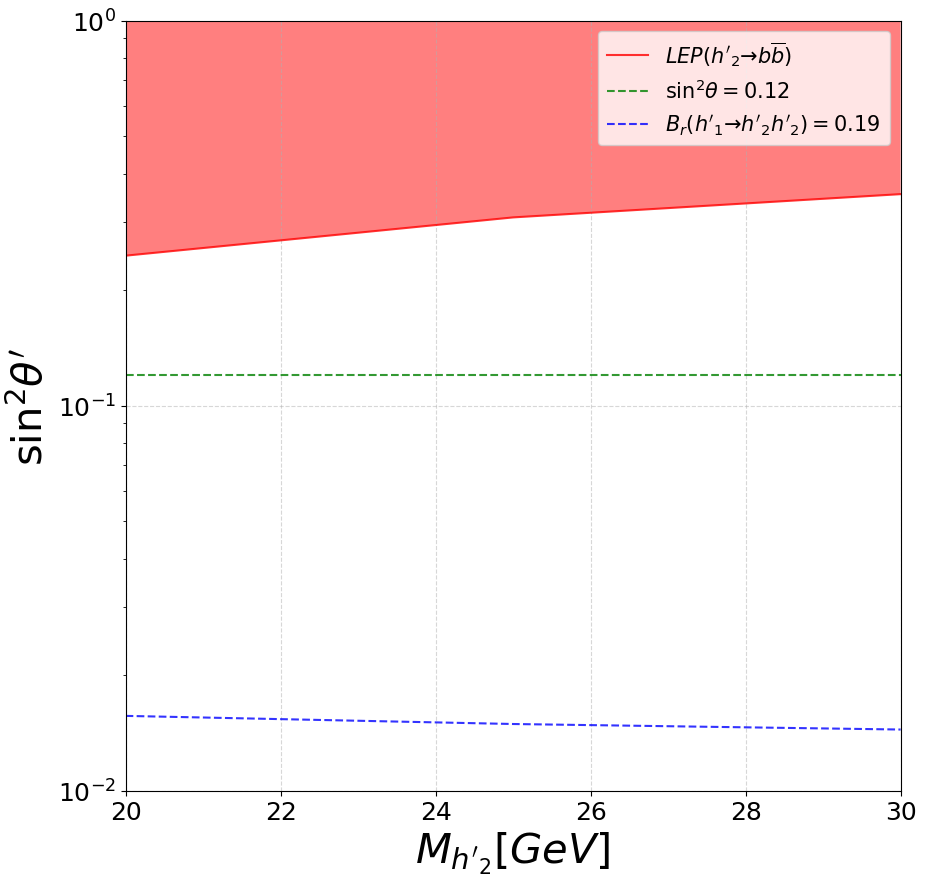}
\includegraphics[height=8cm,width=8cm]{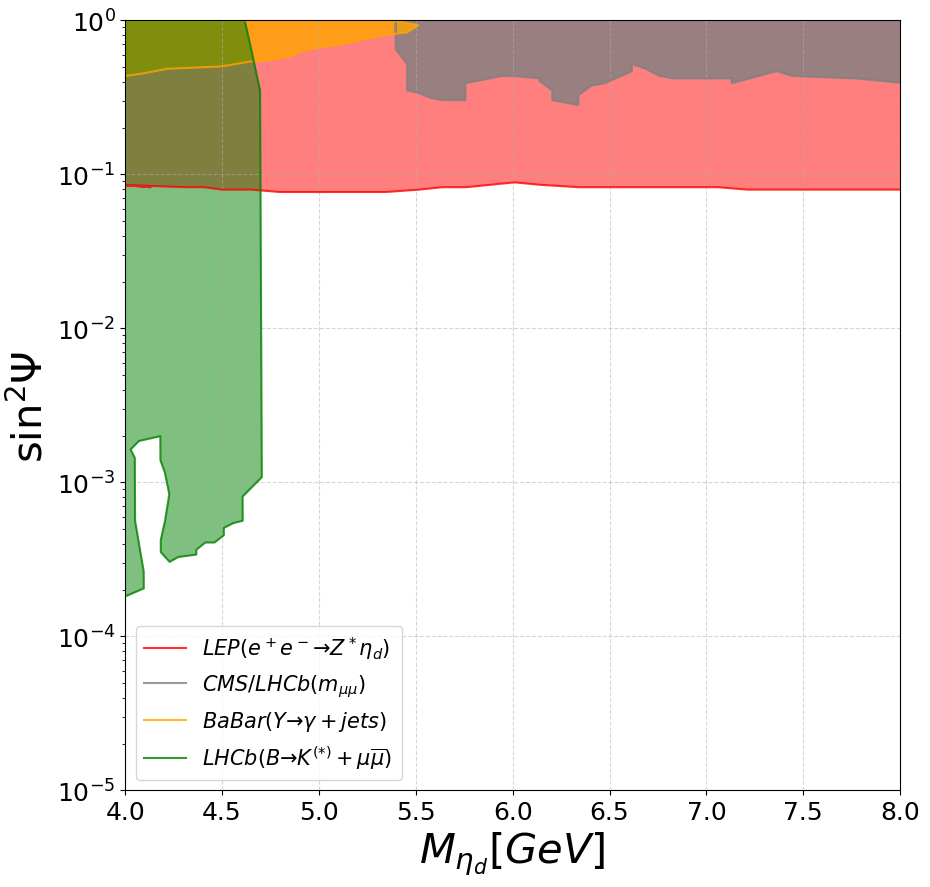}
\caption{Left panel : The existing bounds on the $\left(M_{h'_2},\sin^2\theta'\right)$ plane. The green dotted line represents the constraint on $\sin^2\theta' < 0.12$ at $95\%$ confidence level, the blue dotted line is $Br(h'_1\rightarrow\text{undetected}) < 19\%$ at $95\%$ confidence level, and the red bulk is the constraint on $\sin\theta'$ obtained from the $h'_2\rightarrow b\overline{b}$ decay at LEP. Right panel : The existing bounds on the $\left( M_{\eta_d},\sin^2\Psi\right)$ plane. The yellow bulk is the constraint from $\Upsilon\rightarrow\eta_d\gamma$ at BaBar ,the green bulk is the constraints from $B^{\pm}\rightarrow K^{\pm}\eta_d$ and $B^0\rightarrow K^{\ast 0}\eta_d$ at LHCb, the red bulk is the constraint from $B\rightarrow K\eta_d$, and the gray bulk is the constraint form light $\eta_d$ with precise resonance reconstruction at CMS/LHCb.}
\label{constran on mixing}
\end{figure}

The existing bounds for this model are summarized below. We divide the constraints on SM-like Higgs boson $h'_1$, singlet-like scalar boson $h'_2$ and light dark meson $\eta_d$ into the following two categories : 
\begin{itemize} 
\item The constraints to $h'_1$ and $h'_2$ \\ 
We utilize a $3\times 3$ orthogonal matrix to transform $h_1$, $h_2$, and $\tilde{\eta}$ into the final mass eigenstates $h'_1$, $h'_2$, and $\eta_d$. Through this transformation, we can simplify the complex expression involved into an effective angle, denoted as $\theta'$ for the rotation from the intermediate states $h_1$, $h_2$ to the final states $h'_1$, $h'_2$.
This angle is also a complex function of the model parameters in Eq.~(\ref{eq:potential}).
Based on precision measurements of $h'_1$ at the LHC, $\theta'$ is constrained to $\sin^2\theta' < 0.12$ at $95\%$ confidence level \cite{ATLAS:2015ciy,ATLAS:2016neq}. This constraint is marked by the green dotted line in the left panel of Fig.~\ref{constran on mixing}. There are two types of exotic decays of $h'_1$ in this model: $h'_1\rightarrow h'_2 h'_2$ and $h'_1\rightarrow\varphi_d\varphi^{\dagger}_d$. Therefore, we take $Br(h'_1\rightarrow\text{undetected}) < 19\%$ at $95\%$ confidence level, as reported in Ref.~\cite{ATLAS:2020qdt}, which is represented by the orange dotted line in the left panel of Fig.~\ref{constran on mixing}. Note that there is enough freedom in the parameter space to make one of $Br(h'_1\rightarrow h'_2 h'_2)$ and $Br(h'_1\rightarrow\varphi_d\varphi^{\dagger}_d)$ dominant and the other negligible. Therefore, we set $Br(h'_1\rightarrow h'_2 h'_2) < 19\%$ in Fig.~\ref{constran on mixing}. Finally, we display the constraint on $\sin\theta'$ obtained from the $h'_2\rightarrow b\overline{b}$ decay at LEP~\cite{LEPWorkingGroupforHiggsbosonsearches:2003ing}, which is marked by the red bulk in the left panel of Fig.~\ref{constran on mixing}. As $h'_2$ mainly decays to dark sector particles instead of to $b\overline{b}$ for small $\sin\theta'$, this constraint becomes much weaker than the one shown in Fig.~10 of Ref.~\cite{LEPWorkingGroupforHiggsbosonsearches:2003ing}.
\item The constraints to light $\eta_d$ \\ 
The search for light scalars from $\Upsilon$ decay was first proposed in Ref.~\cite{Wilczek:1977zn}, and it can also be applied to $\eta_d$ in this model. We mark the constraint from $\Upsilon\rightarrow\eta_d\gamma$ at BaBar in the yellow bulk on the right panel of Fig.~\ref{constran on mixing}. Similarly, when $M_{\eta_d} < M_B -M_K$, we can also use $B^{\pm}\rightarrow K^{\pm}\eta_d$ and $B^0\rightarrow K^{\ast 0}\eta_d$ processes to search for a light $\eta_d$~\cite{Batell:2009jf}. We mark the constraints from $B^{\pm}\rightarrow K^{\pm}\eta_d$ and $B^0\rightarrow K^{\ast 0}\eta_d$ at LHCb in the green bulk on the right panel of Fig.~\ref{constran on mixing}. In the same mass range, the constraint from $B\rightarrow K\eta_d$ is much stronger than the one from $\Upsilon\rightarrow\eta_d\gamma$. Since $\tilde{\eta}$ mixes with $h_1$ and $h_2$ to form the mass eigenstate $\eta_d$, the light scalar search from LEP can also be applied to $\eta_d$~\cite{L3:1996ome}. We mark the constraint from LEP in the red bulk on the right panel of Fig.~\ref{constran on mixing}. Finally, searches for the dimuon decay channel from light $\eta_d$ with precise resonance reconstruction can constrain $5.5$ GeV $< M_{\eta_d} < 15$ GeV at CMS/LHCb~\cite{CMS:2012fgd,LHCb:2018cjc,Haisch:2016hzu}. We mark this constraint in the gray bulk on the right panel of Fig.~\ref{constran on mixing}.
\end{itemize} 
We can see from Fig.~\ref{constran on mixing} that these constraints are still relatively weak. Besides, detecting the direct production of $h'_2$ with a mass less than $50$ GeV at the LHC is very challenging~\cite{CMS:2017dcz,CMS:2019emo}. The main reason for this is that we cannot simultaneously find a powerful trigger, such as an initial state radiation photon~\cite{CMS:2019xai}, with a production cross section large enough for light $h'_2$ decaying to heavy-flavor jets in our scenario. Therefore, we will focus solely on the study of $h'_1$ exotic decays at the LHC, which can explore more parameter space during the high luminosity phase.

The study focused on a mass interval of $10$ GeV $< M_{h'_2} \lesssim 60$ GeV, with a minimum requirement of $M_{\eta_d} \geq 2$ GeV. When $M_{\eta_d} \gtrsim 10$ GeV, the primary decay mode of $\eta_d$ is $b\bar{b}$, as previously analyzed in Ref.~\cite{Jung:2021tym}. The benchmark points were chosen as: $M_{h'_2} = 5M_{\eta_d}$, with $M_{\eta_d} = 4$ or $6$ GeV. Additional possibilities, namely $M_{h'_2} = 3M_{\eta_d}$ and $M_{h'_2} = 10M_{\eta_d}$, were selected to comprehensively explore the entire mass interval and examine the impact of varying mass relationships on the results. Furthermore, considering the heavier nature of $\tilde{\omega}$ compared to $\eta_d$, a benchmark point of $M_{\tilde{\omega}}/M_{\eta_d} = 1.8$ was chosen. The subsequent section investigates the effect of different mass ratios, such as $M_{\tilde{\omega}}/M_{\eta_d} = 1.2, 3, 5$, with a detailed discussion on the implications of these choices provided in the subsequent sections.

\section{Framework of Machine Learning }
\label{sec:ml}

\subsection{Pre-selection}
\label{sec:presel}

\begin{figure}[h]
\centering
\includegraphics[height=6cm,width=8cm]{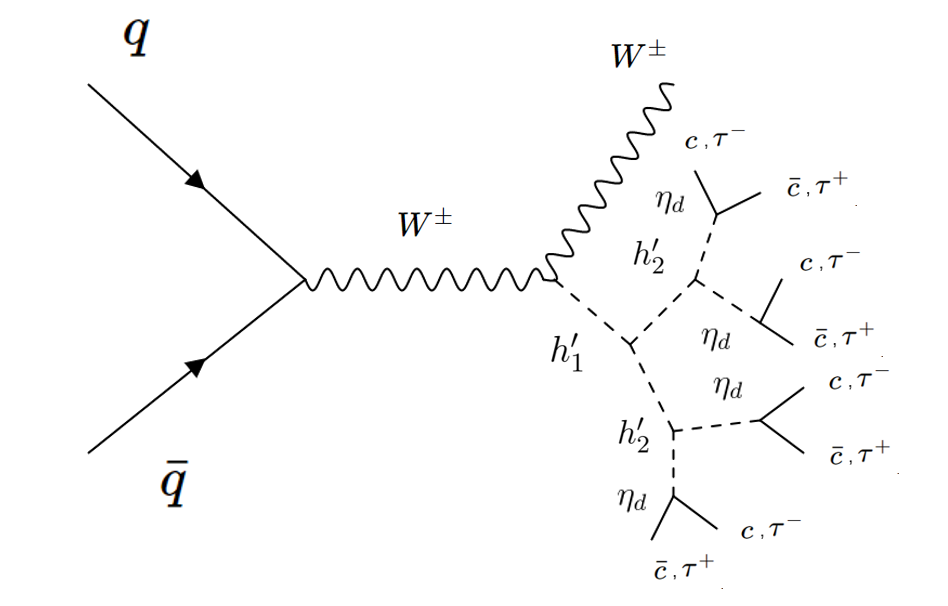}
\includegraphics[height=6cm,width=8cm]{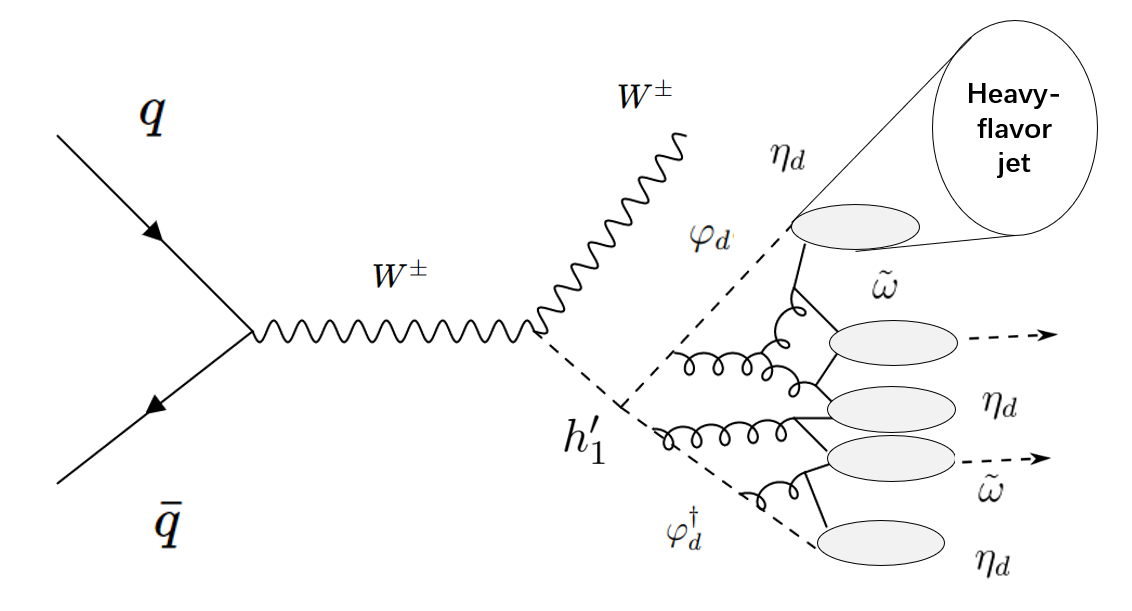}
\caption{Feynman diagrams for different types of signal processes to produce dark mesons at the LHC : The cascade decay (left panel); The dark showers (right panel).}
\label{loss}
\end{figure}

According to the mass spectrum mentioned in Sec.~\ref{sec:model}, there are two ways to produce dark mesons at the LHC either from the cascade decay of $h'_1\rightarrow h'_2 h'_2$ or the dark quark pair produced from $h'_1$ and then undergoing QCD-like showers and hadronization to produce dark mesons. Based on decay modes of $\eta_d$ mentioned in Tab.~\ref{tab:BR-eta}, we first explore collimated multi-heavy-flavor jets from highly boosted $h'_1$ at the LHC for the first kind. The process to generate this signature is 
\begin{equation}
    pp\rightarrow V^{\ast}\rightarrow V h'_1\quad\text{and}\quad h'_1\rightarrow h'_2 h'_2\rightarrow 4\eta_d\rightarrow\text{combinations of heavy-flavor jets},
\label{Eq:production1}
\end{equation}
where $V=W,Z$. On the other hand, since $h'_1$ is much heavier than the dark confinement scale $\Lambda_d$, the $h'_1$ can also decay to a pair of scalar dark quark $\varphi_d$ via the Higgs portal. Note $h'_1$ decay to a pair of dark gluon $g_d$ is from the one-loop process which is much suppressed and can be ignored here. Hence, we can further explore the highly boosted $h'_1$ production and then its decay to a pair of $\varphi_d$ as 
\begin{equation}
    pp\rightarrow V^{\ast}\rightarrow V h'_1\quad\text{and}\quad h'_1\rightarrow\varphi_d\varphi^{\dagger}_d\rightarrow\text{dark showers and hadronization}.
\label{Eq:production2}    
\end{equation} 
This is similar to the ordinary QCD processes at the LHC. Finally, the $\eta_d$ mainly decays back to $c\overline{c}$ and $\tau^+\tau^-$, while $\tilde{\omega}$ is stable. Therefore, this process causes the novel semi-visible heavy-flavor jet mentioned in Ref.~\cite{Cohen:2015toa,Beauchesne:2022phk}. We then use the Hidden Valley module~\cite{Carloni:2010tw,Carloni:2011kk} in Pythia8~\cite{Sjostrand:2014zea} to simulate effects from dark showers, hadronization, and dark meson decays. Feynman diagrams for these processes are depicted in Fig.~\ref{loss}.

In this study, two primary reference points were investigated: (i) $M_{h'_2} = 30$ GeV, $M_{\eta_d} = 6$ GeV, and (ii) $M_{h'_2} = 20$ GeV, $M_{\eta_d} = 4$ GeV, along with discussions for other possibilities in Sec.~\ref{sec:result}. The value of $\sin\theta'$ was fixed at $0.1$ to emphasize novel characteristics, while other model parameters were considered irrelevant. It should be noted that the choice of $\sin\theta'$ does not directly impact the pre-selections or machine learning analysis. However, it does affect the overall cross section. Hence, as long as the chosen $\sin\theta'$ satisfies the established constraints, its selection does not affect the analysis.  The signal process in Eq.~(\ref{Eq:production1}) is studied, along with potential SM backgrounds, including $W^{\pm}(l^{\pm}\nu_{l})j$, semi-leptonic decay of $t\overline{t}$, and $W^{\pm}(l^{\pm}\nu_{l})h(b\overline{b})$ processes, with the primary contribution originating from $W^{\pm}(l^{\pm}\nu_{l})j$\footnote{Here we only focus on the signal process $pp\rightarrow W^{\pm}h'_1$ in this study since the cross section of $pp\rightarrow W^{\pm}(l^{\pm}\nu)h'_1$ is larger than the cross section of $pp\rightarrow Z(l^{+}l^{-})h'_1$ at $\sqrt{s}=14$ TeV.}. To construct the scalar-mediated dark QCD model and generate the UFO model file, we utilize FeynRules~\cite{Alloul:2013bka,Darme:2023jdn,Degrande:2011ua}. We use MadGraph5 aMC@NLO~\cite{Alwall:2014hca} with NN23LO1 PDF set~\cite{NNPDF:2017mvq} to simulate the leading order (LO) of both signal and background processes at the LHC with a center-of-mass energy of $\sqrt{s}=14$ TeV. To improve event generation efficiency while maintaining generality, we require at least one parton-level jet in the final state with transverse momentum $P^j_{T} > 100, 50 $ GeV in the $W^{\pm}j$ and $t\overline{t}$ processes, respectively. This criterion allows us to select the leading jet candidate.

The next-to-leading order (NLO) K-factor for high-$P_T$ $W^{\pm}h$ production is estimated to be approximately $1.5$~\cite{Butterworth:2008sd,Ellis:1998fv,Campbell:2003hd}. Similarly, for the primary background processes, the NLO K-factors for $W^{\pm}j$ and $t\overline{t}$ at the next-to-next-to-leading order (NNLO) are approximately $1.5$ and $1.6$, respectively~\cite{Lindert:2017olm,Czakon:2011xx,Czakon:2013goa,Czakon:2012pz,Czakon:2012zr,Barnreuther:2012wtj,Cacciari:2011hy}. The subsequent results will be calculated using the cross-sections corrected to these K-factors. The total production cross sections for both signal and background processes can be found in Tab.~\ref{basiccut1},~\ref{basiccut2}, and~\ref{basiccut3} below.

The generated events are then subjected to parton showering and hadronization via Pythia8. 
The fast detector simulation is conducted utilizing the CMS template implemented in Delphes3~\cite{deFavereau:2013fsa}.

For the isolation of the reconstructed charged leptons, we employ the following criteria. We adjusted the electron and muon isolation parameters in the Delphes CMS template as follows: DeltaRMax = 0.3, PTMin = 1.0, and PTRatioMax = 0.12 for electrons (0.25 for muons). 
Moreover, in order to achieve successful reconstruction of electrons and muons, it was necessary to satisfy the criteria of $P^{\ell}_T > 20$ GeV and $|\eta_{\ell}|<2.5$. In our jet reconstruction approach, we employed exclusively photons and hadrons, which include both charged hadrons and neutral hadrons. Following the particle-flow algorithm~\cite{CMS:2017yfk,deFavereau:2013fsa}, the deposits in ECAL and HCAL are utilized to form jets by clustering EflowPhotons, EflowNeutralHadrons, and ChargedHadrons using the anti-kt algorithm~\cite{Cacciari:2008gp}. The jet clustering procedure is implemented using FastJet~\cite{Cacciari:2011ma}. The fat-jet ($J$) candidates are clustered utilizing the anti-kt algorithm with a jet cone size of $R_J=1.5$. Conversely, the small-radius jet candidates, such as $b$-jets, are clustered using the anti-kt algorithm with a jet cone size of $R_j = 0.4$. For the b-tagging process, we adopted an efficiency of 0.77, accompanied by mis-tag rates of 1/6 for c-jets and 1/134 for light-flavor jets, as reported in Ref.~\cite{ATLAS:2018alq}. Furthermore, to ensure the exclusion of isolated charged leptons from the jets, we removed jets if the opening angle, $\Delta R$, between the jet and charged lepton was smaller than the jet's radius parameter ($R_J=1.5$ for fat-jets and $R_j = 0.4$ for $b$-jets).

For the signal process described in Eq.~(\ref{Eq:production2}), the relevant SM background processes are the same as before. We use MadGraph5 aMC@NLO to simulate this signal process at the parton level with $M_{\varphi_d} = 2.4$ and $1.6$ GeV for the first and second benchmark points, respectively. To simulate dark showers, hadronization, and dark meson decays, we apply the Hidden Valley module in Pythia8. We set $\Lambda_d = 24$ GeV, \textit{pTminFSR} $= 26.4$ GeV for the first benchmark point and $\Lambda_d = 16$ GeV, \textit{pTminFSR} $= 17.6$ GeV for the second benchmark point. The parameter \textit{probVector} $= 0.617$ for both benchmark points is considered. Finally, the same fast detector simulation procedure is employed as before.

The Hidden Valley module is a generic tool used to study radiation, hadronization, and the decays of new particles in various dark sector models. In this study, we apply the number of dark colors $\tilde{N}_C = 3$, the number of dark flavors $\tilde{n}_F = 1$ and certain parameter relations within this module, namely: $M_{\varphi_d} = 0.4M_{\eta_d}$, $\Lambda_d = 4M_{\eta_d}$, \textit{pTminFSR} $= 1.1\Lambda_d$, and \textit{probVector} $= 0.617$. We set $M_{\varphi_d} = 0.4M_{\eta_d}$ to meet the requirement that $M_{\eta_d}$ must be greater than twice $M_{\varphi_d}$. Additionally, due to the significantly smaller mass of the lightest dark meson, $\eta_d$, compared to $\Lambda_d$, we assign $\Lambda_d = 4M_{\eta_d}$. To ensure a minimum allowable transverse momentum for dark quark emission, we apply $\textit{pTminFSR} = 1.1\Lambda_d$. Furthermore, \textit{probVector} is set to a value of $0.617$, representing the probability of occurrence for vector dark mesons within the considered set of dark mesons. Finally, we adopt the default settings for hadronization in the Hidden Valley module. 
Here we make some comments on the changes of $\Lambda_d$, $\tilde{N}_C$, and $\tilde{n}_F$ on our results based on Ref.~\cite{Cohen:2020afv}. When increasing $\Lambda_d$, it leads to an improved discriminative capability between dark QCD and SM QCD. However, this enhanced discrimination is not unlimited and saturates when $\Lambda_d \geq 50$ GeV. On the other hand, as $\tilde{N}_C$ increases, the discriminative power between dark QCD and SM QCD decreases. Lastly, with an increase in $\tilde{n}_F$, the discriminative power between dark QCD and SM QCD is enhanced. However, this relationship no longer holds when $\tilde{n}_F > \frac{11}{4}\tilde{N}_C$.

In this study, we classify the signal signatures into three scenarios. We begin with the simplest scenario, where the first signal process is $pp\rightarrow W^{\pm}h'_1\rightarrow\left( l^{\pm}\nu\right)\left( h'_2 h'_2\right)$ with $h'_2\rightarrow 2\eta_d\rightarrow 4c\overline{c}$. As there are eight charm quarks in the final state, we denote it as $s_{8c}$. Next, we extend to the real but more complicated situations in the second process, $pp\rightarrow W^{\pm}h'_1\rightarrow\left( l^{\pm}\nu\right)\left( h'_2 h'_2\right)$ with $h'_2\rightarrow 2\eta_d\rightarrow 4c\overline{c}, 4\tau^+\tau^-, 2c\overline{c}2\tau^+\tau^-$. This scenario has a mixture of charm quarks and tau leptons in the final state, so we label it as $s_{c\tau}$. Finally, the third signal process $pp\rightarrow W^{\pm}h'_1\rightarrow\left( l^{\pm}\nu\right)\left(\varphi_d\varphi^{\dagger}_d\right)$ involves the energetic scalar dark quark $\varphi_d$ undergoing dark showers and hadronization. We denote this type of signal signature as $s_{DS}$.

Prior to subjecting the detector-level events to machine learning analysis, pre-selection criteria are applied to define the trigger, identify the signal signature and reduce the impact of relevant SM backgrounds. 
First, for the signal $s_{8c}$ ($pp \rightarrow W^{\pm}h'_1 \rightarrow \left(l^{\pm}\nu\right)\left(h'_2 h'_2\right)$ with $h'_2\rightarrow 2\eta_d\rightarrow 4c\overline{c}$) and the related SM background, we apply certain selection criteria. A charged lepton, and at least one jet, are needed\footnote{Jets satisfy $P^j_T > 20$ GeV and $|\eta_j| < 2.5$, while electrons and muons need to meet $P_{T}^{e} > 20$ GeV and $|\eta_e| < 2.5$, $P_{T}^{\mu} > 20$ GeV and $|\eta_\mu| < 2.5$ criteria, respectively.}. We then impose additional requirements on the leptons: $P_T^l > 25$ GeV and $|\eta_l| < 2.5$. 
Moreover, the observable $P_T^{l+\slashed{E}_T}$ is defined as the transverse component of the four-vector obtained after the vectorial sum of the charged lepton and the missing transverse energy. 
We impose the requirement of $P_T^{l+\slashed{E}_T} > 200$ GeV to indicate the condition of a highly boosted $W$ boson. 
 
Moving on, the transverse mass of charged lepton and $\slashed{E}_T$ is defined as follows:
\begin{equation}
   M_{T} (l^{\pm},\slashed{E}_{T}) = \sqrt{2 P_{T}^{l} \slashed{E}_{T} (1-\cos{\Delta \phi})}.
\end{equation} 
The azimuthal angle between the charged lepton and $\slashed{E}_{T}$, denoted by $\Delta\phi\equiv \Delta\phi({l}^{\pm},\slashed{E}_{T})$, is used to calculate $M_{T} (l^{\pm},\slashed{E}_{T})$. To suppress the background of $t\overline{t}$, the event selection $M_{T} (l^{\pm},\slashed{E}_{T}) < 100$ GeV is applied. 

\begin{table}[h]
\small
    \centering
    \begin{tabular}{|c|c|c|c|c|c|}
    \hline    
   Cross section (fb) & ~$W^{\pm}(l \nu_{l})h$~ & ~$W^{\pm}(l \nu_{l})j$~ & ~$t\overline{t}$~ & ~signal (1)~ & ~signal (2)~ \\
   \hline
     \makecell[c]{   Generator  }&  219.93 & $2.75\times10^{5}$ & $6.95\times10^{3}$& 234.45&232.13 \\
    \hline
    \makecell[c]{ one fat-jet and one lepton  }&125.05& $1.95\times10^{5}$ &$4.92\times10^{3}$ & 141.77&132.48 \\
    \hline
    \makecell[c]{$P_{T}^{l} > 25 $ GeV, \\ $|\eta_{l_{}}|<2.5$,  $P_{T}^{l+\slashed{E}_{T}} > 200 $ GeV, \\ $M_{T} (l^{\pm},\slashed{E}_{T}) < 100$ GeV \\}& 7.19& $2.23 \times10^4$ &668.93&8.15 &8.24 \\
    \hline
    \makecell[c]{  $ 200< P_{T}^{J}< 500$ GeV,  \\ $|\eta_{J_{}}|<2.5$  }&5.30& $1.71 \times10^4$ &575.46 &6.57 &6.60 \\
    \hline
    \makecell[c]{veto on b-jet\\}  &2.46 &$1.65\times10^4$&232.48&6.43&6.13   \\
    \hline
    \makecell[c]{$ 100 < M_{J} < 150 $ GeV }&0.96&$3.89 \times10^3$&15.64&2.46&2.22 \\
    \hline
    efficiency&0.44$\%$& 1.42$\%$& 0.23$\%$&1.05$\%$& 0.96$\%$\\
    \hline
    \end{tabular}
        \caption{The pre-selection cut-flow cross sections (in fb) of the signal $s_{8c}$ and the corresponding backgrounds at 14 TeV LHC. Two signal benchmark points, signal (1) :  $M_{h'_2}=20$ GeV, $M_{\eta_d} = 4$ GeV and signal (2) : $M_{h'_2}=30$ GeV, $M_{\eta_d} = 6$ GeV in the signal $s_{8c}$ are displayed. We assume $\text{Br}(h'_1\rightarrow h'_2 h'_2)=\text{Br}(h'_2\rightarrow \eta_d\eta_d)=\text{Br}(\eta_d\rightarrow c\overline{c})=100\%$ for both signal (1) and signal (2) in this table.} 
    \label{basiccut1}
\end{table}

\begin{figure}[h]
\centering
\includegraphics[height=8cm,width=8cm]{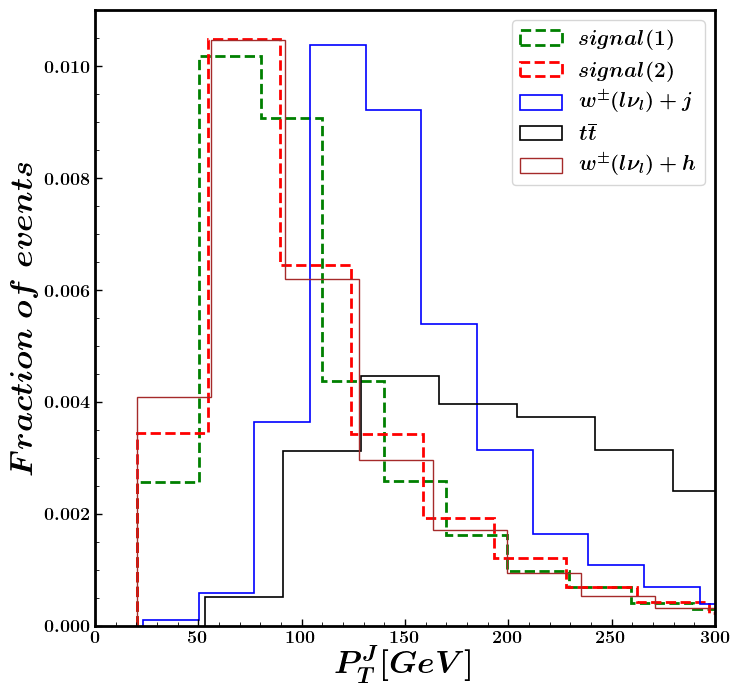}
\includegraphics[height=8cm,width=8cm]{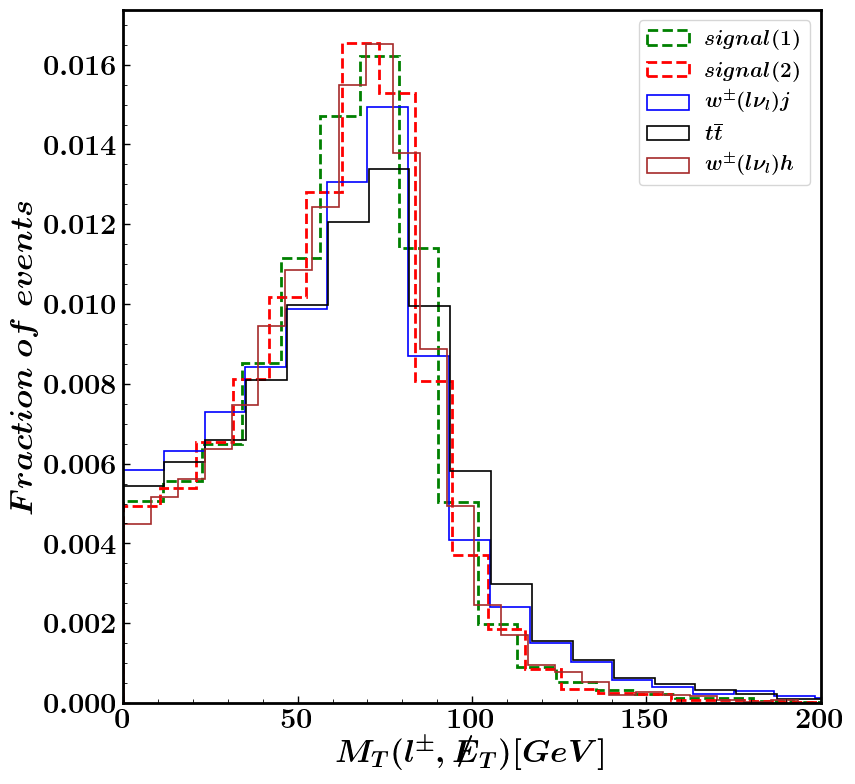}
\includegraphics[height=8cm,width=8cm]{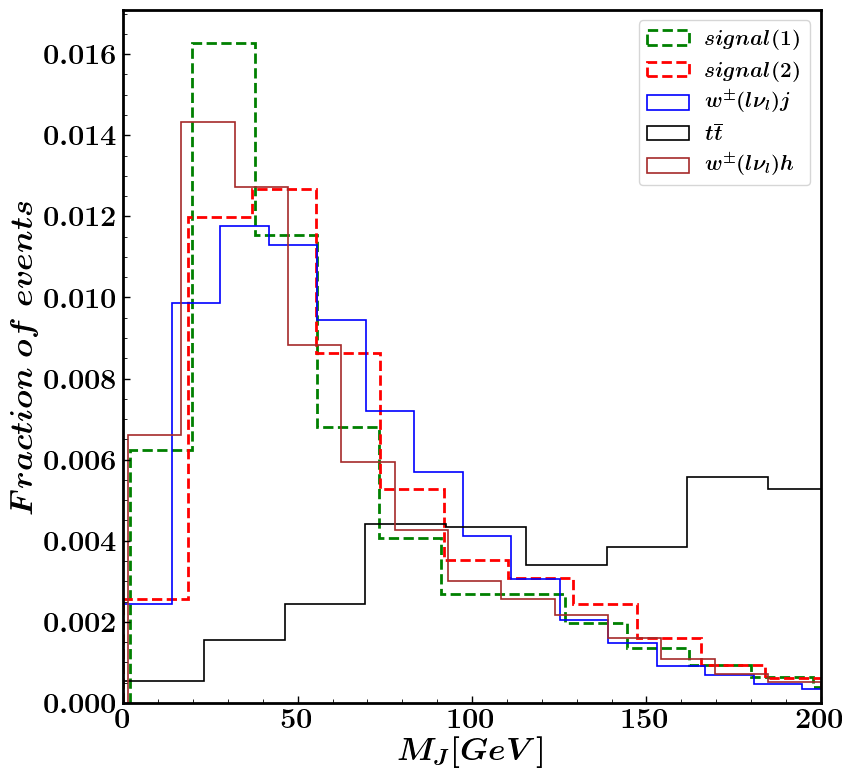}
\includegraphics[height=8cm,width=8cm]{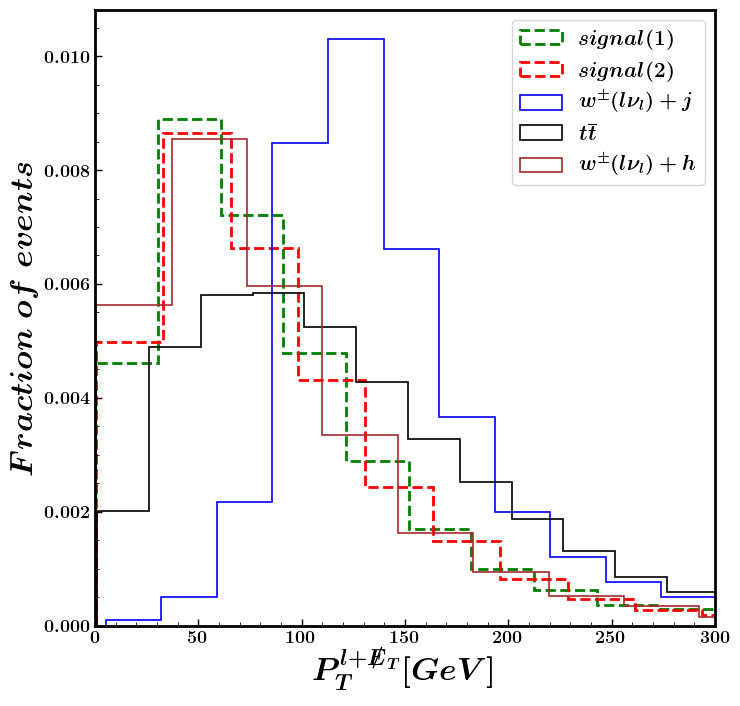}
\caption{The kinematic distributions of $P^J_T$ (upper-left), $M_T (l^{\pm},\slashed{E}_{T})$ (upper-right), $M_J$ (lower-left), and $P_{T}^{l+\slashed{E}_{T}}$ (lower-right) for the signal $s_{8c}$ and relevant SM backgrounds at 14 TeV LHC.}
\label{distributions8c}
\end{figure}

To ensure the presence of at least one energetic fat-jet in the central region, we also require $200$ 
GeV $< P_{T}^{J} < 500$ GeV and $|\eta_{J}| < 2.5$.
 
On the other hand, to suppress the contribution of the $t\overline{t}$ process, we implement an event veto for the presence of accompanying $b$-jets with transverse momentum $P_{T}^{b} > 20$ GeV and pseudorapidity $|\eta_{b_{}}|<2.5$. Furthermore, to mitigate the impact of the $W^{\pm}j$ and $t\overline{t}$ processes, we impose a requirement on the leading fat-jet mass, which must fall within the mass window characteristic of $h'_1$: $100$ GeV $< M_{J} < 150$ GeV. 

The cut-flow table for signal and background processes with these pre-selections is presented in Tab.~\ref{basiccut1}. Some related kinematic distributions for signal and background processes are shown in Fig.~\ref{distributions8c}.
We require at least one parton-level jet in the final state with $P^j_{T} > 100$ GeV for $W^{\pm}j$ and $P^j_{T} > 50$ GeV for $t\overline{t}$. The boosted final state enhances the charged lepton identification efficiency, yielding a significant number of events fulfilling the criteria for a charged lepton and at least one jet. Notably, signal and $W^{\pm}h$ events are unfiltered during generation, resulting in lower lepton identification efficiency. As a consequence, the number of events satisfying the requirement of a charged lepton and at least one jet is relatively small.

\begin{table}[h]
\small
    \centering
    \begin{tabular}{|c|c|c|c|c|c|}
    \hline
   Cross section (fb) & $W^{\pm}(l \nu_{l})h$ & $W^{\pm}(l \nu_{l})j$ & $\quad$$t\overline{t}$$\quad$& signal (1)&  signal (2)\\
    \hline
     \makecell[c]{  Generator  }&  219.93 & $2.75\times10^{5}$ & $6.95\times10^{3}$& 103.61&180.76 \\
    \hline
    \makecell[c]{ one fat-jet and one lepton  }&125.05& $1.95\times10^{5}$ &$4.92\times10^{3}$ &58.86&102.56 \\
    \hline
    \makecell[c]{$P_{T}^{l} > 25 $ GeV, \\ $|\eta_{l_{}}|<2.5$,  $P_{T}^{l+\slashed{E}_{T}} > 200 $ GeV \\}& 7.80& $2.45 \times10^4$ &739.83 &2.20 &4.87 \\
    \hline
    \makecell[c]{  $ 150< P_{T}^{J}< 500$ GeV, \\ $|\eta_{J_{}}|<2.5$  }&6.76&$2.21\times10^4$&650.52&1.92& 4.44  \\ 
    \hline
    \makecell[c]{veto on b-jet\\}&3.41 & $2.12\times10^4$& 260.63& 1.89&4.17 \\
    \hline
    \makecell[c]{$ 100 < M_{J} < 150 $ GeV }&1.22&$4.85 \times10^3$&20.50&0.73& 1.54\\
    \hline
    efficiency&0.55$\%$&1.76$\%$& 0.30$\%$&0.70$\%$& 0.85$\%$\\
    \hline
    \end{tabular}
        \caption{The pre-selection cut-flow cross sections (in fb) of the signal $s_{c\tau}$ and the corresponding backgrounds at 14 TeV LHC. The signal benchmark points are the same as Tab.~\ref{basiccut1} with the $\eta_d$ decay branching ratio in Tab.~\ref{tab:BR-eta}. We assume $\text{Br}(h'_1\rightarrow h'_2 h'_2)=\text{Br}(h'_2\rightarrow\eta_d\eta_d)=100\% $ for both signal (1) and signal (2) in this table.}    
    \label{basiccut2}
\end{table}

\begin{figure}[h]
\centering
\includegraphics[height=8cm,width=8cm]{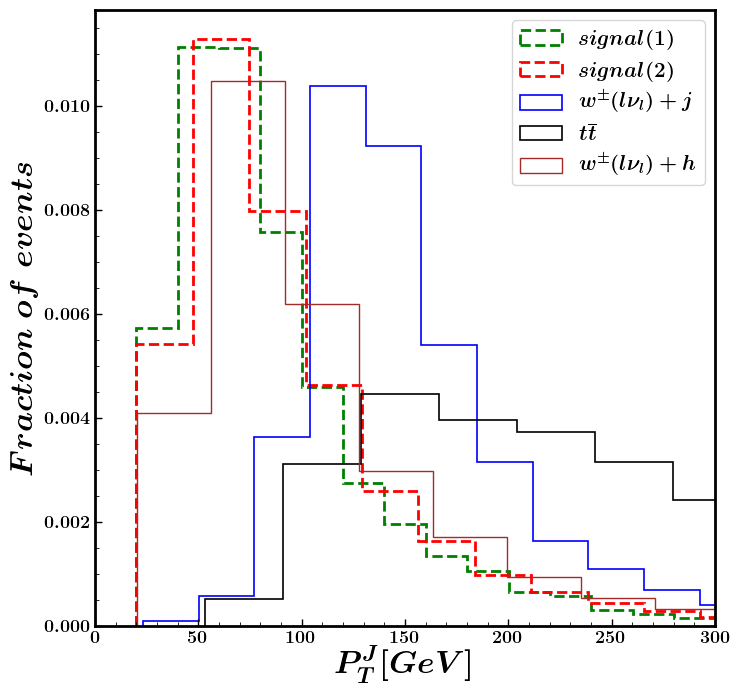}
\includegraphics[height=8cm,width=8cm]{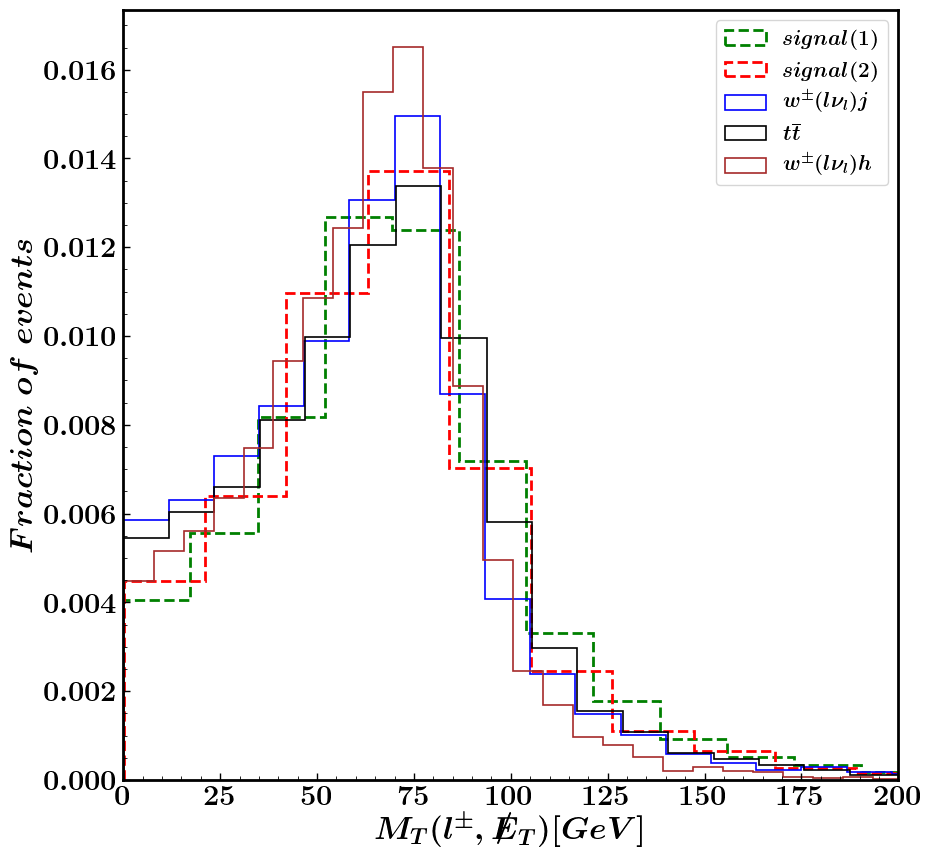}
\includegraphics[height=8cm,width=8cm]{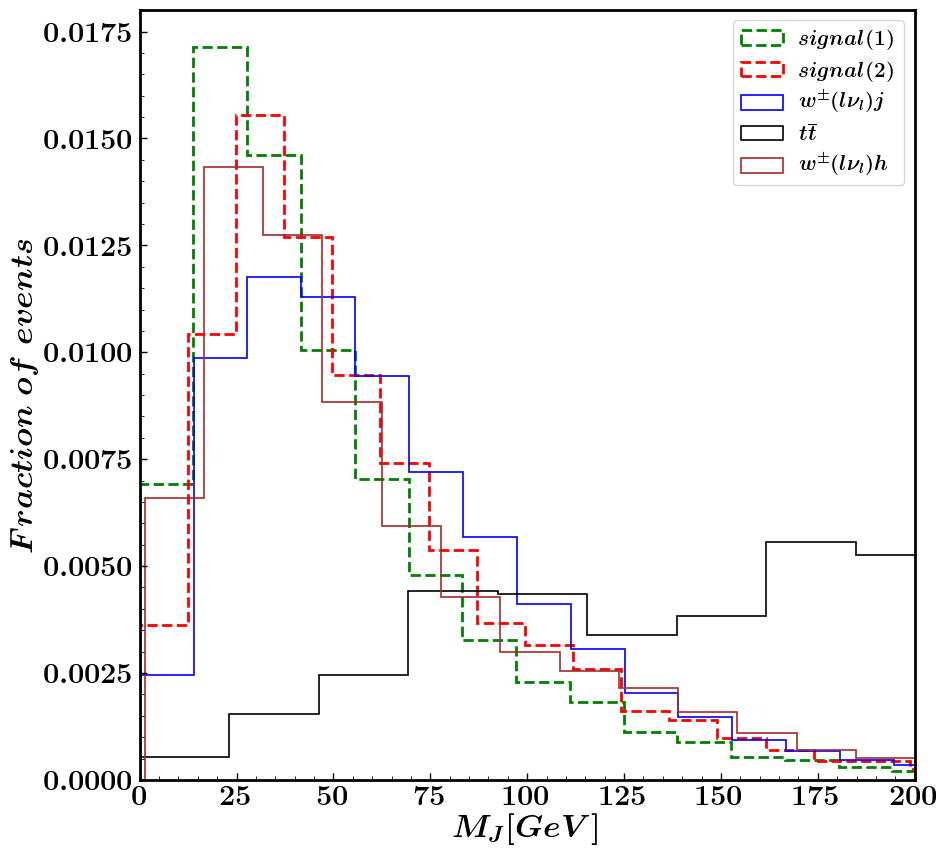}
\includegraphics[height=8cm,width=8cm]{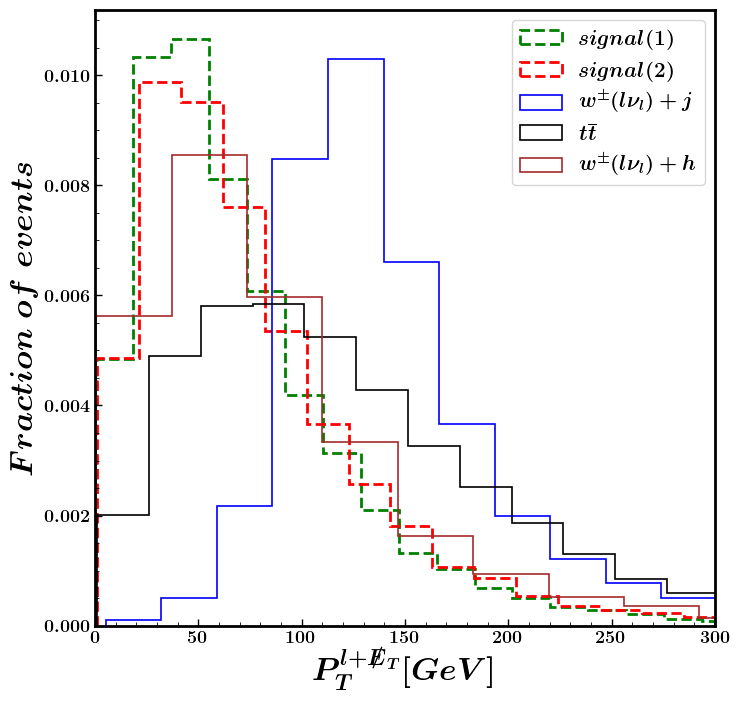}
\caption{The kinematic distributions of $P^J_T$ (upper-left), $M_T (l^{\pm},\slashed{E}_{T})$ (upper-right), $M_J$ (lower-left), and $P_{T}^{l+\slashed{E}_{T}}$ (lower-right) for the signal $s_{c\tau}$ and relevant SM backgrounds at 14 TeV LHC.}
\label{distributions8ct}
\end{figure}

Similarly, Tab.~\ref{basiccut2} shows the pre-selection criteria applied to the signal $s_{c\tau}$ ($pp\rightarrow W^{\pm}h'_1\rightarrow\left( l^{\pm}\nu\right)\left( h'_2 h'_2\right)$ with $h'_2\rightarrow 2\eta_d\rightarrow 4c\overline{c}, 4\tau^+\tau^-, 2c\overline{c}2\tau^+\tau^-$) and its corresponding SM backgrounds. According to Tab.~\ref{tab:BR-eta}, the decay branching ratios for $h'_2\rightarrow 2\eta_d\rightarrow 4c\overline{c}, 4\tau^+\tau^-, 2c\overline{c}2\tau^+\tau^-$ are $42.05\%$ for signal (1) and $77.35\%$ for signal (2), respectively. We also display some kinematic distributions for signal and background processes in Fig.~\ref{distributions8ct}. Unlike the signal $s_{8c}$, the signal $s_{c\tau}$ contains $\tau$ leptons, which results in a lower transverse momentum for the fat-jet. Therefore, we impose an alternative selection on the transverse momentum of the fat-jet to ensure an adequate yield of signal events, which is $150$  GeV $< P_{T}^{J} < 500$ GeV. For the same reason, we exclude the selection $M_T (l^{\pm},\slashed{E}_{T}) < 100$ GeV in this analysis. Other event selections are identical to those in the signal $s_{8c}$ analysis.

After performing the preselection criteria, we employed deep learning and BDT-based jet substructure analysis techniques to discriminate between signal and background events. We will see that the BDT-based jet substructure analysis is worse than the deep learning ones. Therefore, we focus on the deep learning analysis in the main text and carry out a BDT-based jet substructure analysis on signals $s_{c\tau}$ and $s_{8c}$, and compared the effectiveness of BDT and machine learning methods in Appendix~\ref{sec:app2}.

\begin{table}[h]
\small
    \centering
    \begin{tabular}{|c|c|c|c|c|c|}
    \hline
    Cross section (fb) & $W^{\pm}(l \nu_{l})h$ & $W^{\pm}(l \nu_{l})j$ & $\quad$$t\overline{t}$$\quad$& signal (1)&  signal (2)\\
    \hline
     \makecell[c]{   Generator  }&  219.93 & $2.75\times10^{5}$ & $6.95\times10^{3}$& 191.56&191.72 \\
    \hline
    \makecell[c]{ one fat-jet and one lepton  }&125.05& $1.95\times10^{5}$ &$4.92\times10^{3}$ & 111.63&111.01 \\
    \hline
    \makecell[c]{$P_{T}^{l} > 25 $ GeV,  \\$|\eta_{l_{}}|<2.5$,   $P_{T}^{l+\slashed{E}_{T}} > 200 $ GeV,  \\$M_{T} (l^{\pm},\slashed{E}_{T}) > 140$ GeV}& 0.11 &48.67&5.02 &4.22 &3.94  \\
    \hline
    \makecell[c]{   $  P_{T}^{J}>50$ GeV,  \\ $|\eta_{J_{}}|<2.5$  }&0.11 & 46.20& 4.96&2.96&2.89  \\
    \hline
    \makecell[c]{ $\slashed{E}_{T}> 100$ GeV \\  }& $1.34\times10^{-2}$  &2.45&0.53&2.90& 2.81  \\
    \hline
    \makecell[c]{ veto on b-jet\\ }& $3.73 \times10^{-3}$ &1.65&0.23& 2.80&2.68  \\
    \hline
    \makecell[c]{ $\slashed{E}_{T}/M_{T} (J,\slashed{E}_{T}) > 1.22$ }& 0 &0.14& 0&1.78& 1.51 \\
    \hline
    efficiency& 0 & $5.00 \times10^{-7}$ &~0~&0.93$\%$&0.79$\%$\\
    \hline
    \end{tabular}
        \caption{The pre-selection cut-flow cross sections (in fb) of the signal $s_{DS}$ and the corresponding backgrounds at 14 TeV LHC. The signal benchmark points are the same as Tab.~\ref{basiccut1} with the assumption $\text{Br}(h'_1\rightarrow \varphi_d\varphi^{\dagger}_d) = 100\%$. }    
    \label{basiccut3}
\end{table} 

\begin{figure}[htbp]
\centering
\includegraphics[height=7cm,width=7cm]{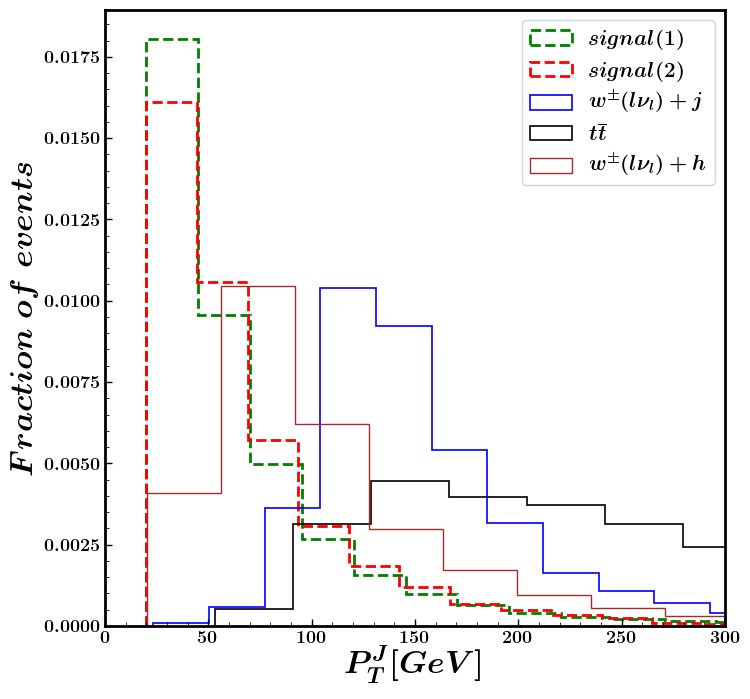}
\includegraphics[height=7cm,width=7cm]{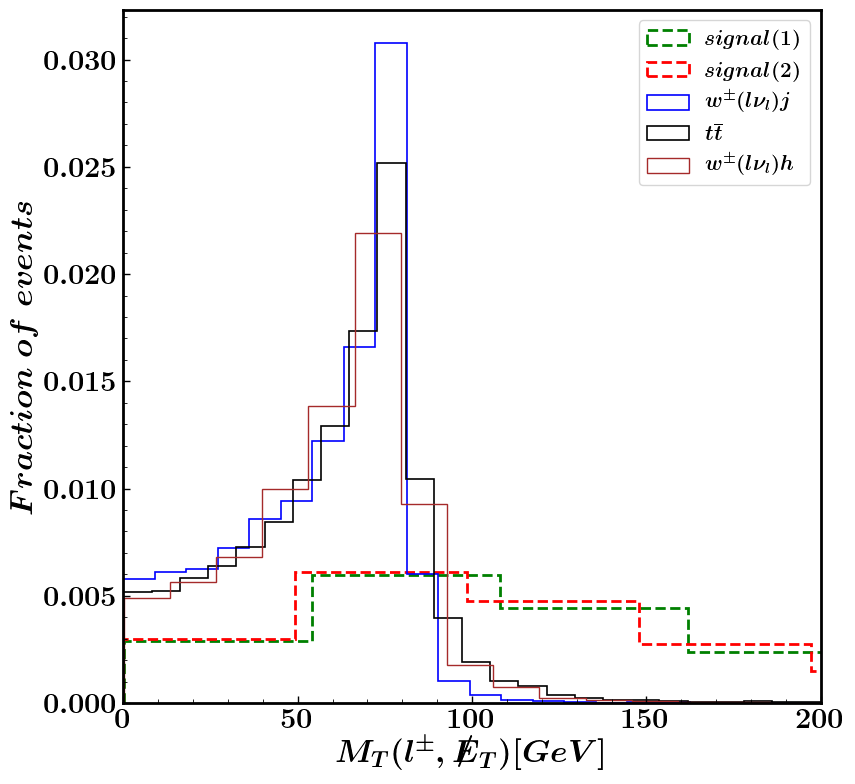}
\includegraphics[height=7cm,width=7cm]{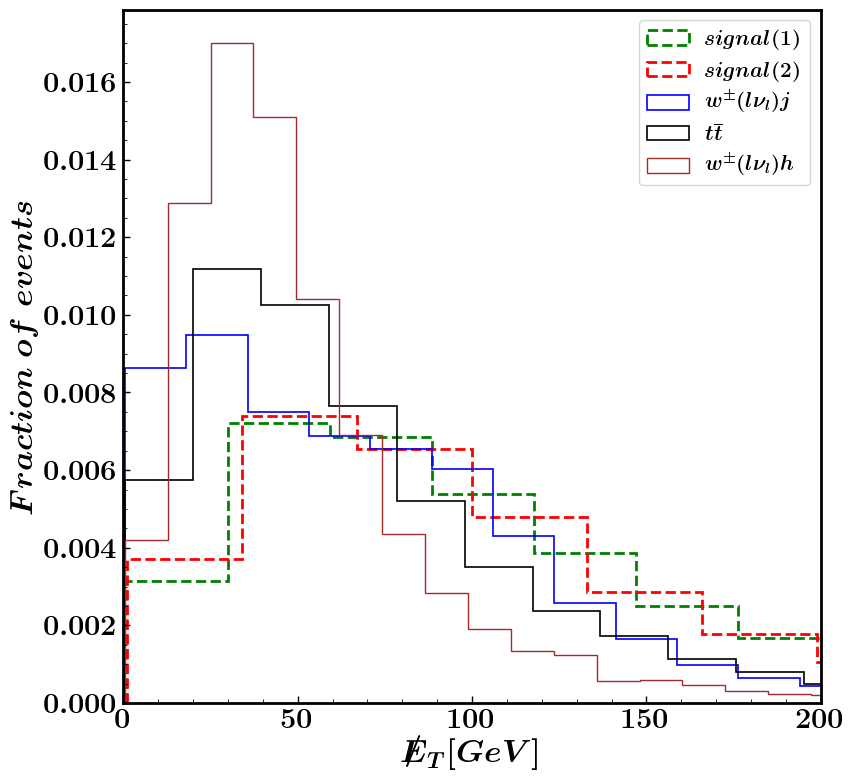}
\includegraphics[height=7cm,width=7cm]{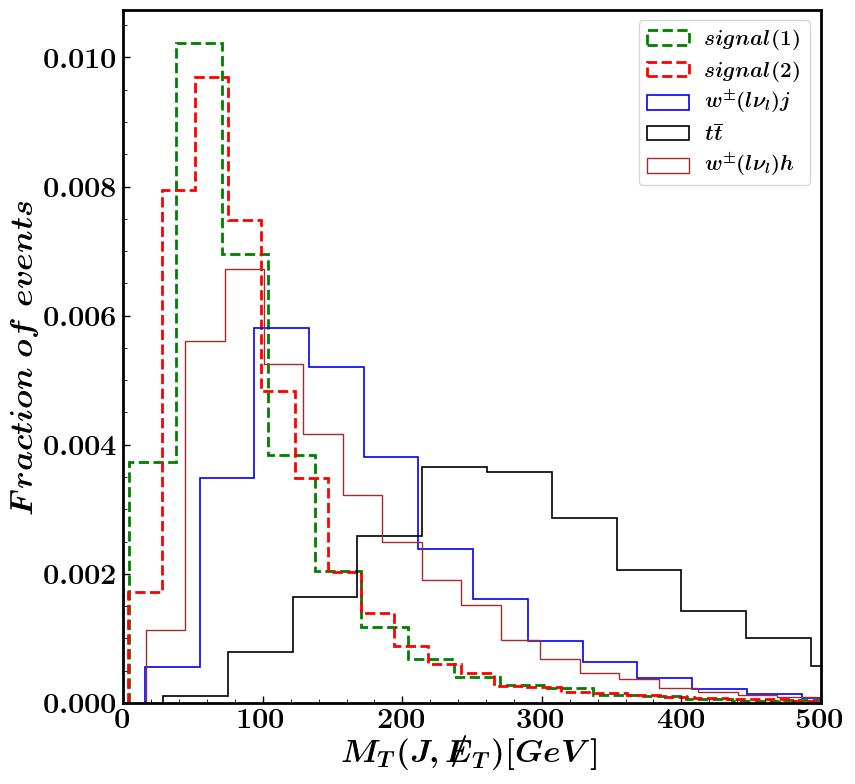}
\includegraphics[height=7cm,width=7cm]{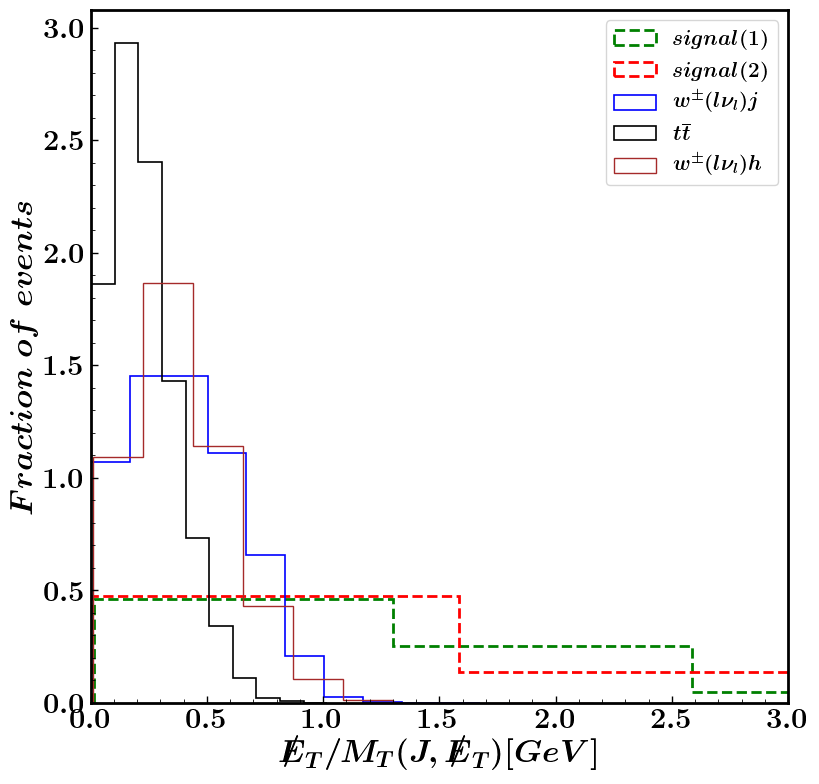}
\includegraphics[height=7cm,width=7cm]{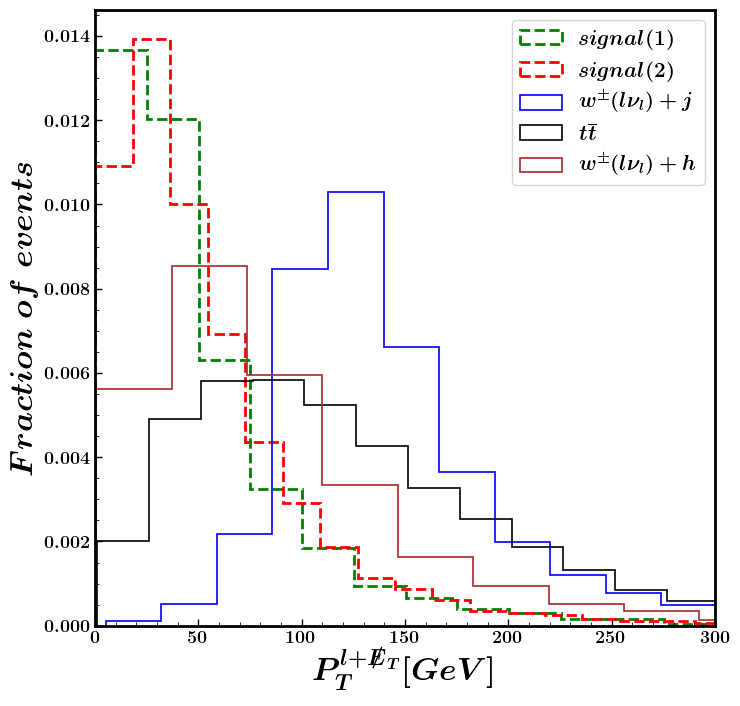}
\caption{The kinematic distributions of $P^J_T$ (upper-left), $M_T (l^{\pm},\slashed{E}_{T})$ (upper-right), $\slashed{E}_{T}$ (middle-left), $M_{T} (J,\slashed{E}_{T})$ (middle-right), $\slashed{E}_{T}/M_{T} (J,\slashed{E}_{T})$ (lower-left), and $P_{T}^{l+\slashed{E}_{T}}$ (lower-right) for the signal $s_{DS}$ and relevant SM backgrounds at 14 TeV LHC. }
\label{distributiondark}
\end{figure}

Finally, Tab.~\ref{basiccut3} displays the pre-selection cut-flow for the signal $s_{DS}$ ($pp\rightarrow W^{\pm}h'_1\rightarrow\left( l^{\pm}\nu\right)\left(\varphi_d\varphi^{\dagger}_d\right)$), as well as the background processes. As the final state of the signal $s_{DS}$ contains invisible dark vector mesons $\tilde{\omega}$, we choose the pre-selection $P_{T}^{J} > 50$ GeV and $|\eta_{J_{}}|<2.5$ of the fat-jet to ensure a sufficient yield of signal events. It is crucial to note that the signal $s_{DS}$ corresponds to a semi-visible jet resulting from dark showers, which induces a distinct $\slashed{E}_{T}$ distribution differing from the backgrounds such as $W^{\pm}j$, $t\overline{t}$, and $W^{\pm}h$. To effectively discriminate the signal $s_{DS}$ from these relevant backgrounds, we adopt the methodology outlined in Ref.~\cite{Cohen:2015toa}, utilizing the transverse mass $M_{T}(J,\slashed{E}_{T})$ with the following definition:
\begin{equation}
   M_{T}(J,\slashed{E}_{T}) = \sqrt{M_{J}^{2}+2( \sqrt{M_{J}^2+P_{T{J}}^{2}}\slashed{E}_{T}-P_{T_{J}}\slashed{E}_{T})}.
\end{equation} 
Based on the features of semi-visible jet, we select $\slashed{E}_{T}/M_{T}(J,\slashed{E}_{T}) > 1.22$ and $\slashed{E}_{T}> 100$ GeV. Since the final state of signal $s_{DS}$ has larger $\slashed{E}_{T}$ than those SM backgrounds, which can be seen in Fig.~\ref{distributiondark}, we choose $M_T (l^{\pm},\slashed{E}_{T}) > 140$ GeV to suppress contributions from relevant backgrounds.  
The effectiveness of these event selections in removing these SM backgrounds can be observed in Fig.~\ref{distributiondark}.

\subsection{Method-1: Convolutional Neural Network } 

By combining event features and the unique advantages of Convolutional Neural Networks (CNNs) in image recognition, we can use jet images of events as input data, providing the network with all the necessary event features. To construct the jet image, we extract the transverse momentum $P_{T}$, pseudorapidity $\eta$, and azimuthal angle $\phi$ of each Eflow object in the detector, and integrate this information provided by ECAL and HCAL into a digital image~\cite{deOliveira:2015xxd,Komiske:2018lor}. In this work, the jet image is set to a $128\times128$ square grid centered around the leading jet with radius $R_{J} =1.5$. Each pixel on the grid in the $(\Delta \eta, \Delta \phi)$ plane represents the sum of the $P_{T}$ of all Eflow objects falling within that grid.

To enable CNN to focus on the essential characteristics of the jet while disregarding irrelevant information like the jet's position and center of gravity, we conducted preprocessing on all jet images. This involved applying translation, rotation, and normalization techniques. By doing so, we ensured that the leading jets of all images were positioned at the center of the jet images, while the "center of gravity" of the jet images consistently pointed towards the 12 o'clock direction. Furthermore, the pixel values were normalized to fall within the interval of $[0, 1]$ as shown in Ref.~\cite{Ren:2021prq, Lv:2022pme}. The first is the translation process. In the detector, we assume a new $(\eta^{'}, \phi^{'})$ plane, and the coordinates of the leading jet are $(\eta, \phi)$. Now we translate the leading jet at the origin of the coordinate system by $\eta_{i}^{'}= \eta_{i}-\eta_{0},\ \phi_{i}^{'}=\phi_{i}-\phi_{0}$, where the index $i$ is the number of each pixel. Then there is a rotation process around the center of the jet image. We let the ``center of gravity'' in the jet image be expressed as $\eta_{C}=\frac{1}{\sum_{i=1}^{n}p_{Ti}}\sum_{i}^{n}\eta_{i}^{'}p_{Ti}, \ \phi_{C}=\frac{1}{\sum_{i=1}^{n}p_{Ti}}\sum_{i}^{n}\phi_{i}^{'}p_{Ti}$, which represents the weighted sum of the transverse momentum of all particles in the jet and $p_{Ti}$ represents the transverse momentum of particles in the jet. Then we rotate the ``center of gravity'' of the jet image to 12 o'clock around the coordinate origin as, $\eta_{i}^{''}= \eta_{i}^{'}\cos{\Theta}-\phi_{i}^{'}\sin{\Theta}, \ \phi_{i}^{''}=\phi_{i}^{'}\cos{\Theta}+\eta_{i}^{'}\sin{\Theta}$, with $\cos{\Theta}=\frac{\eta_{C}}{\sqrt{\eta_{C}^{2}+\phi_{C}^{2}}}, \ \sin{\Theta}=\frac{\phi_{C}}{\sqrt{\eta_{C}^{2}+\phi_{C}^{2}}}$. Finally, we use a normalization method to map the pixel values in the jet image to the $[0, 1]$ interval, thereby speeding up the gradient descent to find the optimal solution and improving the accuracy to a certain extent. We implement the normalization method through a linear transformation as, $P_{T}^{norm} = \frac{p_{Ti}}{\sum _{i} p_{Ti}}$. 

\begin{figure}[h]
\centering
\includegraphics[height=7cm,width=8cm]{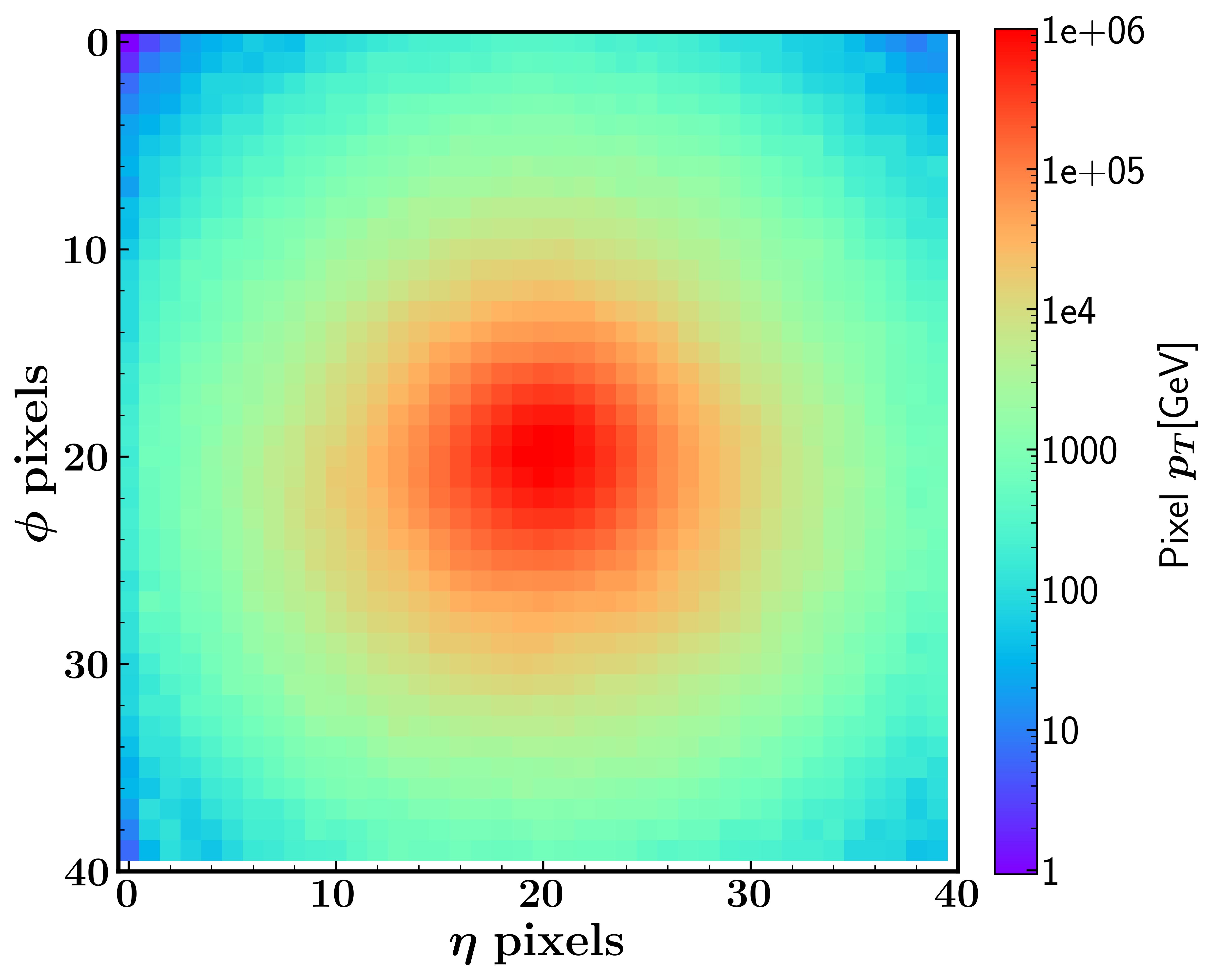}
\includegraphics[height=7cm,width=8cm]{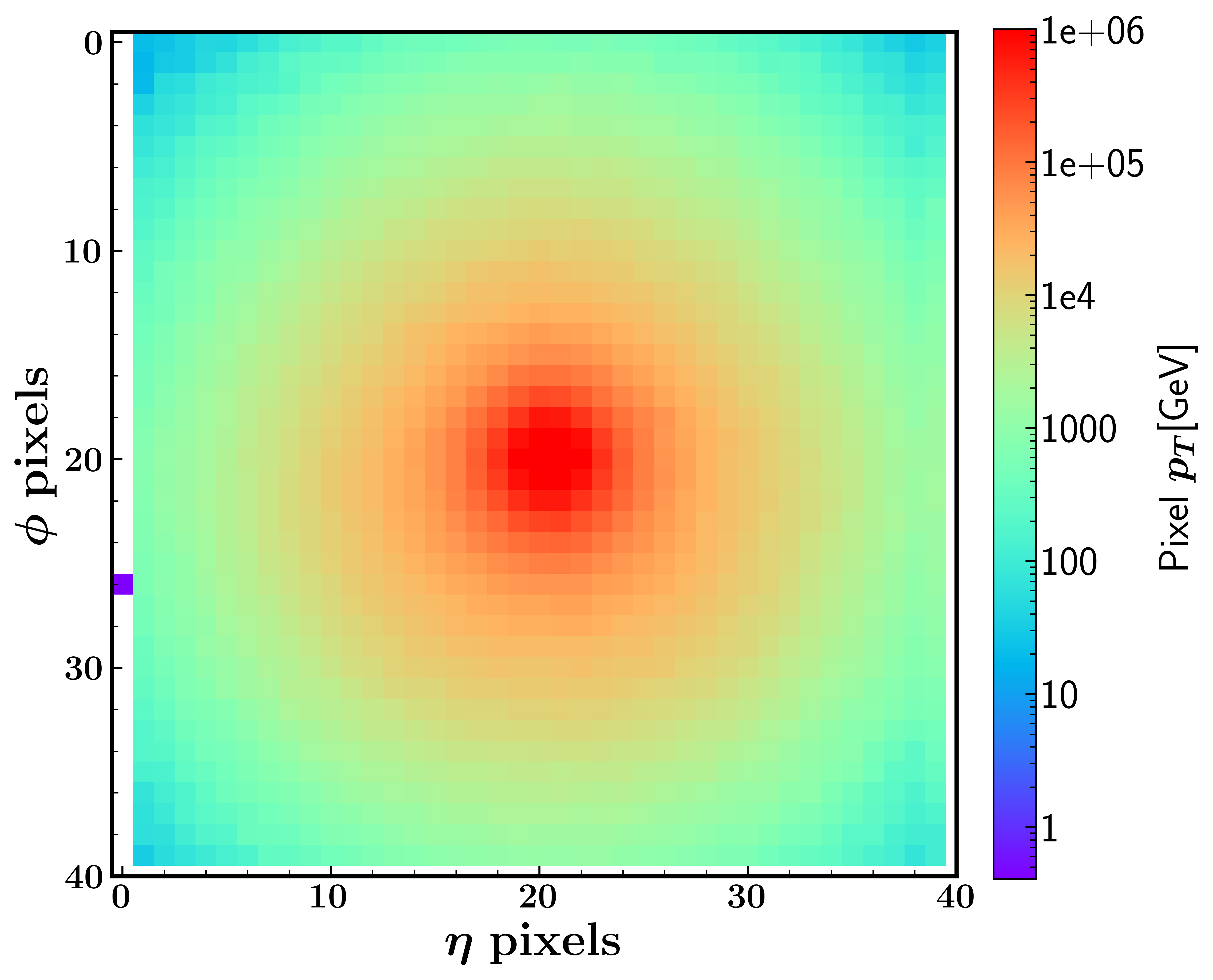}
\caption{The jet images of the leading jet in the signal $s_{8c}$ (left panel) and total backgrounds (right panel) after the translation and rotation.}
\label{pixell1}
\end{figure}

\begin{figure}[h]
\centering
\includegraphics[height=7cm,width=8cm]{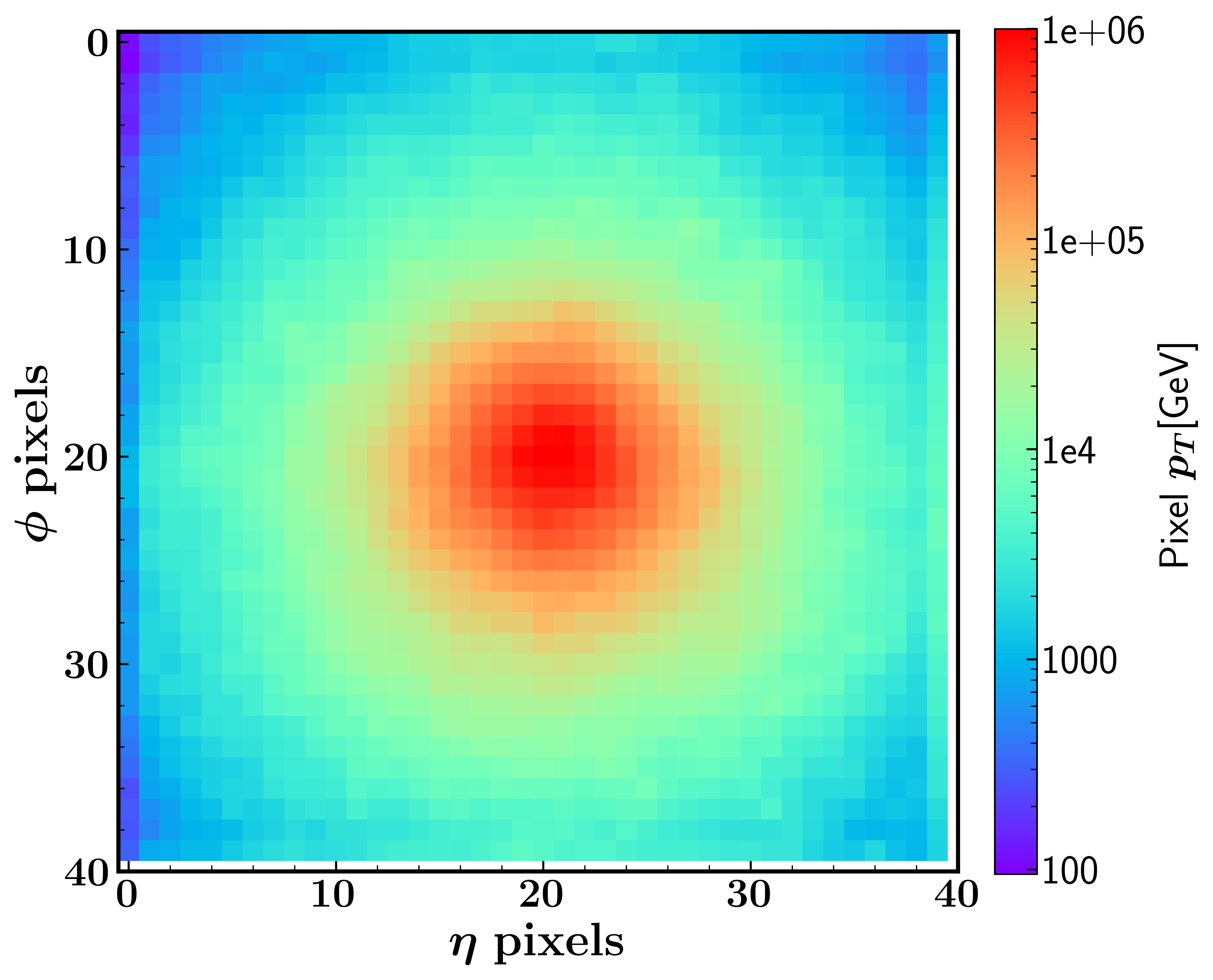}
\includegraphics[height=7cm,width=8cm]{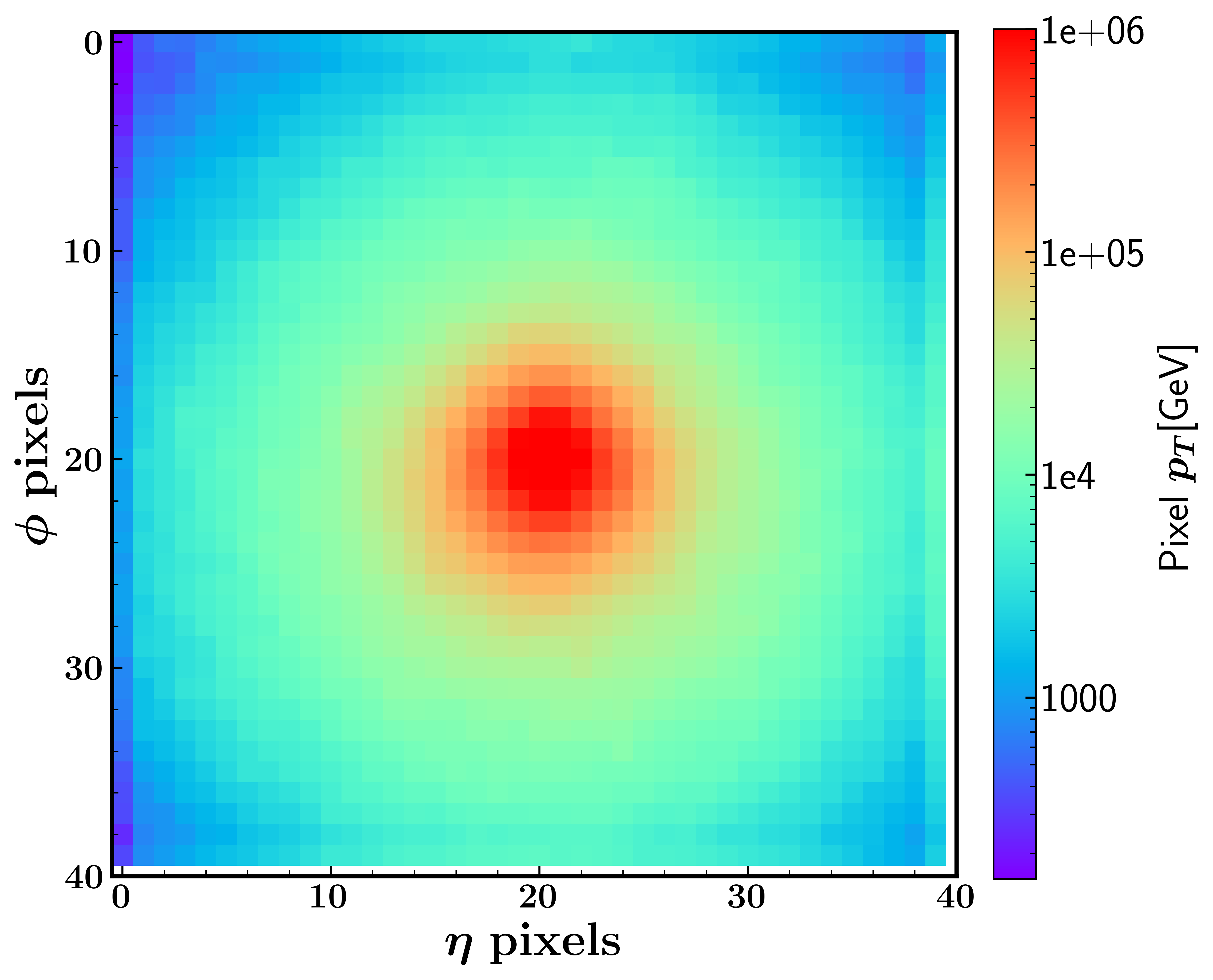}
\caption{The jet images of the leading jet in the signal $s_{c\tau}$ (left panel) and total backgrounds (right panel) after the translation and rotation.}
\label{pixell2}
\end{figure}

\begin{figure}[h]
\centering
\includegraphics[height=7cm,width=8cm]{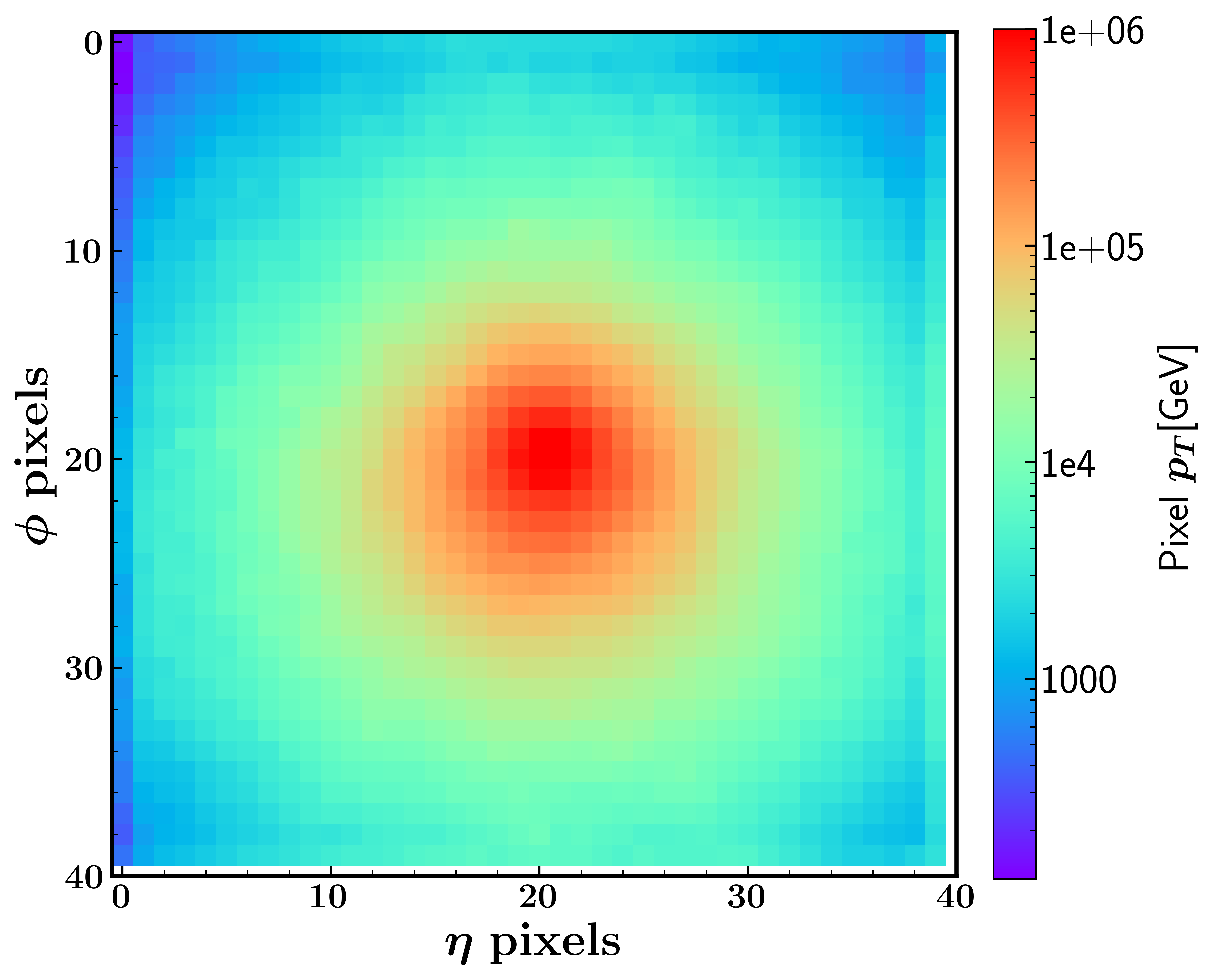}
\includegraphics[height=7cm,width=8cm]{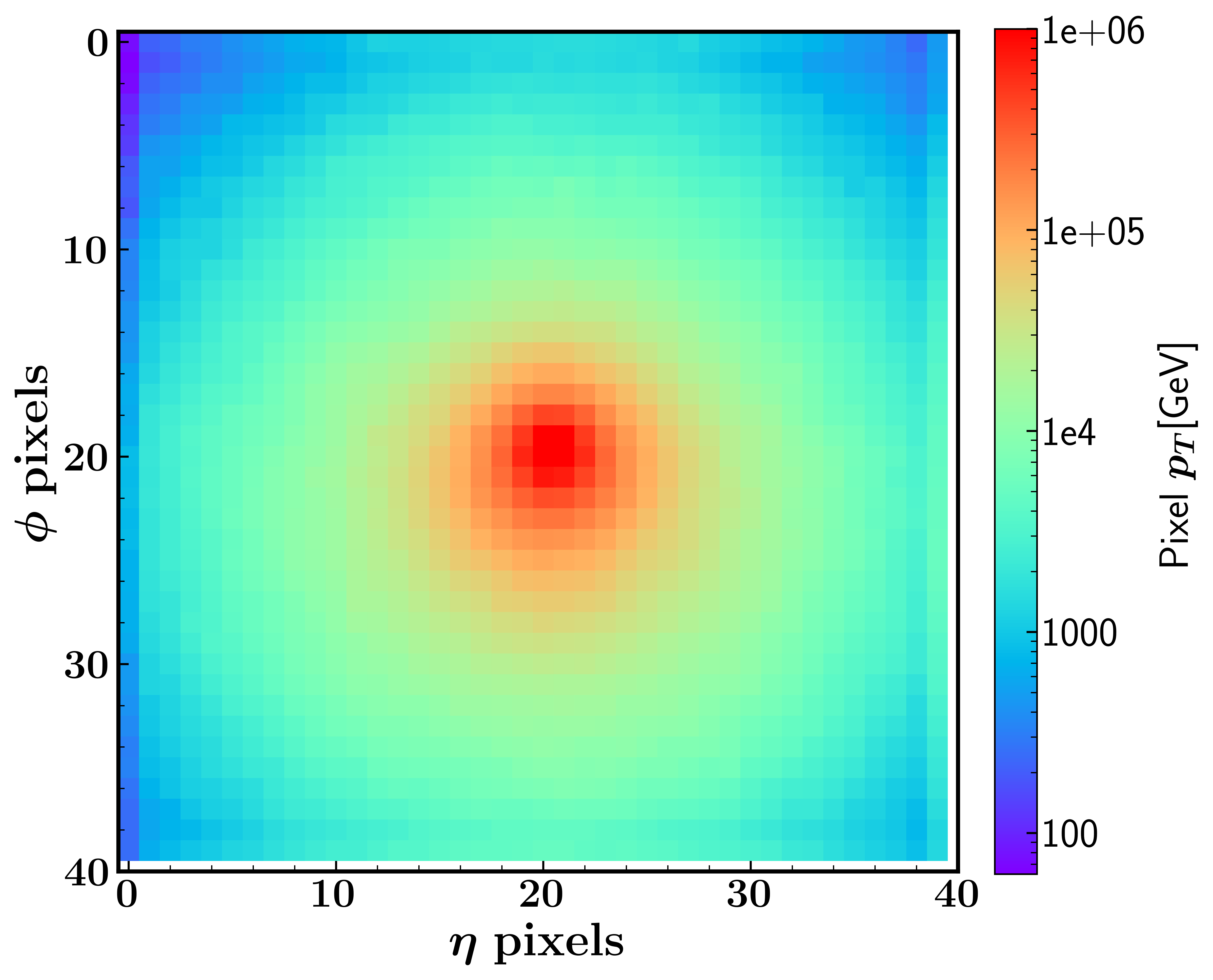}
\caption{The jet images of the leading jet in the signals $s_{c\tau}$ (left panel) and $s_{DS}$ (right panel) after the translation and rotation.}
\label{pixell3}
\end{figure}

After pre-selection in Tab.~\ref{basiccut1} and this preprocessing, we get the jet images of signal $s_{8c}$ and it's total backgrounds as shown in Fig.~\ref{pixell1}. We observe that the signal pixels are broader compared to the background ones. This discrepancy arises due to the presence of a fat-jet in the signal's final state, which contains a greater number of boosted subjets compared to the SM backgrounds. In addition, we also show the jet images of signal $s_{c\tau}$ and its total backgrounds after pre-selection in Tab.~\ref{basiccut2} and the preprocessing as shown in Fig.~\ref{pixell2}. We can find that the pixels of the signal are also wider than those of the backgrounds, which is also caused by more boosted subjects in the signal fat-jet. Moreover, comparing the right panels in Fig.~\ref{pixell1} and Fig.~\ref{pixell2}, it can be found that the jet image of the background of the signal $s_{8c}$ has wider pixels than the background of the signal $s_{c\tau}$. This is because the former added the option $200$ GeV $< P_{T}^{J}< 500$ GeV in the pre-selection but the option of the latter is $150$ GeV $< P_{T}^{J}< 500$ GeV, making the former contain more events with the hard fat-jet candidate. Furthermore, we compare the jet images of signals $s_{c\tau}$ and $s_{DS}$ after pre-selection in Tab.~\ref{diffsignal} of Appendix~\ref{app:2sig} and the preprocessing as shown in Fig.~\ref{pixell3}. We can see that the pixels of the jet image of signal $s_{c\tau}$ are wider than those of signal $s_{DS}$ because the final state of the latter contains invisible particles $\tilde{\omega}$ as shown in the right panel of Fig.~\ref{loss}. It can be seen that if these two signals appear at the same time in the final state, CNN is useful to distinguish them, and the results will be shown in Fig.~\ref{diffsignal} of Appendix~\ref{app:2sig}.

Our convolutional neural network was built with Keras package~\cite{chollet2015keras}. We use the GPU to accelerate the training of the classifier. In the CNN, We use Xception~\cite{chollet2017xception} as the backbone network to extract input jet image features and generate feature maps for our networks. The input is a  $128\times 128$ jet image with one channel. Followed by the input layer, a convolutional layer extracts information, a Batch Normalization layer speeds up the training convergence speed, and then a ReLu activation function layer increases the nonlinear relationship between the layers of the neural network. Then, add a depth separable convolutional layer, SeparableConv, to reduce the number of parameters and computations and improve the efficiency of convolution operations after the aforementioned processes. After repeatedly stacking the above network layers in the backbone network to fully extract signal features, we expand the resulting feature map into a vector with a single dimension of 2048 using a MaxPooling layer. We then add a dropout layer to mitigate overfitting and, finally, the weight of the predicted event is obtained as a signal ($s$) through the sigmoid function:
\begin{equation}
S = \frac{1}{1+\exp^{-s}}.
\label{sigmoid}
\end{equation} 
The output $S$ is a real number between $0$ and $1$, which can be interpreted as the probability of the event occurring given the input.

In addition to the above-mentioned feedforward process, we also need a feedback process to adjust the parameters and optimize the classification accuracy of the model. We use a binary cross-entropy loss function $\cal L$ to measure the gap between the output of the model and the real one, and point out the direction for the optimization of the model,
\begin{equation}
\mathcal{L}=-\frac{1}{N}\sum_{i=1}^{N}(y_{i}\log{p_{i}}+(1-y_{i})\log{(1-p_{i}))},
\label{cross_entropy loss}
\end{equation}
where $N$ is the number of events, $y$ and $p$ are the true and predicted category of the sample. In this work, the label of the signal is denoted by one and the label of the background is set to zero. We take the Adam optimizer with a learning rate of $0.001$~\cite{Kingma:2014vow}.

The total number of events used for machine learning is 800k, representing the total number of events with the remaining signal and background after preprocessing. However, for signals and backgrounds, the actual number of generated events should be 400k, divided by their respective efficiency.
We generate a total of 80k events of signal and background, of which $70\%$ are used for training, $20\%$ for validation, and $10\%$ for testing. The number of batch sizes we use in training is $100$, and the maximum epoch is set to $100$. Each epoch will use all the data once, and in order to prevent unnecessary information such as learning the order of the data during the training process, we scrambled the data in advance. Our validation set loss function reaches the minimum value at the 52nd epoch, after which we will learn some information that is not conducive to our classification, making the classification even worse. To avoid overfitting, we set an early stop-to-end training when the epoch is $52$. Finally, we apply a model with the least loss function in the validation dataset to classify our data.

\subsection{Method-2: Energy Flow Network} 

\begin{figure}[h]
\centering
\includegraphics[height=8cm,width=8cm]{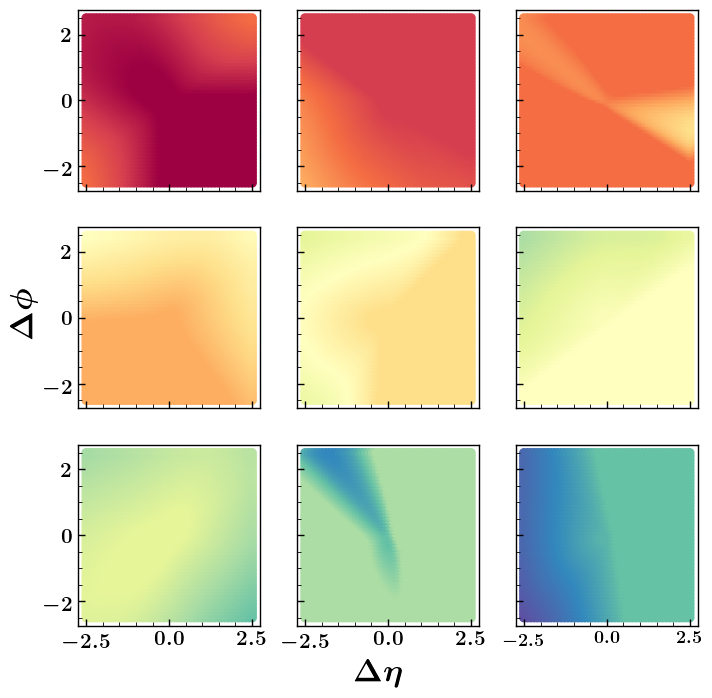}
\includegraphics[height=8cm,width=8cm]{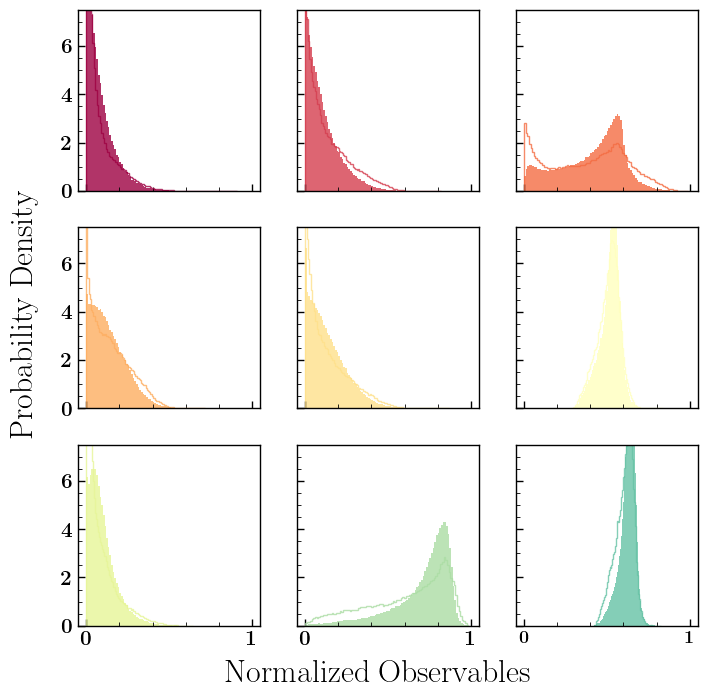}
\caption{The left panel displays the visualization of $9$ out of the $512$ filters, illustrating the model's learned "pixelated" representation resembling an image in the rapidity-azimuth plane. The right panel showcases the corresponding normalized observable distribution obtained through the energy flow network learning. In this distribution, the signal's observable measurements are depicted by separate lines, while the background's observables are represented by the filled area.} 
\label{response}
\end{figure} 

Although jets are naturally represented as images in calorimeters, jet images are not efficient as inputs to deep learning analysis when the finer resolution or sparser distribution appears. A point cloud is suitable for processing the unordered variable-length set of data points, providing a more direct representation of the kinematic and other additional properties of constituents in jets. Inspired by the Energy Flow Polynomials (EFPs)~\cite{Komiske:2017aww} decomposition of the jet substructure, an energy-weighted deep sets algorithm Energy Flow Network (EFN)~\cite{Komiske:2018cqr} is proposed to extract information from point cloud representation of jets, which satisfy both permutation symmetry and infrared and collinear safety (IRC-safety). EFN and variants (Particle Flow Network (PFN), PFN-ID~\cite{Komiske:2018cqr})  as fast and IRC-safe backbone networks have been used in top jet tagging~\cite{Komiske:2018cqr}, di-Higgs classification~\cite{Tannenwald:2020mhq} and BSM anomaly detection~\cite{Ostdiek:2021bem}. EFN is based on a theorem that any IRC-safe observable can be approximated arbitrarily well as:
$
\mathcal{O}(\{p_{1},...,p_{M}\})=F(\sum_{i}^{M}p_{T,i}^{norm}\Phi(\Delta\eta_{i},\Delta\phi_{i})),
$
where we use the same kinematic information and preprocessing method in CNN, the index $i$ is the number of each constituent, and $\Phi: \mathcal{R}^{2}\rightarrow \mathcal{R}^{l}, F: \mathcal{R}^{l}\rightarrow \mathcal{R}$ can be parameterized by multi-layer perceptrons and trained as normal DNN architectures does, jet structure in latent space can be visualized directly by showing the filter $\Phi$ in the rapidity-azimuth plane as shown in Fig.~\ref{response}. 

\begin{figure}[h]
\centering
\includegraphics[width=10cm]{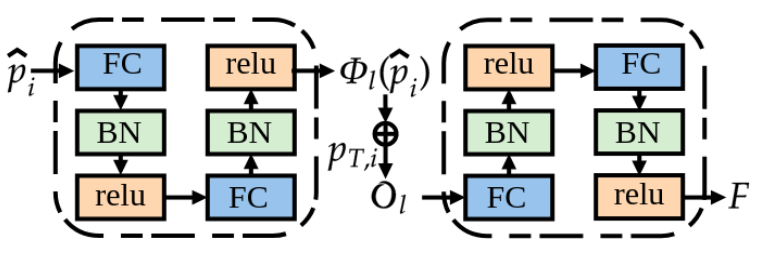}
\caption{Energy flow network architectures using FC-BN-relu structure, where energy-weighted summing operator keeps permutation symmetry and IRC-safety.}
\label{efn}
\end{figure}

We trained EFN as a comparison to the Xception model and visualized the filters $\Phi$ and the corresponding observable distributions of signal and background jets learned by the machine. To parametrize $\Phi$ and $F$, we used two Fully-Connected Batch-Normalization relu (FC-BN-relu) structures depicted in Fig.~\ref{efn} with $512$ hidden units, and a two-layer discriminator with sigmoid activation was added for the last score output. The dataset and dataset segmentation followed the same setting as method-1. We used a larger batch size of $1024$ since the trainable parameters and the floating point operations (FLOPs) of EFN are much less than the Xception model used in method-1. We optimized $\Phi$ and $F$ using the Adam optimizer with an initial learning rate of $0.01$ and a cosine decay learning rate scheduler over $100$ epochs, following the same early stop setting as in method-1. The jet observables distribution learned by the machine and the classification performance of the test dataset is shown in Figs.~\ref{response} and~\ref{results}.

\section{Numerical Results and Discussions} 
\label{sec:result}

\begin{figure}[h]
\centering
\includegraphics[height=8cm,width=8cm]{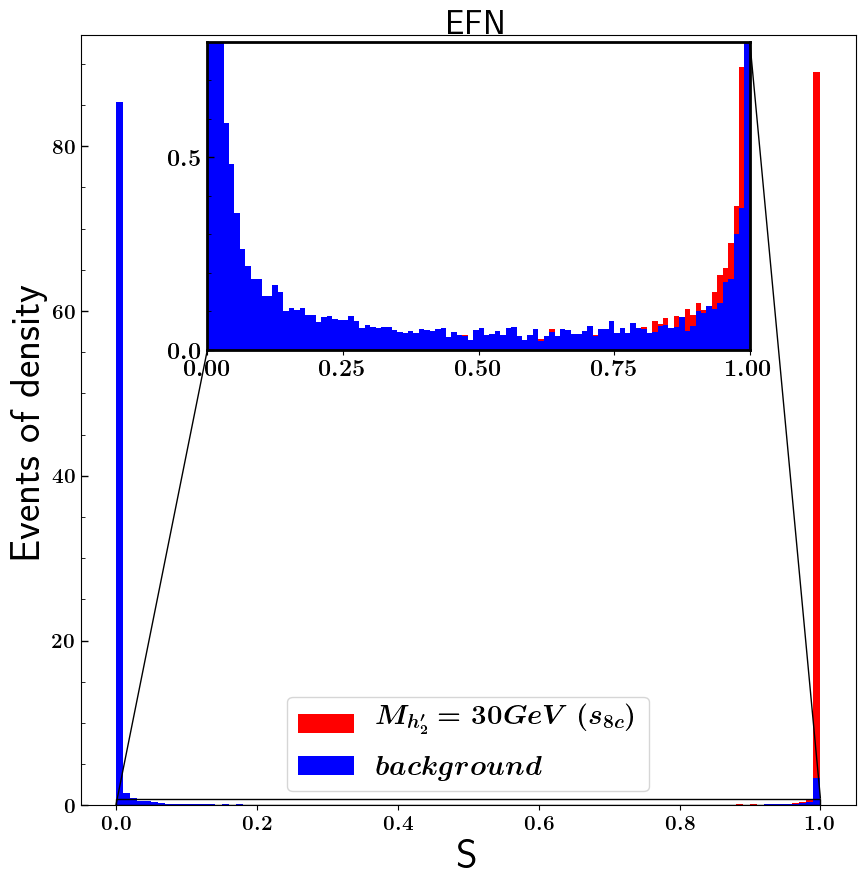}
\includegraphics[height=8cm,width=8cm]{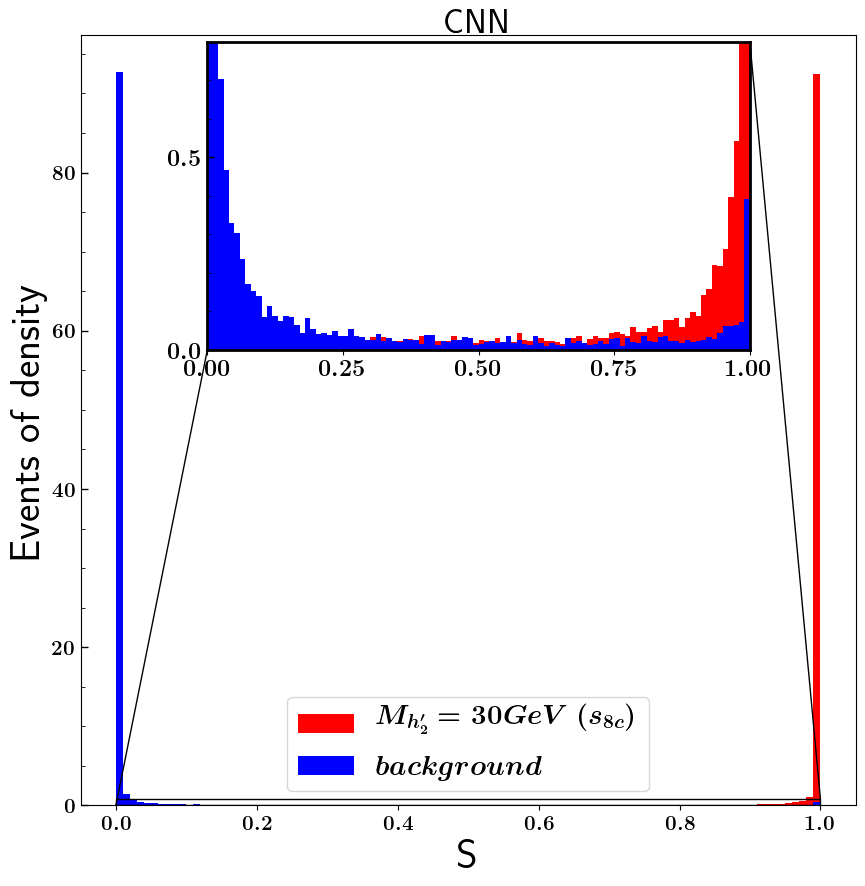}
\caption{ The classification performance of the test set was evaluated for the signal $s_{8c}$ ($M_{h'_2} = 30$ GeV) and background events obtained from EFN (left panel) and CNN (right panel). The dataset used for both signal and background followed a $1:1$ ratio, with a total of 400k events each. }
\label{results}
\end{figure}

The classification performance of the trained EFN and CNN on the test dataset for the signal $s_{8c}$ and total backgrounds are depicted in Fig.~\ref{results}. The horizontal axis denotes the probability of an event being classified as a signal ($S$), while the vertical axis represents the event density. The distribution reveals that the background (represented in blue) is predominantly concentrated around zero, while the signal (represented in red) is mainly concentrated near one. These outcomes showcase the exceptional classification performance of our model.
 
In addition, receiver operating characteristic (ROC) curves can be obtained by setting a threshold $s_0$ between zero and one on the abscissa axis of Fig.~\ref{results} and plotting the probability densities of the signal and background in the $s>s_0$ region. The ROC curves are represented by the horizontal and vertical coordinates, which indicate the signal efficiency ($\varepsilon_S$) and background rejection ($1-\varepsilon_B$), respectively. The signal efficiency can be calculated as follows:
\begin{equation}
   \varepsilon_S =\frac{N_{\text{signal}}^{(s > s_{0})}}{N_{\text{signal}}^{(\text{total})}}. 
\end{equation} 
The background efficiency is calculated by summing the weighted efficiencies of all backgrounds, 
\begin{equation}
   \varepsilon_B =\frac{\sigma_{b_{1}}N_{b_{1}}^{(s > s_{0})}+\sigma_{b_{2}}N_{b_{2}}^{(s > s_{0})}+\sigma_{b_{3}}N_{b_{3}}^{(s > s_{0})}}{\sigma_{b_{1}}N_{b_{1}}^{(\text{total})}+\sigma_{b_{2}}N_{b_{2}}^{(\text{total})}+\sigma_{b_{3}}N_{b_{3}}^{(\text{total})}},
\end{equation}
where $\sigma_{b_{1}}$, $\sigma_{b_{2}}$ and $\sigma_{b_{3}}$ are the weights of three backgrounds, respectively. Three main backgrounds, namely $W^{\pm}(l\nu_{l})h$, $W^{\pm}(l\nu_{l})j$, and $t\overline{t}$, are denoted as $b_{1}$, $b_{2}$, and $b_{3}$, respectively. 
The above two equations show that $\varepsilon_S$ and $\varepsilon_B$ correspond to the proportion of events that pass the selection criteria among the initial signal and background, respectively.
 \begin{figure}[h]
\centering
\includegraphics[height=8cm,width=8cm]{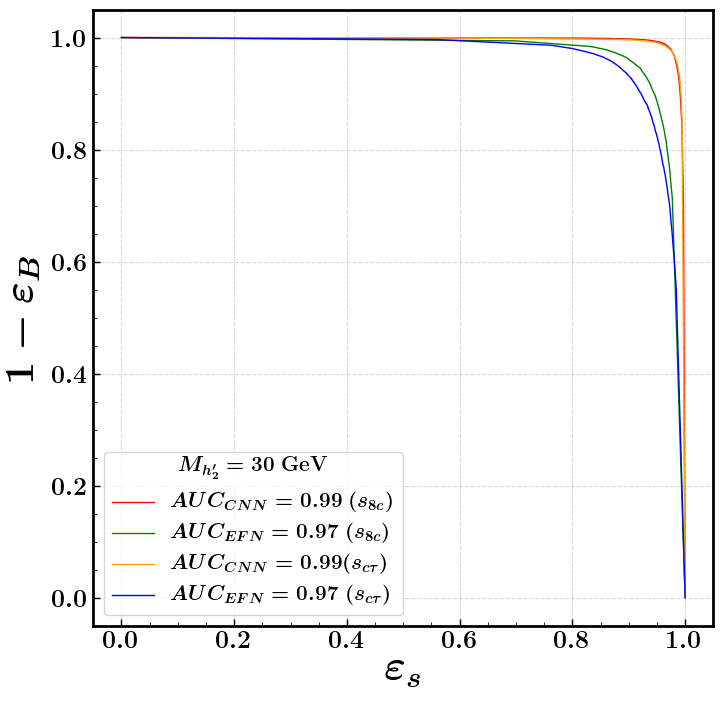}
\includegraphics[height=8cm,width=8cm]{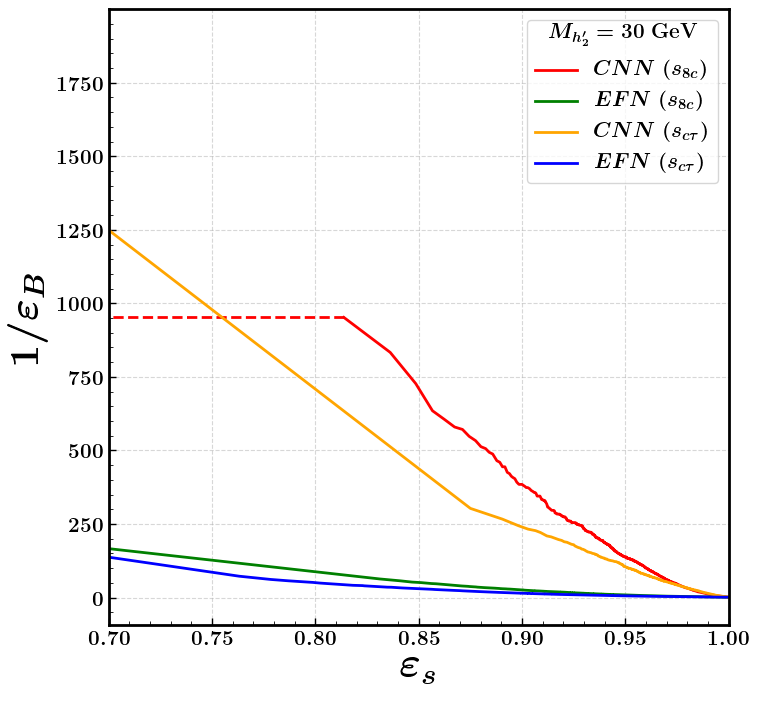}
\includegraphics[height=8cm,width=8cm]{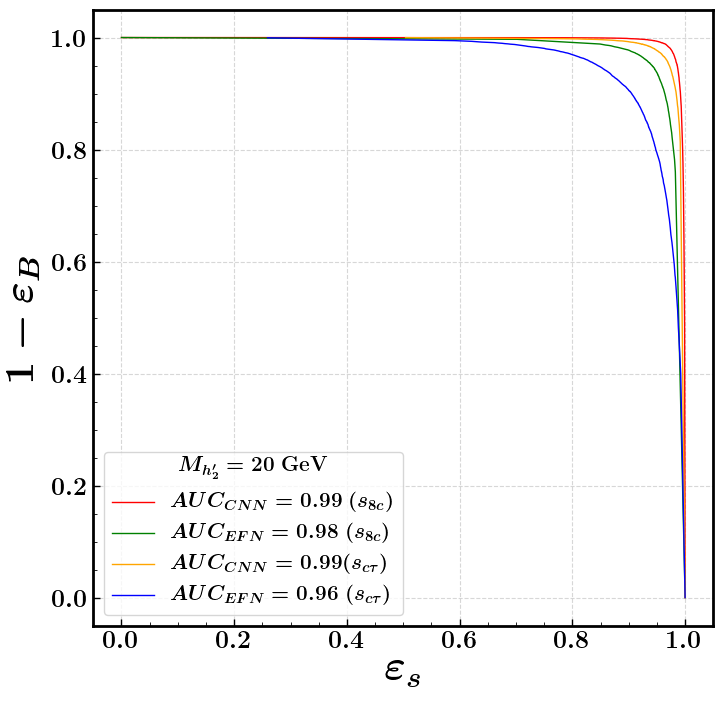}
\includegraphics[height=8cm,width=8cm]{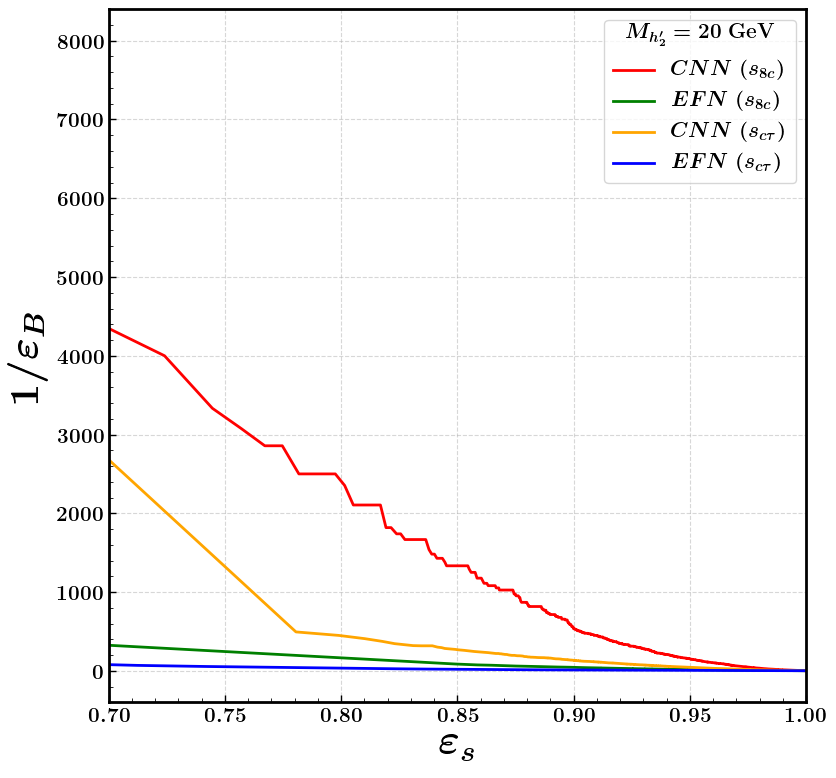}
\caption{The left panels show the ROC curves and corresponding AUC values of CNN and EFN for $M_{h'_2}$ values of $30$ GeV (top) and $20$ GeV (bottom). In the right panels, we depict the relationship between $\varepsilon_S$ and $\frac{1}{\varepsilon_B}$ for the same $M_{h'_2}$ values where the dashed line represents the assumption of background-free region. }
\label{roc}
\end{figure}
The ROC curve is a widely used metric for evaluating the performance of a classification model, and the area under the curve (AUC) is a crucial indicator of its effectiveness.

Fig.~\ref{roc} presents the ROC plots for $M_{h'_2}=20$ GeV and $30$ GeV, along with the corresponding AUC values (left panel) for both the EFN and CNN networks, as well as the relationship between $\varepsilon_S$ and $\frac{1}{\varepsilon_B}$ (right panel). As we approach an $s_0$ cut of $1$, the background becomes negligible due to limited statistics. In order to maintain the completeness of our image, we extend the graph with a dotted line starting from the last data point where the background is still measurable. 
We can find when $\varepsilon_S = 0.7$ for both $M_{h'_2}=20$ GeV and $30$ GeV, the CNN and EFN networks yield $\frac{1}{\varepsilon_B}$ values of $10^2$ and $10^3$, respectively, indicating that utilizing the ($P_{T}$, $\eta$, $\phi$) information of signal and background events can improve the identification of the signal over backgrounds for both networks. However, due to its deeper network structure and more parameters for extracting local features of events, the CNN exhibits  superior performance compared to the EFN in distinguishing the signal from the background, resulting in a better classification effect. Although the EFN has a shorter training time, its classification effect is inferior to that of CNN, and hence we focus on the results of CNN in the subsequent discussion.

\begin{figure}[h]
\centering 
\includegraphics[height=8cm,width=8cm]{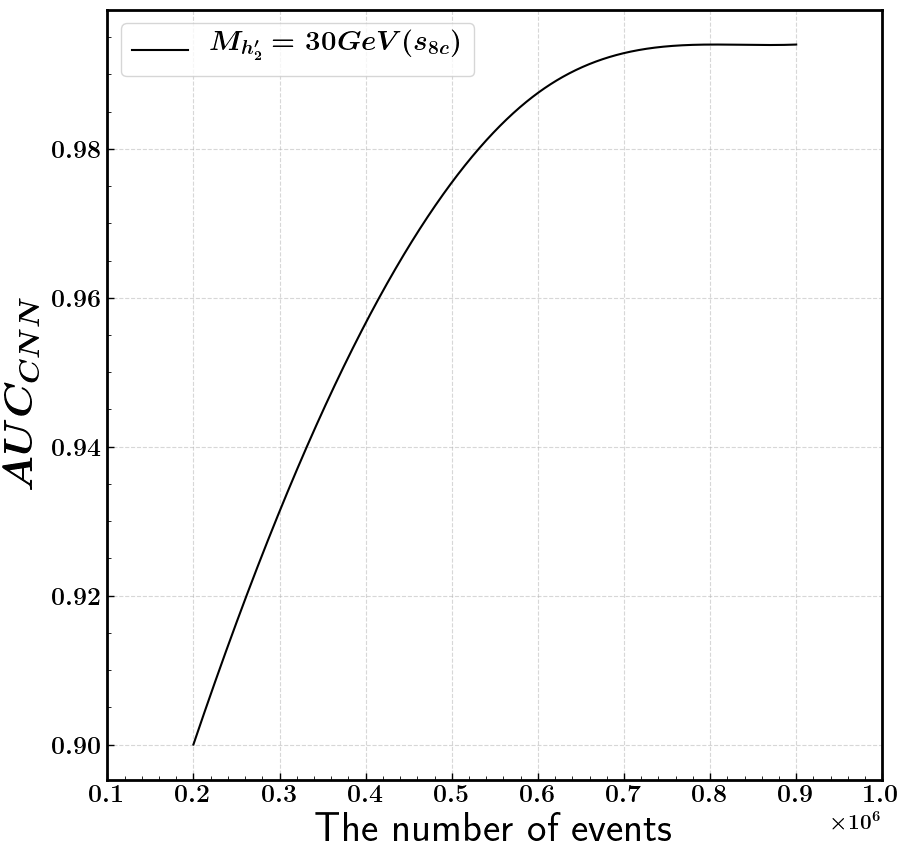}
 \includegraphics[height=8cm,width=8cm]{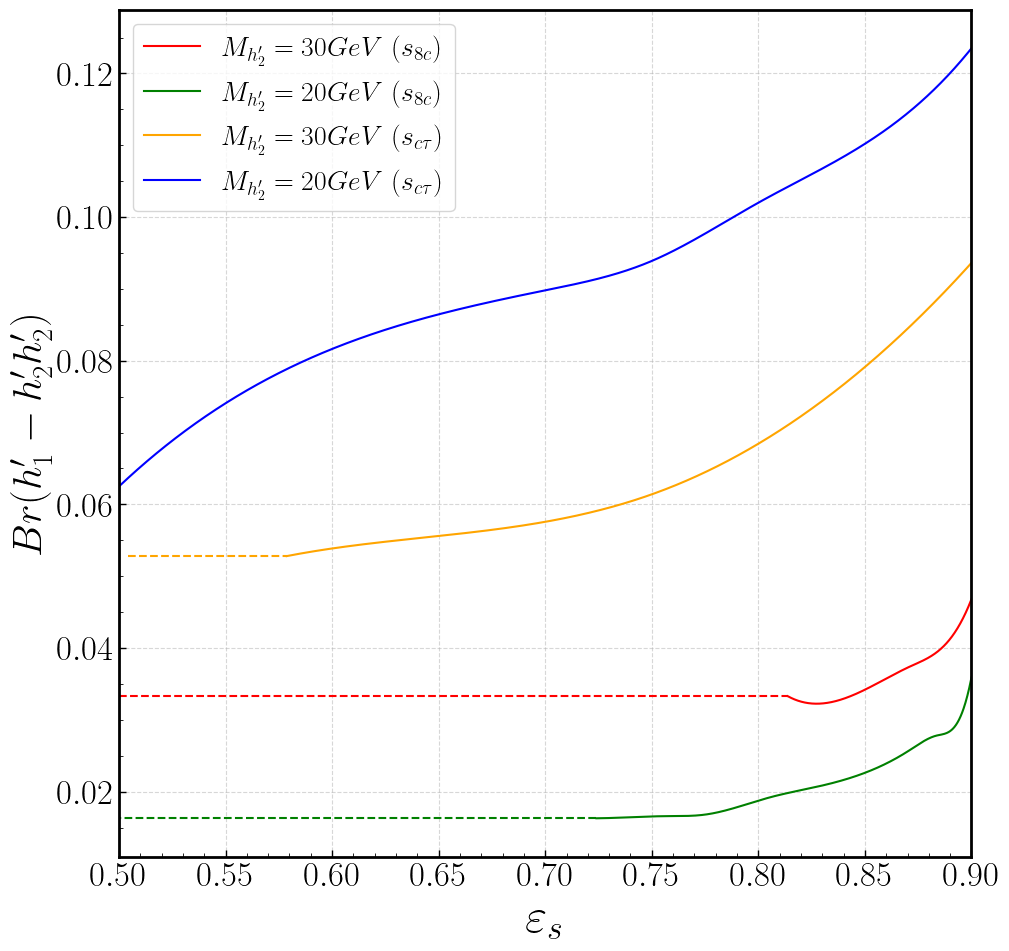}
\caption{The left panel shows the impact of the generated number of events on the classification performance based on AUC of the CNN for the signal $s_{8c}$. The right panel shows the predicted future bounds on the branching ratio $Br(h'_1\rightarrow h'_2 h'_2)$ for the signals $s_{c\tau}$ and $s_{8c}$ at 14 TeV LHC with an integrated luminosity of 3000 $fb^{-1}$, obtained through a CNN. The results are presented for $M_{h'_2}=20$ GeV and $30$ GeV, respectively. The dashed lines indicate the assumption of background-free region. 
}
\label{br}
\end{figure}

The performance of our classifier is quantified by means of the area under the ROC curve, as presented in the left panel of Fig.~\ref{br}, which displays the dependence of the classifier's AUC on the number of events used for training. The classification performance improves gradually with the number of training events but eventually saturates at a maximum value rather than diverging. Our classifier achieves its highest classification performance at 800k training events, and increasing the number of events beyond this threshold does not lead to further improvements. Therefore, in order to enhance the sensitivity of our signal detection, a CNN model trained on a dataset of 800k records is employed. 

Consider the higher-order modifier of the k-factor
\begin{table}[h]
\small
    \centering
    \begin{tabular}{|c|c|c|c|}
    \hline
    \makecell[c]{Future bounds in \\ $Br(h'_1\rightarrow XX)$}& $s_{DS}$ ($ h'_1\rightarrow\varphi_d\varphi^{\dagger}_d$)& $s_{c\tau}$ ($h'_1\rightarrow h'_2 h'_2$) & $s_{8c}$ ($h'_1\rightarrow h'_2 h'_2$) \\
    \hline
    \makecell[c]{ $M_{h'_2} = 20$ GeV \\ }&  $0.62\%$    &  $6.25\% - 12.35\%$ & $1.63\% - 3.59\%$     \\
    \hline
    \makecell[c]{ $M_{h'_2} = 30$ GeV \\  }&  $0.73\%$  & $5.28\% - 9.46\%$ & $3.33\% - 4.68\%$ \\
  \hline
    \end{tabular}
        \caption{The future bounds in $Br(h'_1\rightarrow XX)$ for signals $s_{DS}$ ($ h'_1\rightarrow\varphi_d\varphi^{\dagger}_d$), $s_{c\tau}$ ($h'_1\rightarrow h'_2 h'_2$) and $s_{8c}$ ($h'_1\rightarrow h'_2 h'_2$) at 14 TeV LHC with $\mathcal{L} = 3000 fb^{-1}$. In the final column, $Br(\eta_d\rightarrow c\overline{c})=100\%$ is assumed as an ideal scenario for $s_{8c}$ as a comparison. The intervals in the last two columns represent the future bounds in $Br(h'_1\rightarrow XX)$  when the $\varepsilon_S$ value is from $0.5$ to $0.9$. }    
    \label{restrictions}
\end{table}

To predict future bounds of $h'_{1}$ exotic decay branching ratios, we first define the signal significance $Z$ with a confidence level of $95\%$~\cite{Cowan:2010js} as 

\begin{equation}
Z\equiv\sqrt{2(({N_{S}+N_{B})ln({1+\frac{N_{S}}{N_{B}}})-N_{S})}} = 1.96.
  \label{eq:Z196}
\end{equation}
where the numbers of signal and background events $N_{S} = Br(h'_1\rightarrow XX)\times \sigma_{S} \times \mathcal{L}$ and $N_{B} = \sigma_{B} \times \mathcal{L}$, respectively. The $Br(h'_1\rightarrow XX)$ represents to $Br(h'_1\rightarrow h'_2 h'_2)$ for the signals $s_{8c}$, $s_{c\tau}$ and it represents to $Br(h'_1\rightarrow\varphi_d\varphi^{\dagger}_d)$ for the signal $s_{DS}$. 
The cross sections of signal and total background, obtained after applying pre-selected cuts and machine learning cuts, are denoted as $\sigma_{S}$ and $\sigma_{B}$, respectively. The signal cross section is calculated as $\sigma_{S} = \sigma_{\text{signal}} \times \varepsilon_S \times\varepsilon_{\text{pre}}$, where $\varepsilon_S$ represents the efficiency of machine learning for the signal, and $\varepsilon_{\text{pre}}$ represents the efficiency of the pre-selecting cut. The value of $\sigma_{\text{signal}}$ can be found in Tab.~\ref{basiccut1},~\ref{basiccut2} and~\ref{basiccut3}, which represents the signal cross section in the generator level. The definition of $\sigma_{B}$ is in the same way. 

Our trained model predicts the sensitivity for $h'_{1}$ exotic cascade decay with the benchmark points $M_{h'_2}$ = $20$ GeV and $30$ GeV at $14$ TeV LHC with $\mathcal{L} = 3000 fb^{-1}$ as shown in the right panel of Fig.~\ref{br}. By imposing direct constraints on the coupling coefficient linked to the mixing angles and other model parameters in Eq.~(\ref{eq:potential}), we can effectively restrict the $h'_1$ exotic decay branching ratio. Consequently, we provide a comprehensive summary of forthcoming bounds on $Br(h'_1\rightarrow XX)$ for the following signals: $s_{DS}$ ($ h'_1\rightarrow\varphi_d\varphi^{\dagger}_d$), $s_{c\tau}$ ($h'_1\rightarrow h'_2 h'_2$), and $s_{8c}$ ($h'_1\rightarrow h'_2 h'_2$), anticipated at $14$ TeV LHC with an integrated luminosity of $\mathcal{L} = 3000 fb^{-1}$. The tabulated results can be found in Table~\ref{restrictions}.Note we assume $Br(\eta_d\rightarrow c\overline{c})=100\%$ as an ideal scenario in the final column for $s_{8c}$ as a comparison. 
The intervals given in the last two columns represent the future bounds on $Br(h'_1\rightarrow XX)$ corresponding to the range of $\varepsilon_S$ from $0.5$ to $0.9$ in Fig.~\ref{br}, while the dashed line in  Fig.~\ref{br} represents the assumption of background-free region.

The signal $s_{DS}$ generated through dark showers produces a unique signature, a semi-visible jet, at the LHC due to the stable $\tilde{\omega}$ and prompt decay of $\eta_d$ to $c\overline{c}$ or $\tau^+\tau^-$ in the final state. This signature is specific to the signal $s_{DS}$ and does not appear in backgrounds or other signal types, such as $s_{8c}$ and $s_{c\tau}$, making it distinguishable. Therefore, we simply apply traditional cut-flow methods, and the results are presented in Tab.~\ref{basiccut3}. While machine learning techniques can further improve the signal significance, generating a very large number of Monte Carlo background events (more than $10^{10}$) before the pre-selection shown in Tab.~\ref{basiccut3} is beyond the scope of this work.  

The signals $s_{8c}$ and $s_{c\tau}$ resulting from $h'_1$ exotic cascade decays are challenging to be separated from backgrounds using traditional cut-flow methods as shown in Tab.~\ref{basiccut1} and Tab.~\ref{basiccut2} since these backgrounds can mimic the signal events. However, the jet images of the fat-jet for the signals $s_{8c}$ and $s_{c\tau}$ are still different from the total backgrounds depicted in Fig.~\ref{pixell1} and~\ref{pixell2}. Therefore, sophisticated analysis methods, such as machine learning techniques, are crucial to analyze the fat-jet structure for this problem. Here, we specifically focus on CNN because of its powerful pattern recognition ability to distinguish the signals $s_{8c}$ and $s_{c\tau}$ from the total backgrounds.

We begin by considering the idealized signal $s_{8c}$, which only contains charm quarks inside the fat-jet, and use it as a baseline to establish the best achievable outcome. However, the real signal $s_{c\tau}$ contains a mixture of multiple charm quarks and tau leptons inside the fat-jet, requiring more labeled datasets to cover all possible decay modes. This reduces its signal-to-background discrimination capability with the same amount of training set compared to the signal $s_{8c}$. By analyzing the simple signal $s_{8c}$, appropriate neural network architectures, hyperparameters, and optimization strategies can be selected to process the signal $s_{c\tau}$, thereby improving the ability to distinguish this signal from total backgrounds. On the other hand, in signal $s_{c\tau}$, the tau lepton decays produce neutrinos in the final state, which cannot be detected and lead to missing momentum information inside the fat-jet. Therefore, the transverse momentum and invariant mass distributions of the leading jet in signal $s_{c\tau}$ become relatively smaller than those in signal $s_{8c}$, resulting in poorer performance in signal $s_{c\tau}$ after preprocessing.

We incorporate the systematic uncertainty into the formula for the signal significance $Z$ in Eq.~(\ref{eq:Z196}), which can be represented as follows: 
\begin{equation}
Z\equiv\sqrt{2\left((N_{S}+ N_{B})\ln\left(\frac{\left(N_{S}+ N_{B}\right)\left(N_{B}+\Delta N_{B}^2\right)}{N_{B}^2+\left(N_{S}+ N_{B}\right)\Delta N_{B}^2}\right)-\frac{N_{B}^2}{\Delta N_{B}^2}\ln\left(1+\frac{\Delta N_{B}^2N_{S}}{N_{B}\left(N_{B}+\Delta N_{B}^2\right)}\right)\right)}, 
\end{equation} 
where $\Delta N_{B}$ represents the systematic uncertainty associated with the SM background. 
Our previous results are based on the assumption that systematic uncertainties are well controlled. In the subsequent analysis, we investigate the effect of incorporating systematic uncertainties of $5\%$ and $10\%$.

\begin{figure}[h]
\centering
\includegraphics[height=9cm,width=16cm]{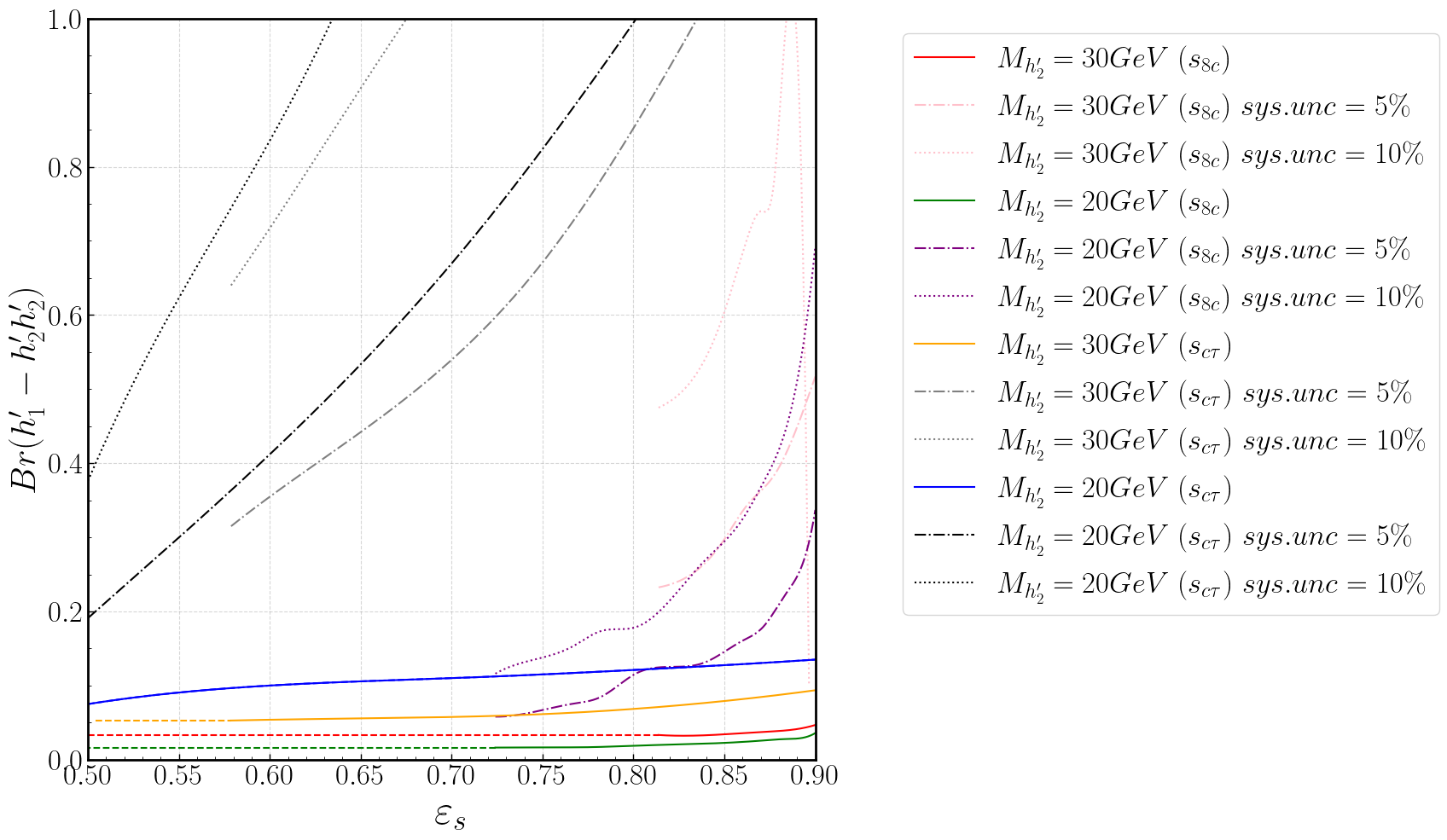}
\caption{
Future bounds of $14$ TeV LHC signals, $s_{c\tau}$ and $s_{8c}$ obtained through CNNs along with the systematic uncertainties of $5\%$ and $10\%$. The bounds are presented for different branching ratios $Br(h'_1\rightarrow h'_2 h'_2)$, with $M_{h'_2} = 20$ GeV and $30$ GeV. The dashed line represents the assumption of background-free region. 
}
\label{seresult}
\end{figure} 

The influence of systematic uncertainties is closely linked to the signal-to-background ratio $N_{S}/N_{B}$. A smaller $N_{S}/N_{B}$ value indicates a higher susceptibility to systematic uncertainties. Therefore, when $N_{S}/N_{B}$ is relatively small, even a slight increase in systematic uncertainties can have a substantial impact on the final results. The $N_{S}/N_{B}$ value is correlated with $\varepsilon_S$ in machine learning analysis, where a smaller $\varepsilon_S$ value corresponds to a larger $N_{S}/N_{B}$ value, resulting in reduced vulnerability to systematic uncertainties. Fig.~\ref{seresult} illustrates the projected future bounds of 14 TeV LHC signals obtained via CNNs for signals $s_{c\tau}$ and $s_{8c}$. A comparison is presented considering a $5\%$ and $10\%$ systematic uncertainties, considering different branching ratios $Br(h'_1 \rightarrow h'_2 h'_2)$ at $M_{h'_2} = 20$ GeV and $30$ GeV. The dashed line represents the assumption of background-free region. On the other hand, for the signal $s_{DS}$, we also consider systematic uncertainties of $5\%$ and $10\%$ for $M_{h'_2} = 20$ and $30$ GeV. 
The results in Tab.~\ref{restrictions} will change from $0.62\%$ to $1.10\%$ and $1.75\%$ for $M_{h'_2} = 20$ GeV. Similarly, when $M_{h'_2} = 30$ GeV, the results in Tab.~\ref{restrictions} will change from $0.73\%$ to $1.19\%$ and $2.06\%$.

In addition, there exists a specific parameter space where both signals $s_{c\tau}$ and $s_{DS}$ can be produced simultaneously. To discriminate between these two signal types, we employed CNN and the results are presented in Fig.~\ref{diffsignal} of Appendix~\ref{sec:app}. It is noteworthy that CNN is powerful enough to distinguish whether the fat-jet originates from cascade decays or dark showers of highly boosted $h'_1$.

Furthermore, we have investigated the effects from pre-selection and machine learning analysis for different mass spectra of $M_{\eta_d}$, $M_{\tilde{\omega}}$ and $M_{h'_2}$ in this model. First, we discuss the relation between $M_{\eta_d}$ and $M_{\tilde{\omega}}$ which will affect the efficiency of signal $s_{\text{DS}}$ in our analysis. 
We use the ratios $M_{\tilde{\omega}}/M_{\eta_d} = 1.2$, $3$ and $5$ as examples. 
Compared to the mass ratio $M_{\tilde{\omega}}/M_{\eta_d} = 1.8$ presented in Table~\ref{basiccut3} for an efficiency of $0.93\%$ ($M_{\eta_d} = 4 $ GeV) and an efficiency of $0.79\%$ ($M_{\eta_d} = 6 $ GeV), the different mass ratios of $M_{\tilde{\omega}}/M_{\eta_d} = 1.2$, $3$, and $5$ exhibit efficiencies of $1.21\%$, $0.55\%$, and $0.23\%$ for $M_{\eta_d} = 4 $ GeV, and $1.17\%$, $0.42\%$, and $0.18\%$ for $M_{\eta_d} = 6 $ GeV.   
These findings indicate that an increase in the ratio $M_{\tilde{\omega}}/M_{\eta_d}$ results in a reduction of the parameter \textit{probVector}, thereby diminishing the probability of $\tilde{\omega}$ appearing in the final state. Consequently, distinguishing the signal $s_{\text{DS}}$ from SM backgrounds becomes more challenging, resulting in a decrease in signal efficiency. 

Second, we explore the relation between $M_{\eta_d}$ and $M_{h'_2}$ which affect the efficiencies of both signals $s_{8c}$ and $s_{c\tau}$ in our analysis. We apply the following two scenarios to study this issue. We fix $M_{\eta_d} = 4$, $6$ GeV and vary the value of $M_{h'_2}$ as a first scenario. Specifically, we investigate two distinct mass relationships $M_{h'_2} = 10M_{\eta_d}$ and $M_{h'_2} = 3M_{\eta_d}$ and compare their results with that obtained for $M_{h'_2} = 5M_{\eta_d}$ using the same machine learning processing method. The performance of the machine learning analysis for classification, as presented in Appendix~\ref{sec:app}, consistently
aligns with the results obtained at $M_{h'_2} = 5M_{\eta_d}$ in the previous text. Our analysis reveals that the classification performance of signals $s_{8c}$ and $s_{c\tau}$ is not sensitive to variations in $M_{h'_2}$. Therefore, this outcome highlights the robustness
and wide applicability of our machine learning analysis for the first scenario. The second scenario is to fix $M_{h'_2} = 20$ and $30$ GeV, but vary the value of $M_{\eta_d}$. Take the same mass relationships $M_{h'_2} = 10M_{\eta_d}$ and $M_{h'_2} = 3M_{\eta_d}$ as examples, we realize that these settings will make $M_{\eta_d} \sim 2.5$ GeV or $M_{\eta_d} \sim 10$ GeV which is beyond the scope of the present study. Therefore, we only focus on the analysis for the first scenario here.

\section{Conclusion} 
\label{sec:conclu} 
The present study aims to explore the feasibility of probing the elusive Dark QCD Sector via the Higgs Portal at the LHC using advanced machine learning techniques. 
We investigate the potential of a singlet scalar mediator with a mass in the tens of GeV range that connects the dark sector and the Standard Model (SM) sector via the Higgs portal which received less attention in the literature. 
Our analysis focuses primarily on two kinds of signal processes: (1) the cascade decay of the Higgs boson into two light scalar mediators, which subsequently decay into four dark mesons, and (2) the Higgs boson decaying into two dark quarks that then undergo QCD-like shower and hadronization, leading to the production of dark mesons.

We focus on the lightest unstable dark meson, $\eta_d$, with a mass around $5$ GeV, which predominantly decays into a pair of charm quarks or tau leptons. Specifically, we investigate two benchmark points at $M_{\eta_d} = 4$ and $6$ GeV. 
We defer the investigation of the scenario where $M_{\eta_d}\sim 2.5$ GeV and the primary decay mode is $\eta_d\rightarrow gg$ to future research. In such a case, the $\eta_d$ would become a long-lived particle, resulting in an emerging jet or a multi-displaced vertex with missing energy at the LHC which needs  special treatment.
For $M_{\eta_d}\gtrsim 10$ GeV, the primary decay mode of $\eta_d$ is $b\bar{b}$. Since the analysis of highly boosted $h'_1$ to multiple $b$ quarks is already presented in Ref.~\cite{Jung:2021tym}, we do not revisit this possibility in our study. 
Therefore, we focus on a highly boosted Higgs boson produced from the above two kinds of signal processes and resulting in multiple charm quarks and tau leptons in the final state, clustered together as a fat-jet. However, distinguishing the signal from SM backgrounds poses a significant challenge, particularly in the case of cascade decays, because of similar production processes and final state characteristics. To address these challenges, we resort to a machine learning approach.

The Convolutional Neural Network (CNN) technique has been successfully applied to analyze jet images and has been found to be more effective than traditional jet substructure observables in certain scenarios. In addition, the Energy Flow Network (EFN) technique is also useful as it considers both local and global features of energy flow patterns within jets, providing a more comprehensive representation of jet structure than traditional methods that only use specific kinematic variables. Therefore, we utilize both CNN and EFN techniques to analyze the fat-jet structure, identify signal signatures, and accurately distinguish signal events from relevant background events with improved accuracy and efficiency. After comparing the CNN and EFN networks, we have observed that different network structures are more effective in extracting dynamic information ($P_T$, $\eta$, $\phi$) of the signal and background. Specifically, CNN outperforms EFN in discriminating between the signal and background due to its deeper network structure and a higher number of parameters for capturing local features of events. However, EFN has faster training and its network can interpret and visualize the contribution of each input point, providing interpretability to model decisions. Hence, EFN may give better results in future studies. Furthermore, with an increase in the number of training events in the CNN network, the classification effectiveness gradually improves, and our classifier achieves maximum classification effectiveness at 800k training events. 

The predictions of future bounds in $Br(h'_1\rightarrow h'_2 h'_2)$ for the signals $s_{c\tau}$ and $s_{8c}$ at 14 TeV LHC with an integrated luminosity of $3000 fb^{-1}$ are obtained using the training results from CNN. 
When the $\varepsilon_S$ value falls between $0.5$ and $0.9$, the projected future bounds for the branching ratio of $h'_1\rightarrow h'_2 h'_2$ in the signal $s_{c\tau}$ at $M_{h'_2}=20$ GeV and $30$ GeV are estimated to be $6.25\%-12.35\%$ and $5.28\%-9.46\%$, respectively. For the ideal signal $s_{8c}$ scenario, the estimated future bounds for the branching ratio of $h'_1\rightarrow h'_2 h'_2$ signal at $M_{h'_2}=20$ GeV and $30$ GeV are in the range of $1.63\%-3.59\%$ and $3.33\%-4.68\%$, respectively, when the $\varepsilon_S$ value falls between $0.5$ and $0.9$. 
On the other hand, the semi-visible jet signature produced by the signal $s_{DS}$ from dark showers, which results from the stable $\tilde{\omega}$ and the prompt decay of $\eta_d$ to $c\overline{c}$ and $\tau^+\tau^-$, is unique to this signal type and is absent in SM backgrounds and other signals, such as $s_{8c}$ and $s_{c\tau}$. Therefore, the application of traditional cut-flow methods allows us to estimate the future bounds for $Br(h'_1\rightarrow\varphi_d\varphi^{\dagger}_d)$ in the signal $s_{DS}$ at $M_{h'_2}$ = $20$ GeV and $30$ GeV. The estimated future bounds are found to be 0.62$\%$ and 0.73$\%$, respectively.

In order to ensure the universality of our findings, we also explored the impacts from various mass spectra in this model. For the signal $s_{\text{DS}}$, an escalating trend in the mass ratio $M_{\tilde{\omega}}/M_{\eta_d}$ gives rise to an increasingly formidable task of discriminating the signal $s_{\text{DS}}$ from the SM backgrounds and causes a substantial reduction in the signal's efficiency. The outcome underscores the robustness and wide applicability of our machine learning algorithm in accurately analyzing both signal and background components in experimental contexts.

In summary, our study demonstrates the effectiveness of using CNN and EFN techniques to enhance the detection accuracy when probing the Dark QCD sector through the Higgs Portal with a singlet scalar mediator in the tens of GeV mass range. We have estimated the future bounds for the branching ratio of $h'_1\rightarrow h'_2 h'_2$ at $14$ TeV and 3000 $fb^{-1}$, which can reach around $10\%$ in this model. In addition, the outcomes of our study reveal that CNN and EFN are proficient in distinguishing the signal arising from cascade decays and that from dark showers. Consequently, our investigation can extend the search for the model parameter space that encompasses both aforementioned signal signatures.

\section*{Acknowledgments} 
We thank Kevin Pedro and Kepan Xie for helpful discussions. This work is supported by the National Natural Science Foundation of China (NNSFC) under grants No. 12275134 and Project 12047503 supported by NSFC.

\appendix

\section{Mass Ratio Effects for the Signal $s_{c\tau}$ and CNN Classification of Signals $s_{c\tau}$ and $s_{DS}$}
\label{app:2sig}

\begin{figure}[h]
\centering
\includegraphics[height=8cm,width=8cm]{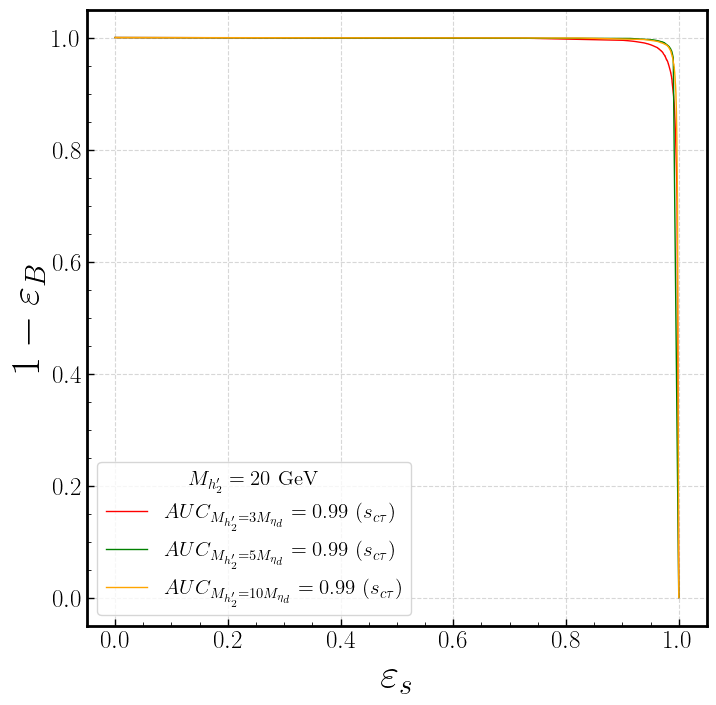}
\includegraphics[height=8cm,width=8cm]{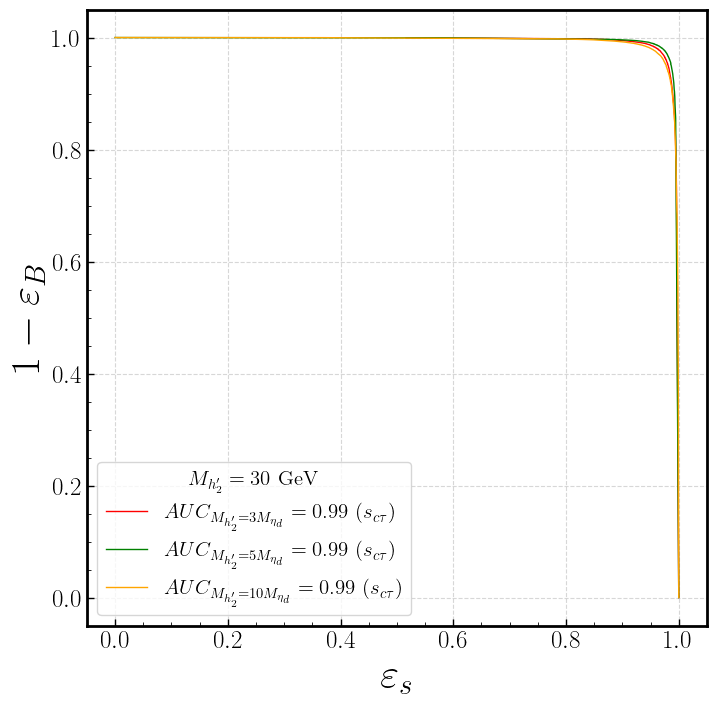}
\caption{
The ROC curves for CNNs with different mass ratios, specifically $M_{h'_2}/M_{\eta_d} = 3$, $5$, and $10$, along with their corresponding AUC values. The plots correspond to $M_{h'_2} = 20$ GeV (left panel) and $M_{h'_2} = 30$ GeV (right panel). 
}
\label{diffmassauc}
\end{figure}

In this Appendix, we first present the analysis in Fig.~\ref{diffmassauc} to reveal a noteworthy consistency in the CNN classification performance across diverse mass ratios, including $M_{h'_2}/M_{\eta_d} = 3$, $5$ and $10$. This suggests that the machine learning classification exhibits a robust behavior and is not excessively affected by variations in the mass ratio. Therefore, our machine learning methods exhibit the potential to deliver enhanced results for signals $s_{8c}$ and $s_{c\tau}$ characterized by different mass ratios, underscoring their broader applicability.

\label{sec:app}
\begin{table}[h]
\small
    \centering
    \begin{tabular}{|c|c|c|}
    \hline
    Cross section (fb) & $s_{DS}$ & $s_{c\tau}$ \\
    \hline
    \makecell[c]{$P_{T}^{l} > 25 $ GeV,  \\$|\eta_{l_{}}|<2.5$, $P_{T}^{l+\slashed{E}_{T}} > 200 $ GeV } &28.98 & 6.92 \\
    \hline
    \makecell[c]{   $  P_{T}^{J}>50$ GeV,  \\ $|\eta_{J_{}}|<2.5$  }& 14.73&6.46  \\
    \hline
    efficiency & 7.35$\%$ & 3.58 $\%$\\
    \hline
    \end{tabular}
        \caption{The pre-selection cut-flow cross sections (in fb) of signals $s_{DS}$ and $s_{c\tau}$ which are mentioned in the main text at $14$ TeV LHC. Here the benchmark point $M_{h'_2} = 30$ GeV, $M_{\eta_d} = 6$ GeV is considered.}     
    \label{siffsignal}
\end{table}

We then summarize some details for the CNN classification effect of signals $s_{c\tau}$ and $s_{DS}$ which are mentioned in the main text. For the same purpose as in Sec.~\ref{sec:presel} and to maximize the number of remaining events for machine learning analysis, we preprocessed both signals $s_{c\tau}$ and $s_{DS}$ and presented the pre-selection cut-flow cross sections in Tab.~\ref{siffsignal}. 
Since the signal $s_{DS}$ has much more $\slashed{E}_{T}$ in the final state than the signal $s_{c\tau}$, after the same event selection $P_{T}^{l+\slashed{E}_{T}} > 200$ GeV, the signal $s_{DS}$ has more events left such that the efficiency of the signal $s_{DS}$ is larger than the signal $s_{c\tau}$ in Tab.~\ref{siffsignal}.

\begin{figure}[h]
\centering
\includegraphics[height=8cm,width=8cm]{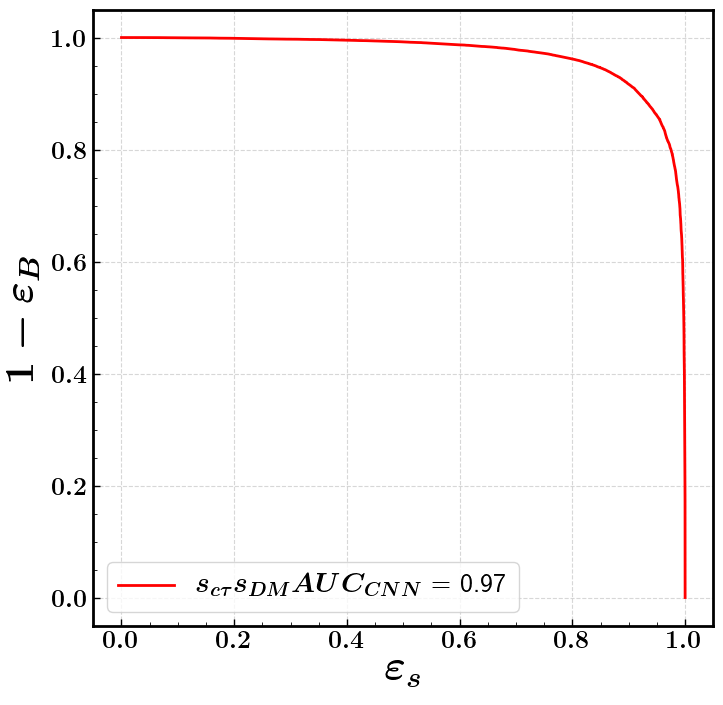}
\includegraphics[height=8cm,width=8cm]{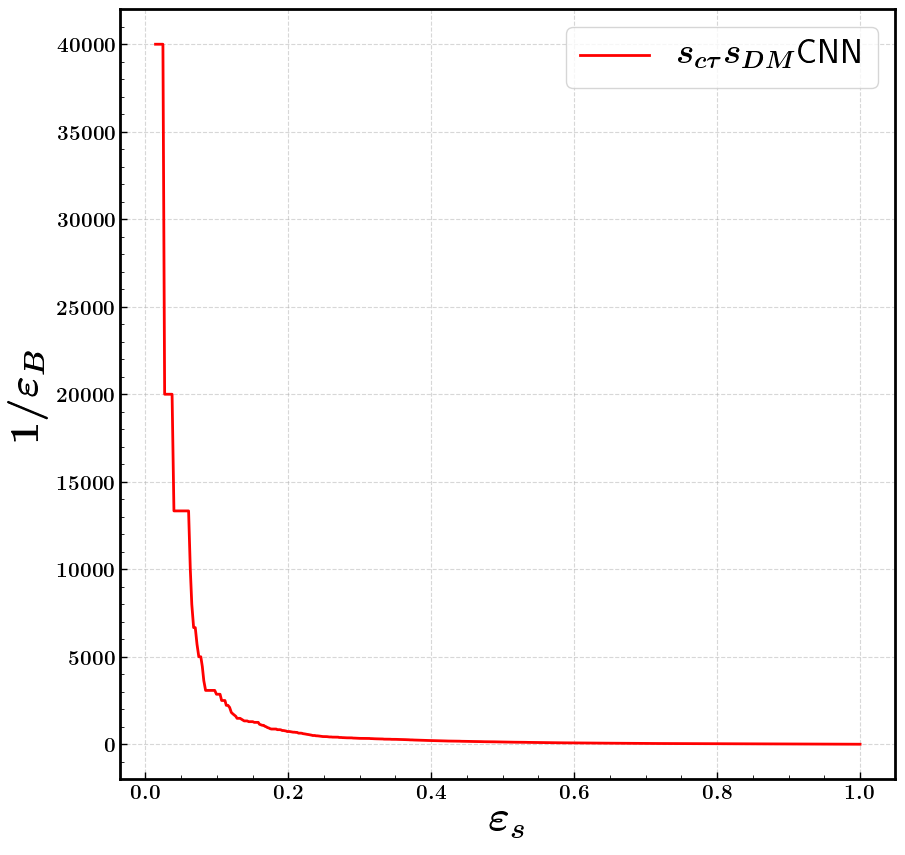}
\caption{The left panel shows the ROC curve and corresponding AUC value of CNN's classification performance in distinguishing between different signal types, namely signals $s_{DS}$ and $s_{c\tau}$, while the right panel displays the relationship between $\varepsilon_S$ and $\frac{1}{\varepsilon_B}$.}
\label{diffsignal}
\end{figure}
 During data processing, we treat the $s_{DS}$ as a background sample and input it into the CNN along with the $s_{c\tau}$ as a signal sample. The jet images of the leading jet for these two processes have been shown in Fig.~\ref{pixell3}. Because of the powerful pattern recognition ability of the CNN, it can accurately identify whether the signal was generated through exotic cascade decays or dark showers from highly boosted $h'_1$. The results are depicted in Fig.~\ref{diffsignal}.  
The ROC curve shown in  Fig.~\ref{diffsignal} demonstrates the effectiveness of our CNN network in distinguishing between $s_{DS}$ and $s_{c\tau}$ events. Specifically, the AUC is $0.97$, indicating a high level of performance. Furthermore, when the signal efficiency ($\varepsilon_S$) is $0.1$, the inverse of the background efficiency ($\frac{1}{\varepsilon_B}$) can reach  $10^3$, indicating a significant reduction in the number of false positives. This demonstrates that our CNN network is capable of effectively distinguishing between two different signal types.

\section{BDT-based Jet Structure Classification Performance for Signals $s_{c\tau}$ and $s_{8c}$ }

In this Appendix, we compare the performance of two machine learning techniques, namely the Boosted Decision Tree (BDT) and Convolutional Neural Network (CNN) methods, for distinguishing between signal and background events. We utilized BDT analysis on the jet structure, using suitable jet substructure observables based on Ref.~\cite{Gallicchio:2012ez,CMS:2021dzg}, and implemented the same preselection cuts described in Sec.~\ref{sec:presel}. For the BDT training, we employed the TMVA package~\cite{Hocker:2007ht} in ROOT, using the "BDTD" option with $220$ trees, a minimum of $2.5\%$ training events in each node, a maximum tree depth of $2$, and default values for other parameters. To ensure the same number of events in each category, we trained and tested the BDT classifier on 8k signal events and 18k background events. We conducted a Kolmogorov-Smirnov test on the BDT analysis and set a threshold of $0.01$ to avoid overfitting~\cite{Wang:2021uyb}.

In our investigation, we utilized a broad range of jet substructure observables, such as the girth of the jet, the fragmentation distribution of the jet ($p_{T}D=\frac{\sqrt{\sum_{i}p_{T,i}^{2}}}{\sum_{i}p_{T,i}}$) in Refs.~\cite{CMS:2013kfa,Lee:2019ssx}, the shape of the jet, jet invariant mass ($M_J$) in Ref.~\cite{Gallicchio:2012ez}, the number of constituent particles $sumnumber$ in a given jet in Ref.~\cite{CMS:2013kfa}, the angle between the jet and the missing transverse momentum ($\Delta\phi(J_{1},\slashed{E}_{T})$) in Ref.~\cite{CMS:2021dzg}, the energy of the monojet ($E_{\text{jet}}$) in Ref.~\cite{CMS:2021dzg}, and N-subjettiness in Refs.~\cite{Thaler:2010tr,Wang:2021uyb}. The definitions of these observables are provided below:
\begin{itemize} 
\item Jet shapes \\ 
Jets possess a conical structure that can be projected onto the 
$(\eta, \phi)$ plane. The width of a jet can be measured by the two principal components of the second moment of the constituent distribution in the $(\eta, \phi)$ plane. The shape of a jet can be approximated by an ellipse, which is described by its major and minor axes as well as the orientation of the major axis in the plane. To construct a $2\times 2$ symmetric matrix $M$, the following elements are used:
\begin{equation}
 M_{11}=\sum_{i}p_{T,i}^{2}\Delta\eta_{i}^{2},\quad M_{22}=\sum_{i}p_{T,i}^{2}\Delta\phi_{i}^{2},\quad M_{12}=M_{21}=-\sum_{i}p_{T,i}^{2}\Delta\eta_{i}\Delta\phi_{i}.
\end{equation} 
In the above equations, the variables $\Delta\eta$ and $\Delta\phi$ represent the disparities in the pseudorapidity and azimuthal angle, respectively, between each constituent particle and the axis of the jet. The eigenvalues ($\lambda_1$ ,$\lambda_2$) of the matrix $M$ can be used to determine the major and minor axes of the jet, denoted by $\sigma_1$ and $\sigma_2$, respectively~\cite{CMS:2013kfa}. These are given by 
\begin{equation}
 \sigma_{1}=\sqrt{\frac{\lambda_{1}}{\sum_{i}p_{T,i}^{2}}},\quad \sigma_{2}=\sqrt{\frac{\lambda_{2}}{\sum_{i}p_{T,i}^{2}}}.
\end{equation} 

\item Linear Radial Moment (Girth) \\ 
The girth of a jet is computed by summing the transverse momentum deposits within the jet, weighted by their distances from the jet axis~\cite{Gallicchio:2010sw,Gallicchio:2011xq}. Specifically, the girth $g$ is defined as
\begin{equation}
g = \sum_{i \in \text{jet}} \frac{p_{T,i} r_{i}}{p_{T}^{\text{jet}}}, 
\end{equation} 
where $r_{i}$ is the distance of this particle from the jet axis, and $p_{T}^{\text{jet}}$ is the total transverse momentum of the jet. The distance $r_i$ is given by $r_{i} = \sqrt{(\eta_{i}-\eta_{0})^{2}+(\phi_{i}-\phi_{0})^{2}}$, where $(\eta_{0},\phi_{0})$ denotes the jet coordinates in the frame with the interaction point of the proton-proton collision as the origin.

\item N-subjettiness \\ 
The N-subjettiness is a measure that quantifies the degree to which a jet can be decomposed into N subjects. It is given by the formula, 
 \begin{equation}
 \tau_{N}=\frac{\sum_{i}p_{T,i}min \left\{\Delta R_{1,i},\Delta R_{2,i},...,\Delta R_{N,i}\right\}}{\sum_{i}p_{T,i}R_{0}}
\end{equation} 
$\Delta R_{J,i}$ denotes the distance in the rapidity-azimuth plane between a candidate subject $J$ and a constituent particle $i$, and $R_{0}$ is the characteristic jet radius used in the original jet clustering algorithm~\cite{Thaler:2010tr,Wang:2021uyb}.
In this study, we are interested in examining the ratios of N-subjettiness variables, namely $\tau_{21}=\tau_{2}/\tau_{1}$, $\tau_{31}=\tau_{3}/\tau_{1}$, and $\tau_{32}=\tau_{3}/\tau_{2}$, to investigate the properties of the jet substructure~\cite{Thaler:2010tr,Wang:2021uyb}. Specifically, these ratios provide information about the relative separation of the subjects within the jet, as $\tau_{21}$ indicates the degree of two-pronged structure in the jet, $\tau_{31}$ measures the presence of three-pronged structure, and $\tau_{32}$ provides insight into the relative separation among three subjects.
\end{itemize}

\begin{figure}[h]
\centering
\includegraphics[height=8cm,width=8cm]{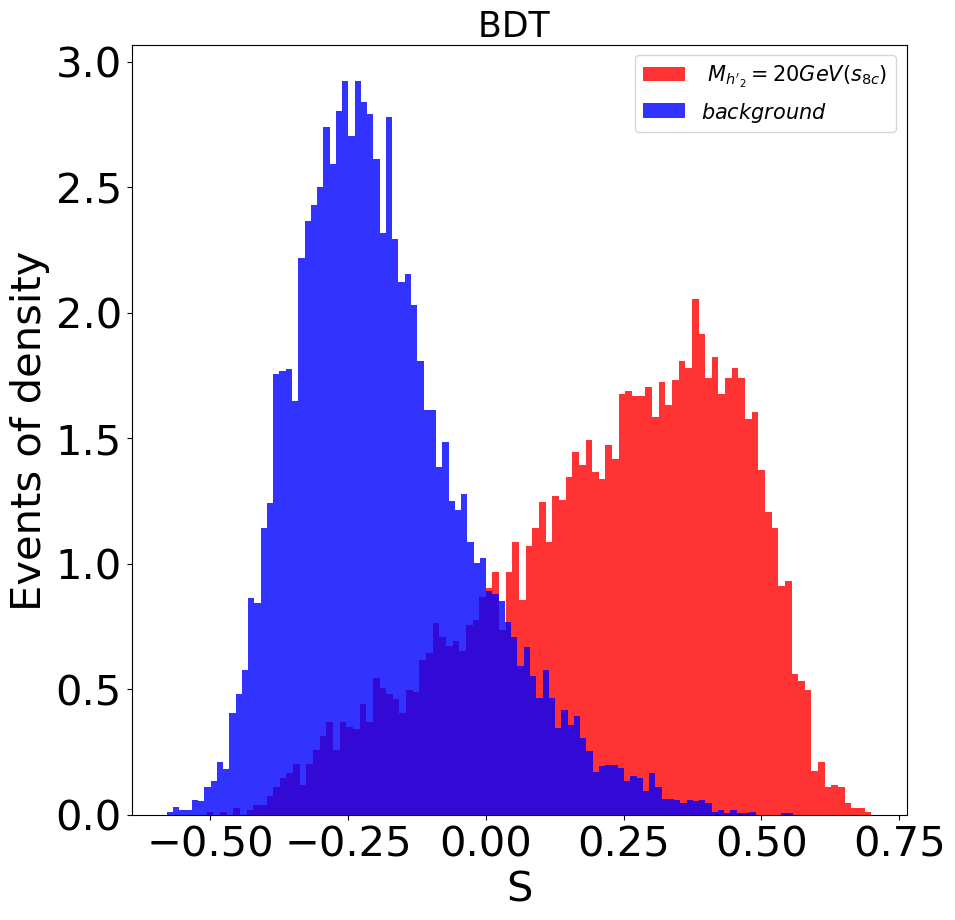}
\includegraphics[height=8cm,width=8cm]{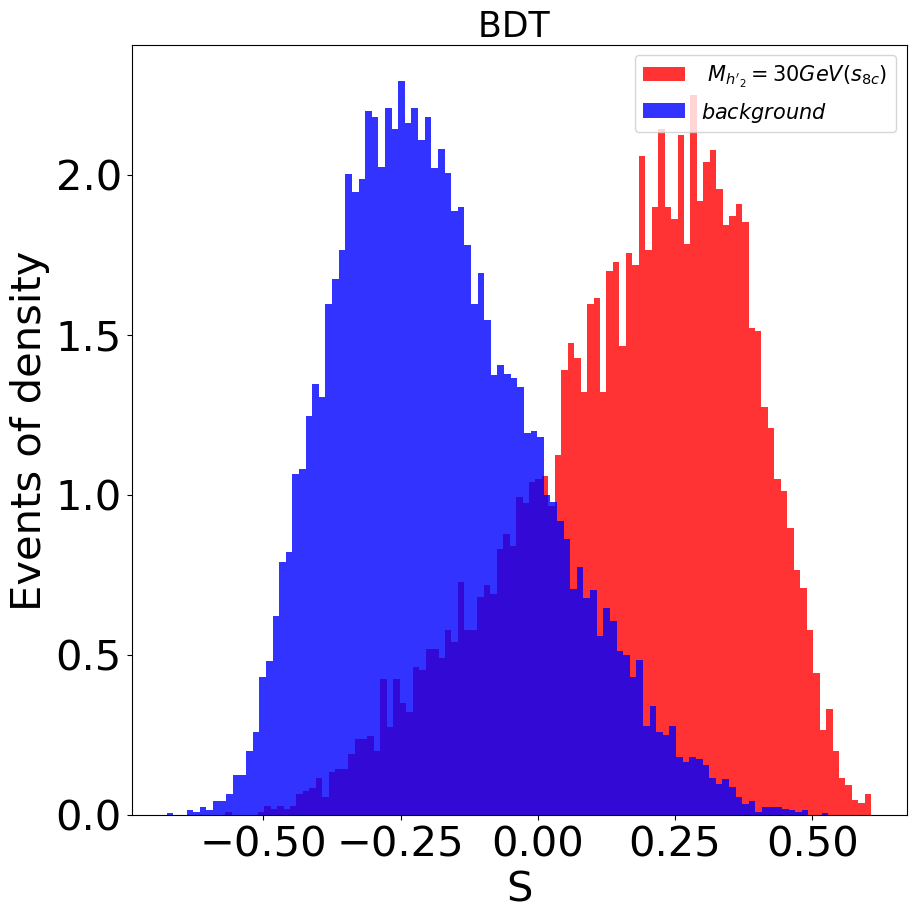}
\includegraphics[height=8cm,width=8cm]{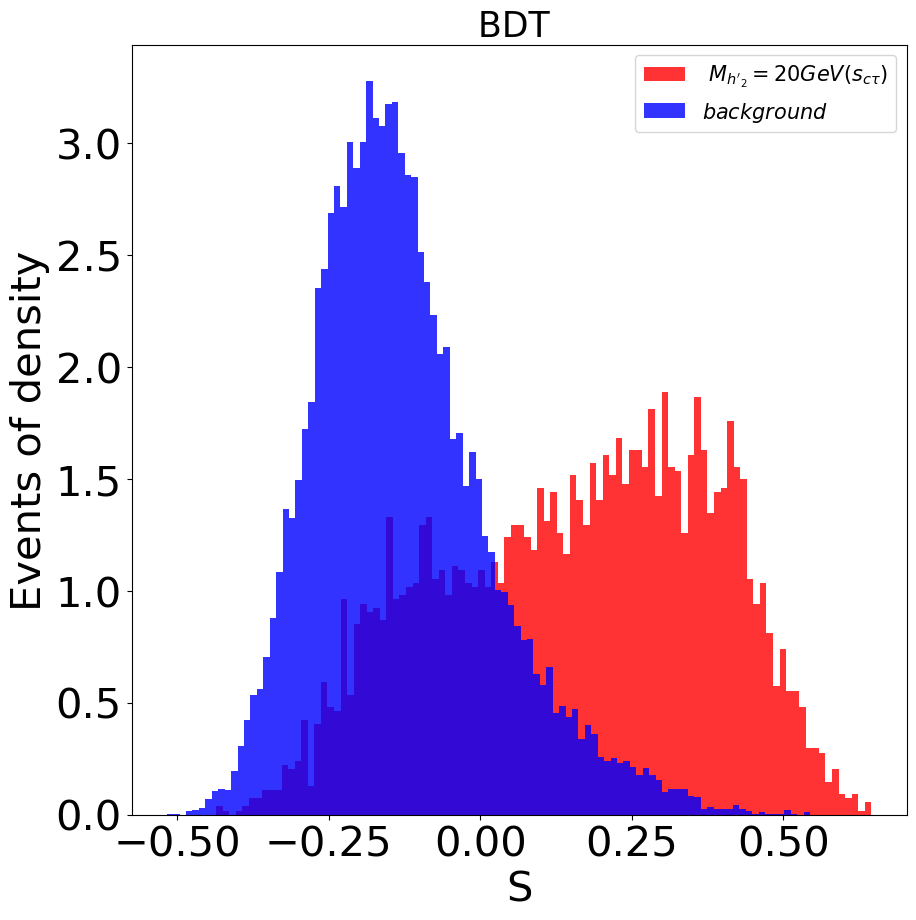}
\includegraphics[height=8cm,width=8cm]{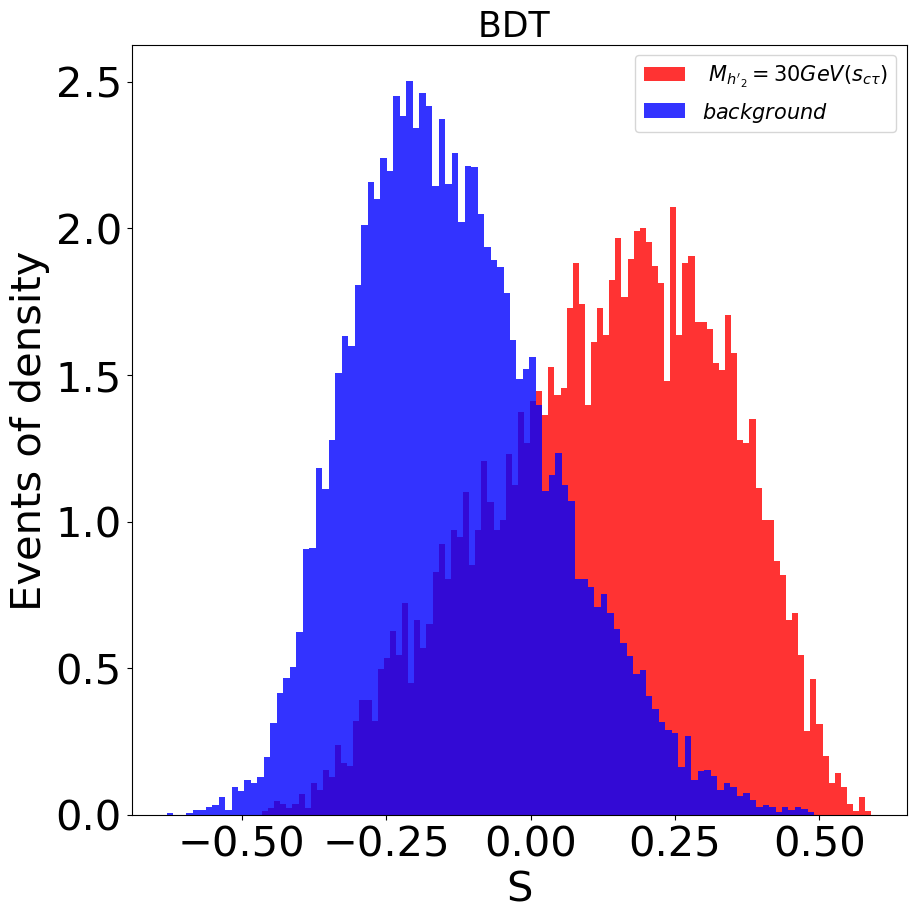}
\caption{This figure displays the classification performance of the BDT for signals $s_{8c}$ (upper), $s_{c\tau}$ (lower) and background test sets, with $M_{h'_2}=20$ GeV on left panels and $M_{h'_2}=30$ GeV on right panels.}
\label{bdtresults}
\end{figure} 

\begin{figure}[htbp]
\centering
\includegraphics[height=4cm,width=4cm]{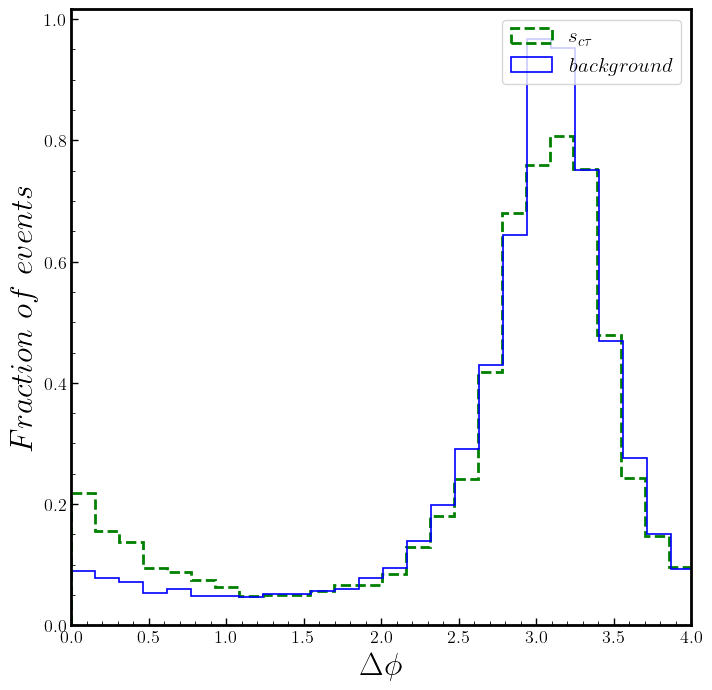}
\includegraphics[height=4cm,width=4cm]{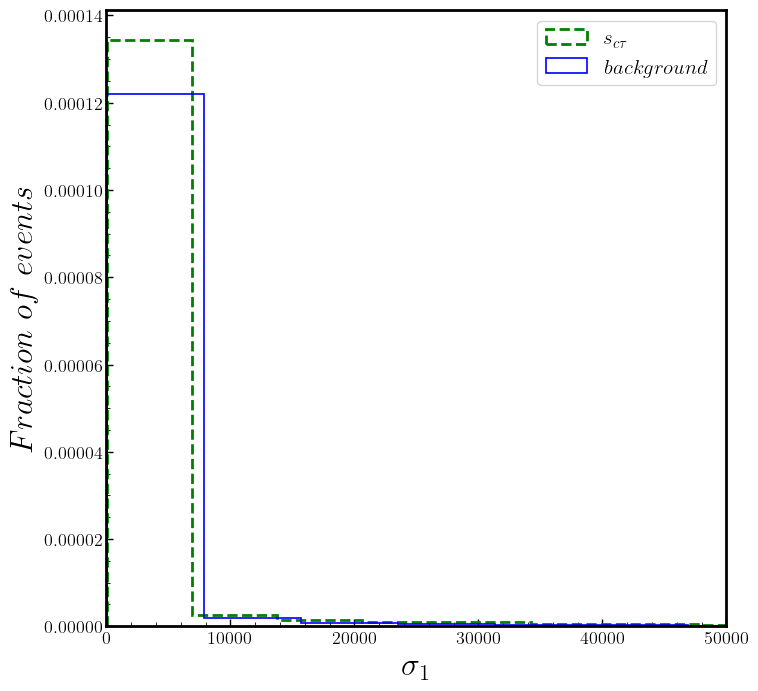}
\includegraphics[height=4cm,width=4cm]{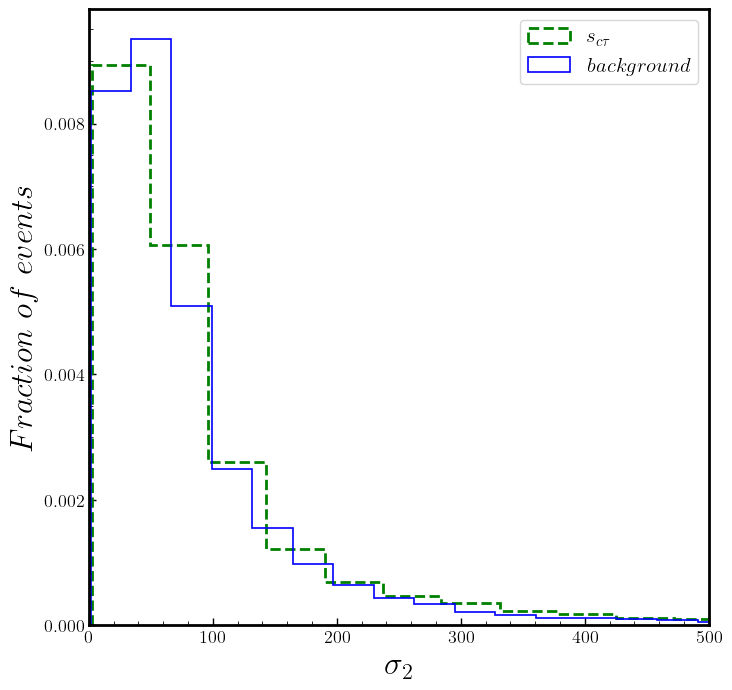}
\includegraphics[height=4cm,width=4cm]{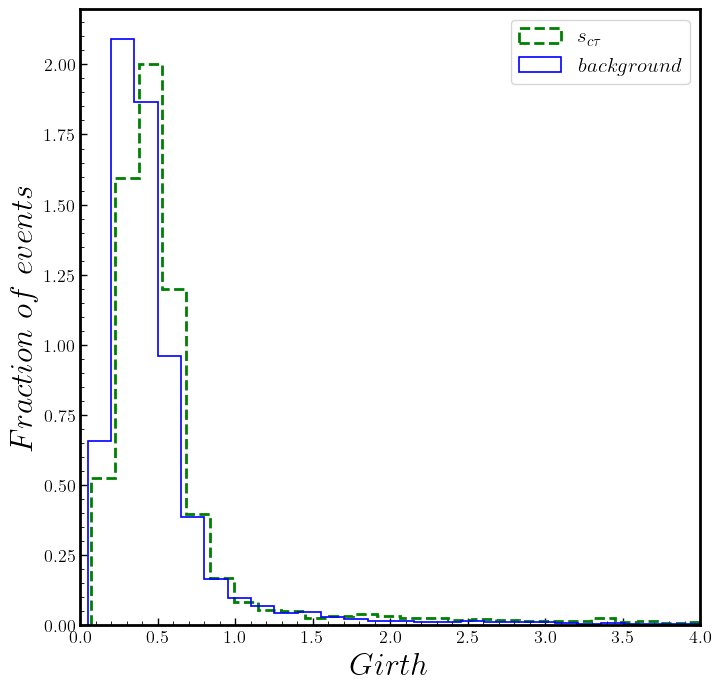}

\caption{The kinematic distributions of $\Delta\phi(J_{1},\slashed{E}_{T})$ (first column), $\sigma_1$ (second column), $\sigma_2$ (third column), and Grith (fourth column) for the signal $s_{c\tau}$ and relevant SM backgrounds at $14$ TeV LHC are presented. These observables are used as inputs to evaluate the classification performance of the BDT method.}
\label{distributionbdt}
\end{figure}

The discriminative performance of the BDT classifier in separating signal and background events is shown in Fig.~\ref{bdtresults}. The BDT response values are plotted on the horizontal axis, with the background and signal events represented by the blue and red regions, respectively. As demonstrated in Fig.~\ref{bdtresults}, the background events exhibit lower response values, while the signal events exhibit higher response values. Moreover, Fig.~\ref{bdtresults} reveals that the signal $s_{8c}$ is more widely distributed compared to signal $s_{c\tau}$, indicating that the BDT discrimination performance of $s_{8c}$ is superior.
A comparison of Fig.~\ref{bdtresults} with Fig.~\ref{results} reveals that the BDT approach based on jet substructure observable variables exhibits inferior performance when compared to CNN and EFN. This outcome may be influenced by several factors. 
Based on the distribution of input variables, we have observed that the observable variables of the jet substructure presented in Fig.~\ref{distributionbdt}, including Grith, $\Delta\phi(J_{1},\slashed{E}_{T})$, $E_{\text{jet}}$, $\sigma_1$ and $\sigma_2$, fail to effectively discriminate between the signal and background. Since it is impractical to exhaustively enumerate all possible observable variables of jet substructure for BDT analysis, and the BDT network is unable to automatically adjust learning parameters like CNN for achieving optimal performance, we have conjectured that CNN's results may outperform those of BDT. As a result, we mainly focus on using the CNN technique for the fat-jet structure in this work. 

\begin{figure}[htbp]
\centering
\includegraphics[height=4cm,width=4cm]{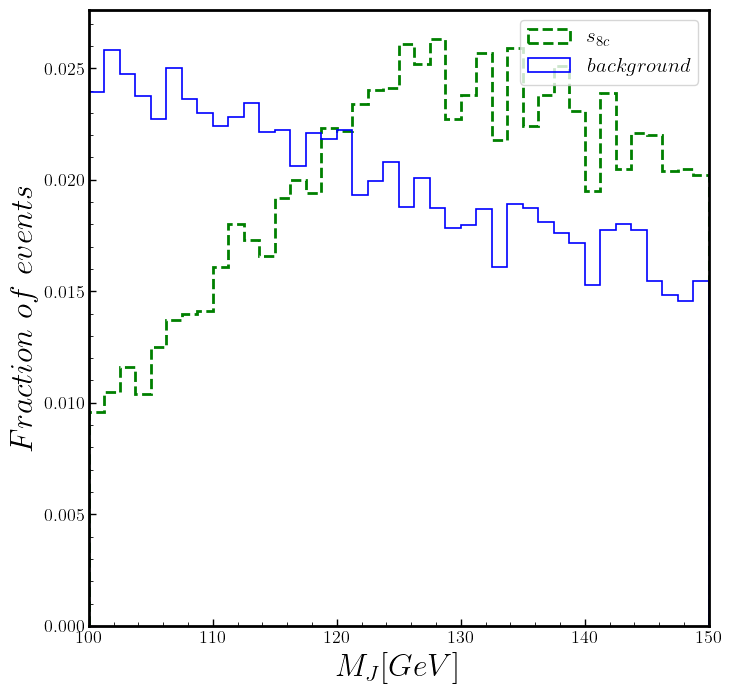}
\includegraphics[height=4cm,width=4cm]{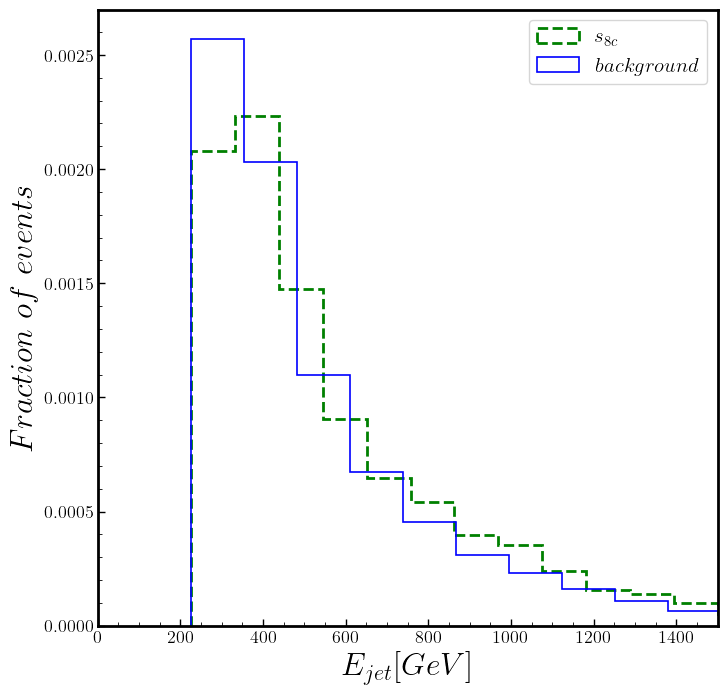}
\includegraphics[height=4cm,width=4cm]{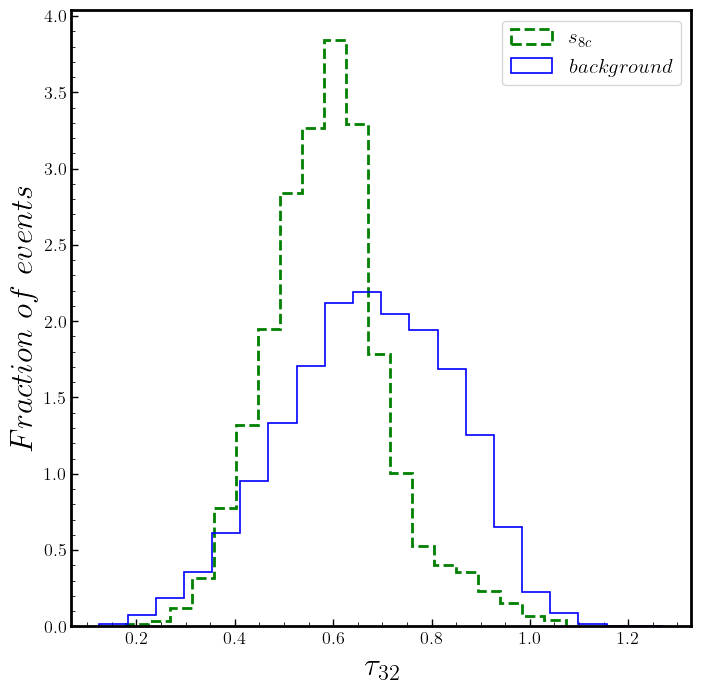}
\includegraphics[height=4cm,width=4cm]{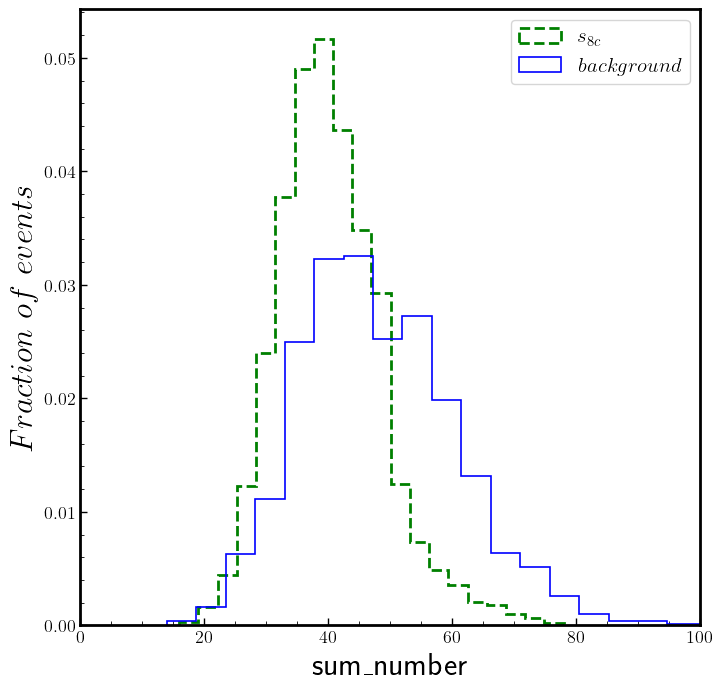}
\includegraphics[height=4cm,width=4cm]{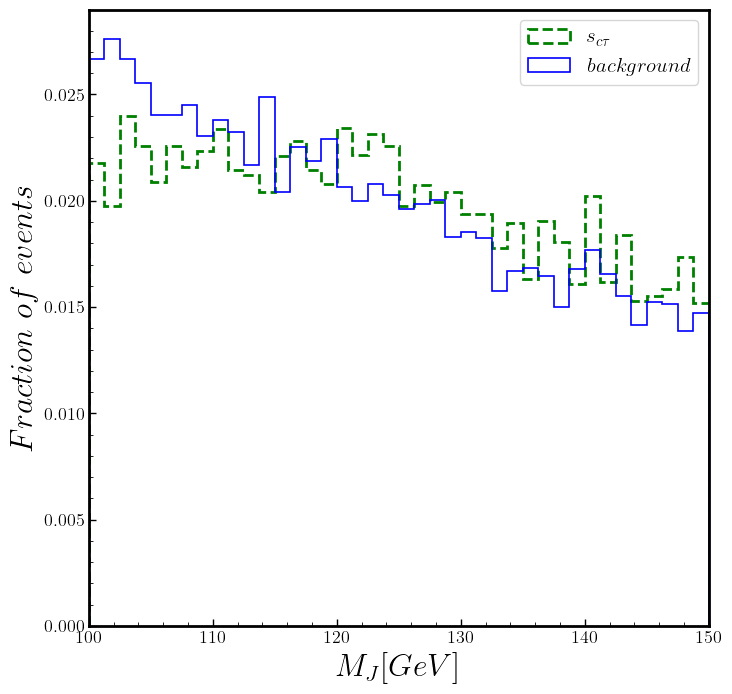}
\includegraphics[height=4cm,width=4cm]{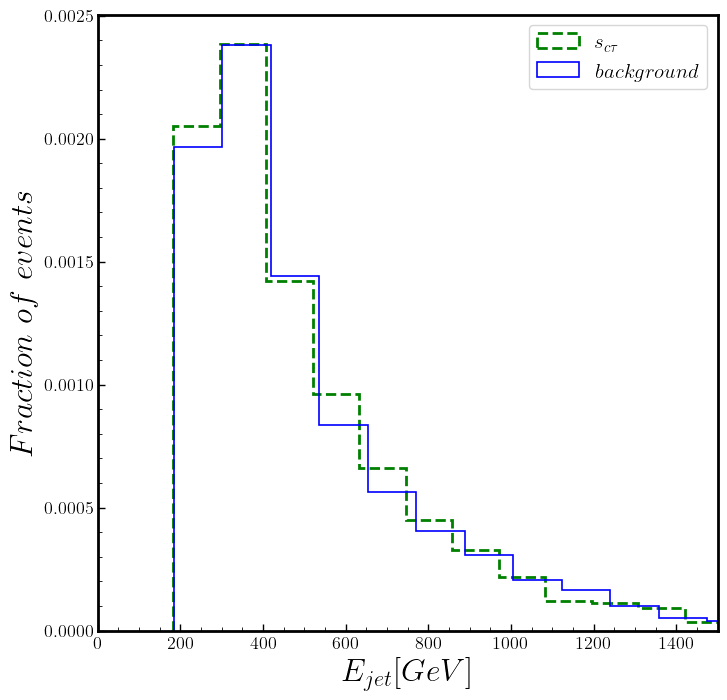}
\includegraphics[height=4cm,width=4cm]{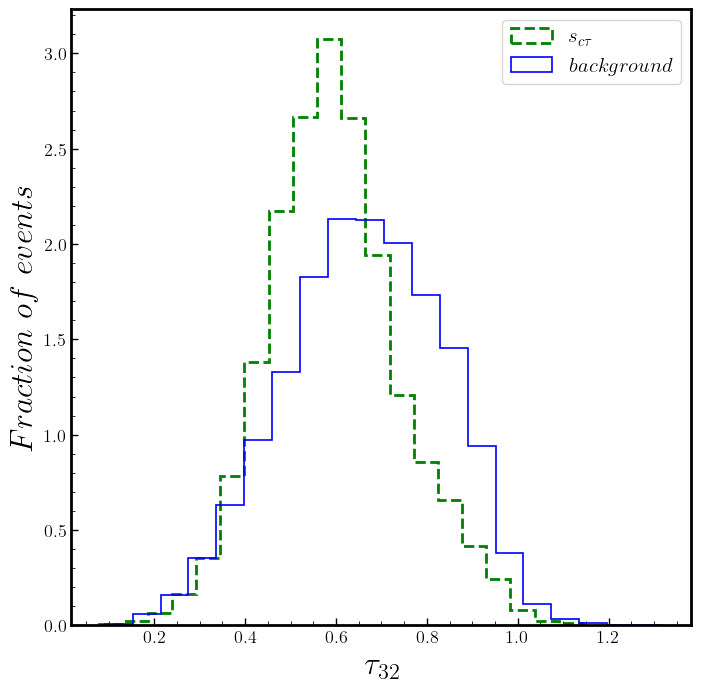}
\includegraphics[height=4cm,width=4cm]{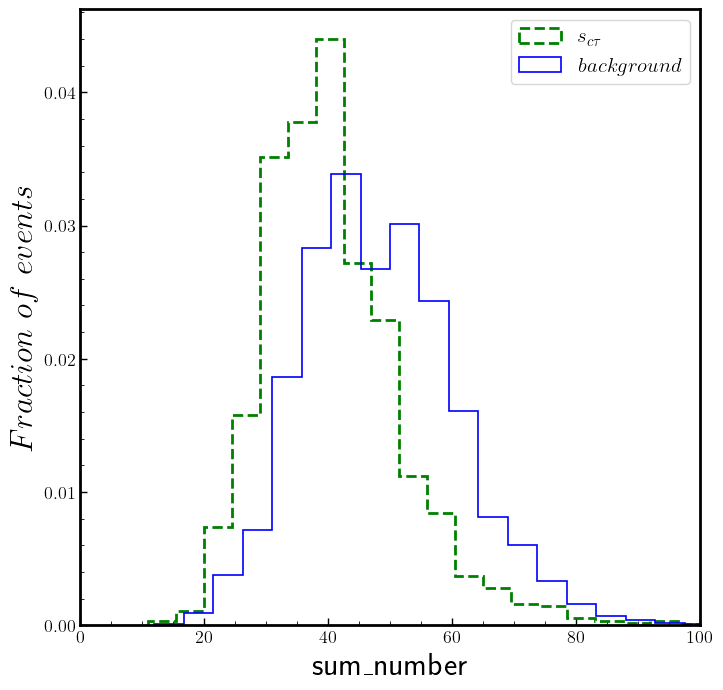}

\caption{We compared the kinematic distributions of $M_J$ (first column), $E_{\text{jet}}$ (second column), $\tau_{32}$ (third column), and $sum number$ (fourth column) for two signals, $s_{c\tau}$ (upper panels) and $s_{8c}$ (lower panels), as well as relevant SM backgrounds at the $14$ TeV LHC. These observables were used as inputs to evaluate the classification performance of the BDT method. }
\label{distributionbdtdiff}
\end{figure}

Additionally, the BDT discrimination performance of $s_{8c}$ is superior compared to $s_{c\tau}$, as revealed by Fig.~\ref{bdtresults}. This is due to the differences in the BDT input variables between the two signals, as depicted in Fig.~\ref{distributionbdtdiff} and explained in Sec.~\ref{sec:result}. The leading jet mass distribution of $s_{c\tau}$ is smaller than that of $s_{8c}$, which weakens the discrimination power between $s_{c\tau}$ and the background as shown in the $M_J$ distribution of Fig.~\ref{distributionbdtdiff}. Moreover, the $\eta_d$ particle in $s_{c\tau}$ can decay into $\tau^+\tau^-$, resulting in the possible production of secondary jets that contain high-$p_T$ $\tau$ leptons. These secondary jets may contain more constituent particles, resulting in a larger number of particles $sum number$ in the jet than in $s_{8c}$. Similarly, the fat-jet from $s_{c\tau}$ may have more substructure, leading to larger values of $\tau_2$ and $\tau_3$, and thus, a larger value of $\tau_{32}$ than that from $s_{8c}$. These differences lead to a weaker discrimination power between $s_{c\tau}$ and the background in the BDT, compared to $s_{8c}$, as shown in Fig.~\ref{bdtresults}.

\label{sec:app2}
\begin{table}[h]
\small
    \centering
    \begin{tabular}{|c|c|c|}
    \hline
    \makecell[c]{Future bounds in \\ $Br(h'_1\rightarrow XX)$}& $s_{c\tau}$ ($h'_1\rightarrow h'_2 h'_2$) & $s_{8c}$ ($h'_1\rightarrow h'_2 h'_2$) \\
    \hline
    \makecell[c]{ $M_{h'_2} = 20$ GeV \\ }&    88.69$\%$ &13.88$\%$     \\
    \hline
    \makecell[c]{ $M_{h'_2} = 30$ GeV \\  }& 59.35$\%$& 23.88$\%$  \\
  \hline
    \end{tabular}
        \caption{The future bounds in $Br(h'_1\rightarrow XX)$ for $s_{c\tau}$ ($h'_1\rightarrow h'_2 h'_2$) and $s_{8c}$ ($h'_1\rightarrow h'_2 h'_2$) at 14 TeV LHC with $\mathcal{L} = 3000 fb^{-1}$ using BDT. In the final column, $Br(\eta_d\rightarrow c\overline{c})=100\%$ is assumed as an ideal scenario for $s_{8c}$ as a comparison. }    
    \label{BDTrestrictions}
\end{table}

Tab.~\ref{BDTrestrictions} shows the projected bounds on the branching ratio of $h'_1\rightarrow h'_2 h'_2$ using the BDT method within the jet structure analysis framework for signals $s_{c\tau}$ and $s_{8c}$. These bounds are based on the expected results from $14$ TeV LHC with an integrated luminosity of $\mathcal{L} = 3000 fb^{-1}$. For the signal $s_{c\tau}$, the expected lower bounds on the branching ratio of $h'_1\rightarrow h'_2 h'_2$ are $88.69\%$ and $59.35\%$ for $M_{h'_2}=20$ GeV and $30$ GeV, respectively. In contrast, for the signal $s_{8c}$, the expected lower bounds on the branching ratio of $h'_1\rightarrow h'_2 h'_2$ are $13.88\%$ and $23.88\%$ for $M_{h'_2}=20$ GeV and $30$ GeV, respectively. Note that the discrimination power of $s_{c\tau}$ is significantly weaker than that of $s_{8c}$ in the BDT approach due to two main reasons. Firstly, the preselection-induced cross-section difference, as discussed in Sec.~\ref{sec:result}, leads to a significant performance gap. On the other hand, the discrimination power of $s_{c\tau}$ with respect to the background is weaker than that of $s_{8c}$ in the BDT input variables, further reducing its discrimination performance.

%\bibliography{sample}
\end{document}